\documentclass[referee]{raa}
\usepackage{CJK}
\usepackage{graphicx,times}
\usepackage{natbib}
\usepackage{amssymb,amsmath}
\usepackage{nameref}
\bibpunct{(}{)}{;}{a}{}{,}
\usepackage[pagebackref=true]{hyperref}
\usepackage{caption}
\usepackage{subcaption}
\usepackage{float}
\usepackage{booktabs}
\usepackage{bm}
\usepackage{threeparttable}

\begin{document}

\begin{CJK*}{UTF8}{gbsn}

\title{LISC Catalog of Open Clusters.III. 83 Newly found Galactic disk open clusters using Gaia EDR3}

%   \volnopage{Vol.0 (20xx) No.0, 000--000}      %%preserved for Editor. DOn't remove!
%   \setcounter{page}{1}          %%starting page, preserved for Editor. DOn't remove!

   \author{Huanbin Chi  ({\CJKfamily{gbsn}迟焕斌)}
      \inst{1,2}
   \and Feng Wang
   ({\CJKfamily{gbsn}王锋)}
      \inst{2,3}
   \and Zhongmu Li
   ({\CJKfamily{gbsn}李忠木)}
      \inst{4}
   }


   \institute{School of Management and Economics, Kunming University of Science and Technology, Kunming, China \\
        \and
             Center for Astrophysics, Guangzhou University, Guangzhou, China\\
        \and Peng Cheng Laboratory, Shenzhen, 518000, China\\
        \and
             Dali University, Institute of Astronomy, Dali, China, {\it zhongmuli@126.com}\\
\vs\no
   {\small Received 20xx month day; accepted 20xx month day}}

\abstract{ As groups of coeval stars born from the same molecular cloud, an Open cluster (OC) is an ideal laboratory for studying the structure and dynamical evolution of the Milky Way.
The release of High-Precision Gaia Early Data Release 3 (Gaia EDR3) and modern machine-learning methods offer unprecedented opportunities to identify OCs. In this study,
we extended conventional HDBSCAN (e-HDBSCAN)  for searching for new OCs in Gaia EDR3. A pipeline was developed based on the parallel computing technique to blindly search for open clusters from  Gaia EDR3 within Galactic latitudes $\left| b  \right|$ $<$25 $^\circ$. As a result, we obtained 3787 star clusters, of which 83 new OCs were reported after cross-match and visual inspection. At the same time, the main star cluster parameters are estimated  by  colour-magnitude diagram fitting. The study significantly increases the sample size and physical parameters of open clusters in the catalogue of OCs. It shows the incompleteness of the census of OCs across our Galaxy.
}

   \authorrunning{Chi et al.}            %author_head in even pages
   \titlerunning{LISC Catalog of Open Clusters.III}  % title_head in odd pages

   \maketitle

\section{Introduction}
\label{sect:intro}

An open cluster (OC) is a star cluster in which all member stars are gravitationally bound to each other.
In the Milky Way, most of such clusters are found in the disc region.
Their formation is thought to be related to the structure of the Galaxy.
The stars of an OC usually form in the same molecular cloud and therefore have approximately the same age and metallicity.
Thus the member stars of an OC distribute on an isochrone in the colour-magnitude diagram (CMD).
This can supply some clues to the studies of the evolution of stars.
Therefore, it is essential to find OCs in the Galaxy and study their stellar properties in detail.

Many efforts have been made to compile some catalogues of OCs, particularly after the publication of the data releases of Gaia (DR2, EDR3, DR3) \citep{Lindegren2018,GaiaCollaboration2021,GaiaCollaboration}.
The data of the Gaia satellite supply the celestial positions, parallaxes, and proper motions ($l$, $b$, $\varpi$, $\mu_\alpha$ and $\mu_\delta$), in three photometric bands (G, G$_{BP}$ and G$_{RP}$) for more than 1.8 billion  sources. It gives us a golden opportunity to search for OCs.
Thousands of new OC candidates are found by different methods, and this enlarges the sample of Galactic OCs significantly. In addition,  they have been also useful to discard several unlikely OCs of them. 
For example, \citet{Cantat-Gaudin2018} discovered 60 new OCs using the data of Gaia DR2.
After that, \citet{Cantat-Gaudin2019,Sim2019,Liu&Pang2019} found 41, 207 and 76 new OCs in 2019, respectively.
\citealt{Castro-Ginard2021,He2021,Ferreira2020, Hunt2021}  identified 582, 74, 34, and 41 new OC candidates respectively.
Recently, \citealt{Hao2022} used a sample-based clustering search method with a high spatial resolution to hunt 704 potential open clusters. \citealt{He2022} and \citealt{He2022ApJS} reported 541 new open cluster candidates and 46 new nearby star clusters  at high-Galactic-latitude regions.
One such typical work is the series works done by \citealt{CG2018}, \citealt{CG2018A}, \citealt{CG2019}, \citealt{CG2019A}, \citealt{CG2020}, \citealt{Cantat-Gaudin2019}, \citealt{CG2020B}, \citealt{CG2022}, which method based on clustering algorithm K-means or DBSCAN  and  more than 3, 000 OCs are reported (hereafter CG series).
However, most of the OCs in the Galaxy is still not discovered, and the results are affected a lot by the method that is used for searching for OCs.
Younger populations often have complex internal structures such that there is no single way to decompose them into clusters according to~\citet{Krumholz2019}. 
It is valuable to try different methods to search the Galactic OCs.

This work aims to identify more OCs by taking an effective method, i.e., e-HDBSCAN. The paper is organized as follows. We introduce data preparation in Sect.~\ref{secdata}. In Sect. ~\ref{sec:method}, we explain the method for searching OCs. The results and discussion are next given in Sect. ~\ref{sec:results}. Finally, we conclude this work in Sect. ~\ref{sec:conclusion}.

\begin{figure*}[h!]
\centering
\includegraphics[scale=0.4]{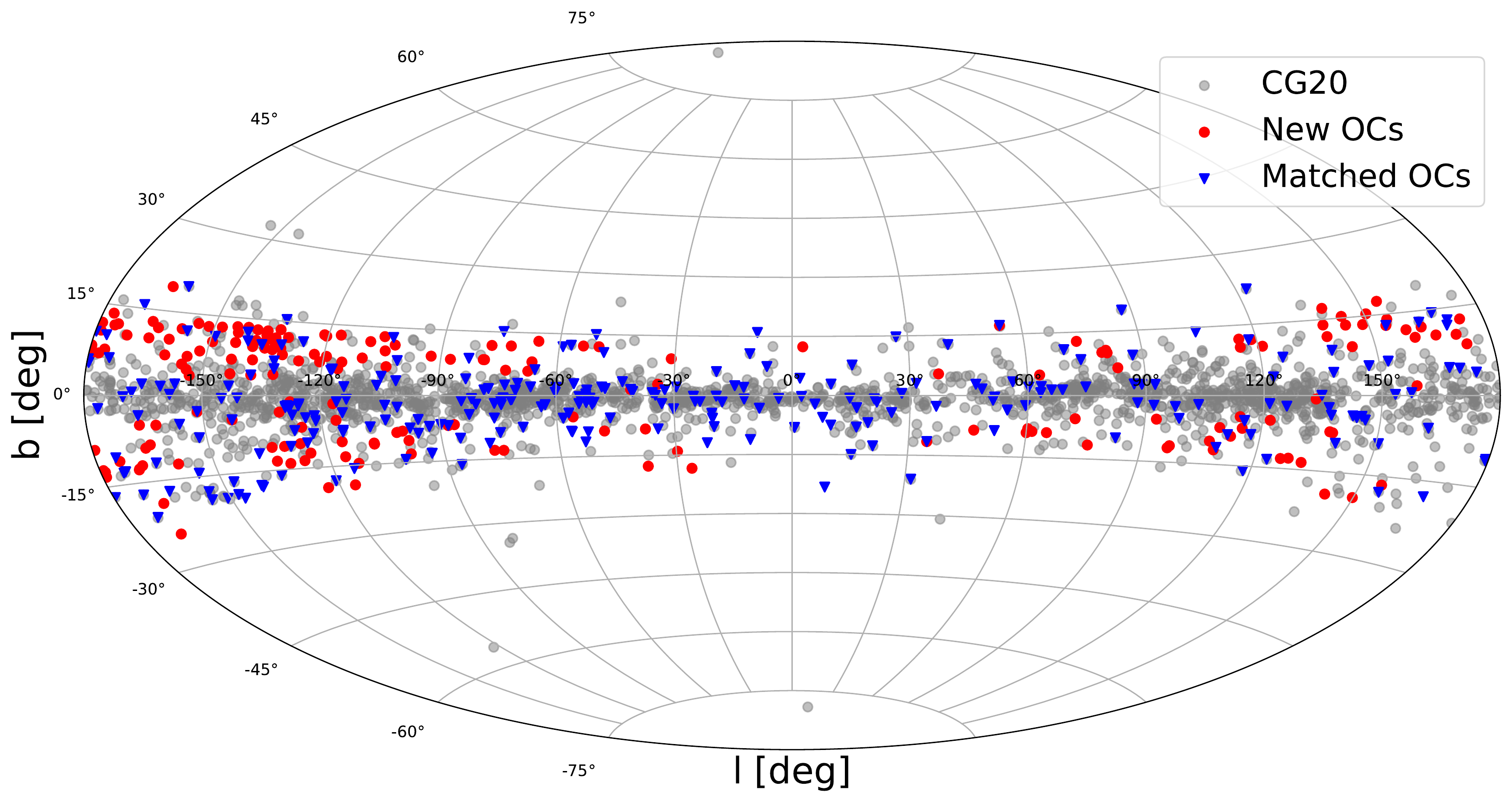}
\caption{Comparison of distribution.}
\label{fig:universe}
\end{figure*}

\section{Data Preparation}\label{secdata}
\subsection{Data Set}
Gaia EDR3, the European Space Agency released early third data releases  in 2020. The position, velocity, and magnitude data of more than 1.8 billion stars, and trigonometric parallax, proper motion, and photometric data of more than 1.3 billion stars are provided in this data release \citep{Cantat-Gaudin2022Universe}.
 Furthermore, the systematic errors in astrometry are reduced by (30 $-$ 40)\% for the parallaxes and by a factor of 2.5 for the proper motions, whereas for photometry, the systematic error is reduced to below 1\% \citep{Riello2021}.

To create the data set, we extracted stars with $\varpi$ $>$ 0.2  mas in Gaia EDR3 firstly because  most OCs are located near the galactic equator $|b|<$ 20 $^\circ$~\citep{Dias2012, Kovaleva}. In addition,  to reduce the pollution of field stars, as same as \cite{Cantat-Gaudin2019}, we chose stars with G $<$ 18 mag because not only the error of Gaia astrometric data increases with the decrease of star brightness but also due to the completeness limit of the Gaia DR2, the sample used by~\citet{Cantat-Gaudin2019}. This described in detail by~\citet{Gaia_Collaboration2018}.

We further filtered the data set with three criteria: 1) $G$ $<$ 18 mag~\citep{Lindegren2018}; 2) 0.2 mas $<$ $\varpi$ $<$ 7.0 mas; 3) $\mu_\alpha \cos\delta$ $<$ 30 mas yr$^{-1}$, $\mu_\delta$ $<$ 30  mas yr$^{-1}$ and $|b|$ $<$25$^\circ$. It should be noted that stars with $G$ $<$ 18 mag and which possess a parallax uncertainty of more than 0.2 mas or better were added to our sample, which is as same as previous works \cite{Cantat-Gaudin2019, Liu&Pang2019, Ferreira2020, Li2022}.

Finally, we obtained a data set of 186,464,070 stars with 5-Dimension feature space, $X=\{l,b,\varpi,\mu_\alpha \cos\delta,\mu_\delta,G, G_{RP}, G_{BP} \}$, 
where $l$ and $b$ denote spatial galactic coordinates,  $\varpi$ is trigonometric parallaxs, $\mu_\alpha \cos\delta$ and $\mu_\delta$ mean  proper motions, and $G, G_{RP}, G_{BP}$ are photometric data .
It is worth to point that  corrections of magnitudes in $G$ is not taken into account
since it  have no impact in the detection of OC
but it can be in the determination of the clusters features from their colour-magnitude diagrams. 

\subsection{Data Preprocessing}

% Identifying clusters in the position and proper motion space of the trigonometric parallax and the two-dimensional proper motion data of each star is a common method widely used.

Same to many previous studies, we selected five parameters, i.e., $\{l, b, \varpi, \mu_\alpha \cos\delta, \mu_\delta\}$, for the identification of open cluster. We constructed a quintet

\begin{center}
$X=\{l,b,\varpi,\mu_\alpha \cos\delta,\mu_\delta \}$
\end{center}
for each star, respectively.

To better facilitate the clustering calculation and improve the clustering effect, referring to \citet{Liu&Pang2019}, we calculated weight ($w$) for each star and further updated the quintet using

\begin{equation}\label{equ1}
x^\prime=\frac{x-min\left(x\right)}{max\left(x\right)-min\left(x\right)}
\end{equation}

\begin{equation}\label{equ2}
    \boldsymbol{w}=\dfrac{(\cos \mathit b, 1,0.5,1,1)}{0.2 \cos \mathit b+0.7}
\end{equation}

\begin{equation}\label{equ3}
    X_d= \boldsymbol{w} * X
\end{equation}

Here, cos $b$ is due to the contraction of $l$ at a given $b$ in spherical geometry.

\section{Method of Identification of Open Clusters}
\label{sec:method}
\subsection{e-HDBSCAN}
The Hierarchical Density-Based Spatial Clustering of Applications with Noise (HDBSCAN) is a clustering algorithm proposed by \cite{ Campello2013,Campello2015}. It is an improvement of the DBSCAN \citep{Ester1996A} algorithm that is density-based and has been widely  to identify OCs.

Compared with DBSCAN,  HDBSCAN can detect overdensities of varying densities in a dataset.
HDBSCAN does not depend on a hyperparameter ($\xi$), which is used in DBSCAN as a particular neighbor radius. Instead, it condenses the minimum spanning tree by pruning off the nodes that do not meet the minimum number of sources in a cluster. It only demands the user to set a minimum cluster size which is easy to do in OCs hunting with astronomical domain knowledge. Some recent works suggest that HDBSCAN can solve many issues encountered by DBSCAN and was more sensitive than the algorithm of GMMs, DBSCAN, which are widely used in search OCs \citep{Hunt2021}.

HDBSCAN is the most sensitive and effective algorithm for identifying open clusters in Gaia data \citep{Hunt2021}. Therefore, we  adopt HDBSCAN algorithm to blind search OCs in the study by extending the conventional HDBSCAN algorithm. This allows one to use a big data environment to search for new OCs in Gaia EDR3.

\begin{figure}
\centering
\includegraphics[width=80mm]{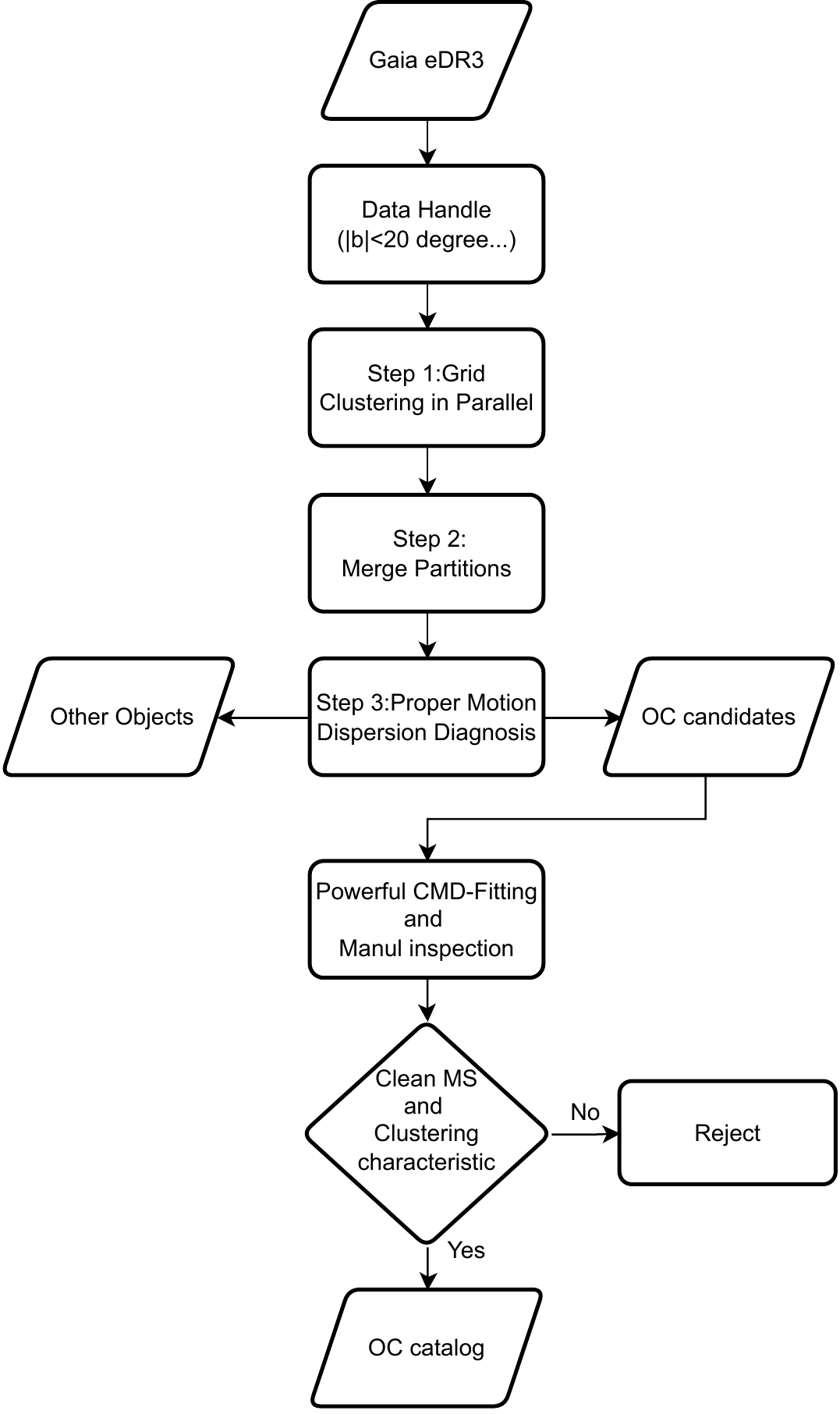}
\caption{Flowchart of e-HDBSCAN method. The process of this schedule consist of 3 stages. Step 1 is Grid clustering in Parallel within Gaia EDR3 to get rough clusters. Then, to merge clusters is performed in step 2. After fusion of those clusters, we implement proper motion dispersion diagnosis for identify of OCs in step 3. Power-CMD fitting and visual inspection is carried out subsequently. }
\label{flowchart}
\end{figure}

 Due to the large number of sources in Gaia EDR3, cluster analysis is hard to be performed on a stand-alone generic computer.
Based on HDBSCAN, we proposed an enhanced Hierarchical Density-Based Spatial Clustering of Applications with Noise (e-HDBSCAN) approach. The e-HDBSCAN is written in Python3 and implemented using Message Passing Interface (MPI), which can run well under a Linux-based high-performance computing cluster, thereby improving the clustering efficiency.
e-HDBSCAN splits the data grid, further merges and fuses the clustering results based on the parallel clustering of the grid data, and eliminates clusters that do not meet the statistical criteria by proper motion judgement on the membership of the obtained clusters. This overcomes the shortcomings of the traditional stellar agglomeration method, which needs multiple  trials and is  time-consuming.
 It is worth pointing out that the extended framework of our proposed algorithm is applicable not only to HDBSCAN but also to other clustering algorithms.
The flowchart is shown in Figure~\ref{flowchart}.
In summary, the biggest value of e-HDBSCAN is that traditional HDBSCAN is enhanced with a parallel split-and-conquer strategy to ensure that large data beyond the memory of a single machine can be processed. 

\subsection{Step 1: Grid clustering in Parallel}
To obtain rough star clusters conveniently, we partitioned the source data into multiple grids first. We then carried out the clustering algorithm in each data grid, as it is too large to read all sources into the computer memory for processing at once.

After extraction of sources within Gaia EDR3 as our source selection criteria as mentioned above, we obtained some grids by splitting sources into gridding data slices uniformly, which got in 3D spatial coordinates $\{l, b, \varpi \}$ using KD\_tree method.
It is worth mentioning that each domain size is larger than two times the 20pc in dimensions of the Galactic latitude and longitude, which is the typical scale of a star cluster \citep{Brown2010MNRAS}.
Besides, to eliminate gridding with high uncertainty, the minimum domain size is set to larger than 0.2 mas in $\{\varpi\}$ dimension space.
Then, we used HDBSCAN clustering algorithm in the weighted 5D parameter space {$X$} calculated as ~\ref{equ1},~\ref{equ2} and ~\ref{equ3}  for each data grid.

\subsection{Step 2: Merging Partitions}

 Many open star cluster candidates appear in more than one domain.
 We adopt a merge strategy to account for clusters in the borders of the grid.
 If more than 50\% of the minimum of members in each pair of star clusters are the same, we merge the clusters.
 What needs to be mentioned is that the value of the parameter minPts in the algorithm is based on the physical prior of star cluster composition, and we set  to 50, which is consistent with ~\citep{Liu&Pang2019} and our previous series work ~\citep{Li2022}.
 To speed up the clustering process, we carry out parallel computing using Mpi4py \citep{2008MPI} that speeds up the computation time and allows us to process a volume of data that does not fit in the memory of a single machine.
 
We adopted a recursive merge strategy to account for clusters at the edge of the data regions. We merge star clusters in two adjacent regions if more than fifty percent of their minimum members are the same.
The overlapping regions in the model were limited to no less than 20 pc. Thus, there is no intersection point where the two clusters share less than 50\%, unless the cluster is larger than 60 pc and symmetrically straddles the established overlap region. In fact, this is rarely the case.

\subsection{Step 3: Proper Motion Dispersion Diagnosis}

It is obvious that the detection of more true positive OCs always also resulted in false positive OCs \citep{Hunt2021}.
On the other hand, HDBSCAN may identify either statistical clusters, asterisms, or other physical groups located in the same field \citep{Tarricq2022}.
In this work, it is suggested that the objects identified by the e-HDBSCAN method with mean proper motions and parallaxes. If they were real open clusters that should be consistent to the value computed by \cite{Castro-Ginard2020}.
That means the mean proper motion dispersion of true positive OC candidates should be compatible with a real OC. Draw on the method of \cite{ Cantat-Gaudin2019,Hao2022}, which used internal proper-motion dispersions to filter unreliable open cluster candidates.
We further diagnosed those hunted candidates in Sect. ~\ref{sec:method} with this method.
To select only high-quality OC nominated objects, we use the rigorous proper-motion criterion  \ref{equ4} and \ref{equ5}, which is adopted by  \cite{Cantat-Gaudin2019}.
\begin{center}
\begin{equation}\label{equ4}
\sqrt{\sigma_{\mu_{a^{*}}}^{2}+\sigma_{\mu_{\delta}}^{2}} \leq 0.5 \mathrm{mas}\ \mathrm{yr}^{-1} \& \text { if } \varpi<1 \text { mas }
\end{equation}

\end{center}
% and
\begin{center}
\begin{equation}\label{equ5}
\sqrt{\sigma_{\mu_{a^{*}}}^{2}+\sigma_{\mu_{\delta}}^{2}}  \leq 5 \sqrt{2} \frac{\varpi}{4.7404} \mathrm{mas} \  \mathrm{yr}^{-1} \& \text {   if } \varpi \geq 1 \text { mas }
\end{equation}

\end{center}

 We filter out most of candidates  that are not real by this criterion. It is worth mentioning that this criterion is necessary for the identification of OCs. It does not imply that clusters with dispersion higher than the criterion in these formulas are real clusters. However, by doing so, we can filter out most of the candidates that do not meet this criterion, obtain high-quality cluster candidates, and reduce the final identification effort.

\section{Results}
 \label{sec:results}
 We extract 186,464,070 sources of Gaia EDR3 using the source selection criteria presented in the previous section and obtain 4091 grids after uniformly gridding data slices in 3D spatial coordinates $\{l,b,\varpi \}$ using KD\_tree method.
 A total of 7756 local clusters are detected in step 2.
 After merging the rough local clusters, 3,787 open clusters are finally found.

 %%%%%%%%%%%%%%%%%%%%%%%%%%% V%%%%%%%%%%%%%%%%%%%%%%%%%%%%%%%%%%%
 Different from other works based HDBSCAN algorithm  \citep{Ye2021,Tarricq2022, Casamiquela202}, in the present work, the e-HDBSCAN method systematically searched nearby ($|b|<25^\circ$) all-sky regions using this algorithm.
 In addition, we strictly limit the number of members of open cluster candidates to be larger than 50. For consistency with our previous series of work \citet{Li2022}, we set the minPts of the HDBSCAN algorithm to 50 in this process while \cite{Castro-Ginard2018A} set this parameter equal to 8 when using density clustering.
%  Another reason is that \cite{Hunt2021}  compared a set of  values  ranged from 20 to 80 for minPts,  and suggest that low values  were found to produce many false positives, also miss some valid members of the cluster and setting the value too large may miss some OCs.
 The another aspect is that we perform this algorithm in a weighted 5D parameter projection space, namely, $\{l,b,\varpi,\mu_\alpha \cos\delta,\mu_\delta\}$ in the 4,091 splitting searching grids as  mentioned above.
Due to the schemes mentioned above, it is possible to find new objects even in regions that have already been searched in previous work.
 %%%%%%%%%%%%%%%%%%%%%%%%%%%%%%%%%%%%%%%%%%%%%%%%%%%%%%%%%%%%%%%%%%%%%%%%%%%%%%%%%%%%%%%%%%%%%%%%%%%%%%%%%%%%%%%%%%%%%%%%%%%
\subsection{Cross Match}
\label{sec:crossmatch}

Many OC candidates were found and reported in previous works.
We collected and compiled those catalogues and cross-matched them with our candidates.
To eliminate as many OCs as possible that have already been found and obtain OC candidates that have not been unnoticed before, we consider an OC to be positionally matched to a catalogued one if their centres lie within a circle of radius r = 0.5 degrees and rest of the astrometric mean parameters are compatible within 5 $\sigma$ (where $\sigma$ is the uncertainties quoted in both catalogues for each quantity) which is consistent with \cite{Hunt2021} and ~\cite{Castro-Ginard2021}.

We first performed cross-matching with the pre-Gaia cluster catalogue.
The pre-Gaia cluster catalogues (MWSC) contained 3006 star clusters gathered by \cite{Dias2012} and \cite{Kharchenko2013},  of which 2976  objects localized at Galactic latitude $<$ 25 degrees, aggregated from various data sources.
Since they do not allow for a sufficiently excellent comparison in the proper motion space, we only performed a 0.5-degree positional cross-match based on sky coordinates.
Second, to serve the following crossover, we collected the best-known catalogues of the identified clusters described above and compiled those as MWSC, CG2017, Hao3794, UBC series, CWNU, Dias1743 and Hao704, respectively.
The CG series contained 2017 OCs provided by \cite {Cantat-Gaudin2019}.
\cite{Hao2021} report a considerably larger catalogue including 3794 OCs (Hao3794) and UBC series reported by CG series, which consist of \cite{CG2018}, \cite{CG2018A}, \cite{CG2019}, \cite{CG2019A},\cite{Cantat-Gaudin2019}, \cite{CG2020},  \cite{CG2020B}, \cite{CG2022}.
In recent times, \cite{He2022} have reported 541 new open cluster candidates found in Galactic Disk Using Gaia DR2/ EDR3 Data (CWNU).
\cite {Dias2021} present a catalogue of updated parameters of 1743 open clusters based on Gaia DR2 in 2021 (Dias1743).
And \cite{Hao2022} recently provided 704 newly detected open clusters in the Galactic disk using Gaia EDR3 (Hao704).
Similarly, we use the same cross-matching method for some previous catalogs, i.e., \cite{Liu&Pang2019}, \cite{Ferreira2020}, \cite{Hao2020PASP}, \cite{Li2021}, \cite{Hunt2021}.
Finally, using the same method, we  cross-matched 46 newly reported clusters of high Galactic latitudes, which were found in the region of $|b|>20^\circ$ by \cite{He2022ApJS}.

As a result, 530 candidates are not included in those published catalogues, implying that they are new potential OCs that require further identification.

\subsection{Qualification}

Since members of OCs have similar motions and ages, they should exhibit clustering characteristics in spatial distribution position and kinematic space and show a clean  in color-magnitude diagram (CMD).

\begin{figure*}[ht]
\centering
	\subfloat{\includegraphics[width=6.0in,height=1.7in]{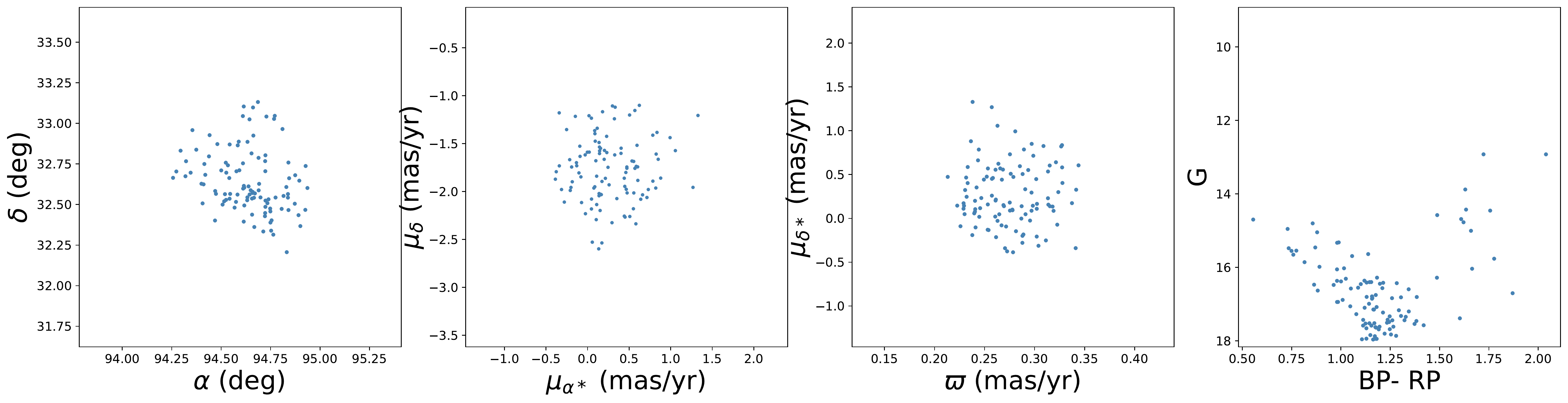}}
	\hfill
	\subfloat{\includegraphics[width=6.0in,height=1.7in]{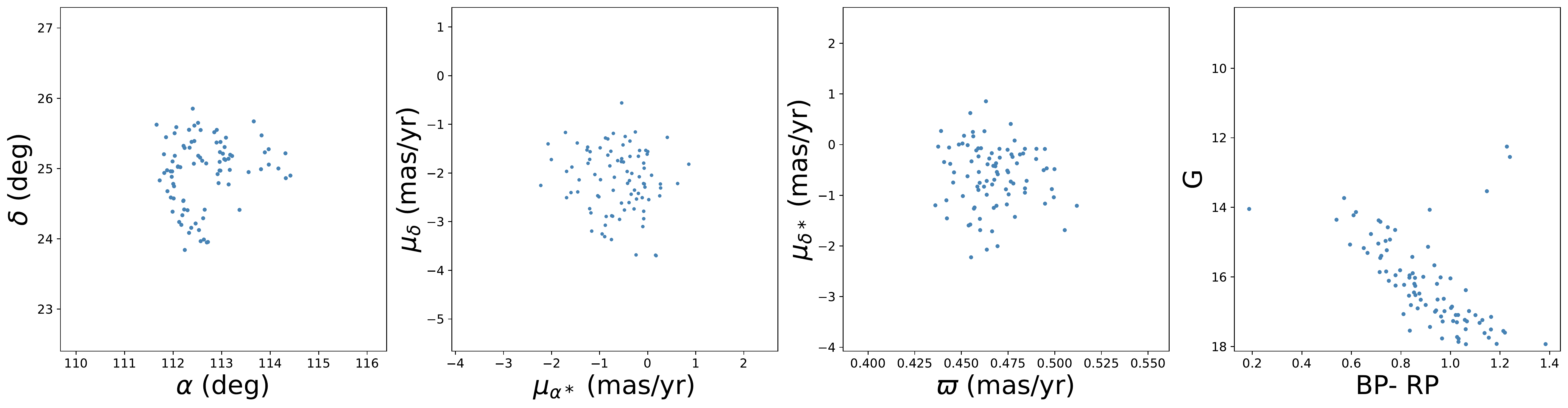}}
	\hfill
	\subfloat{\includegraphics[width=6.0in,height=1.7in]{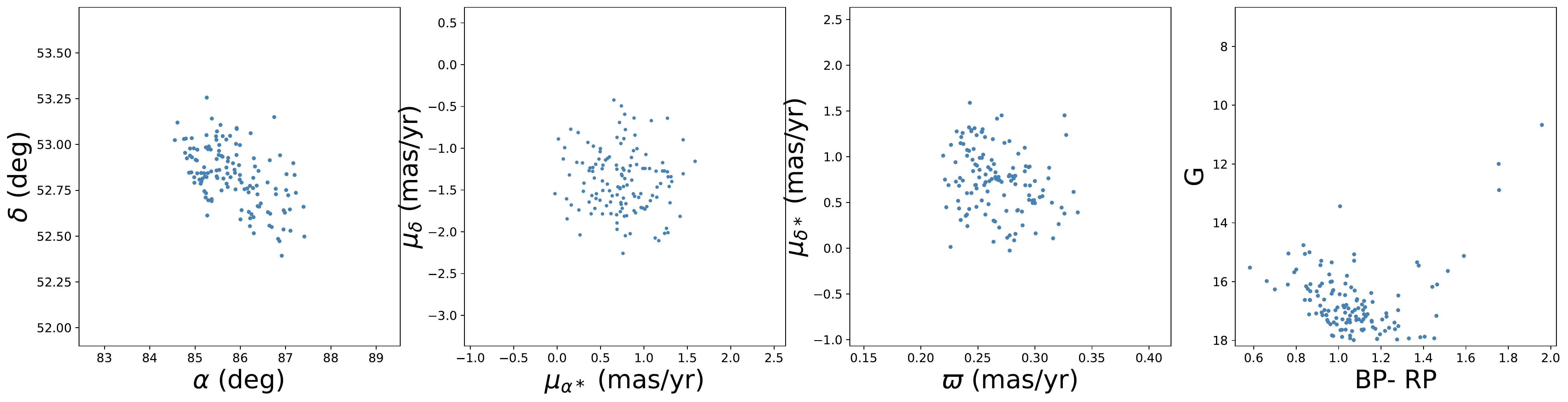}}
	\hfill
	\subfloat{\includegraphics[width=6.0in,height=1.7in]{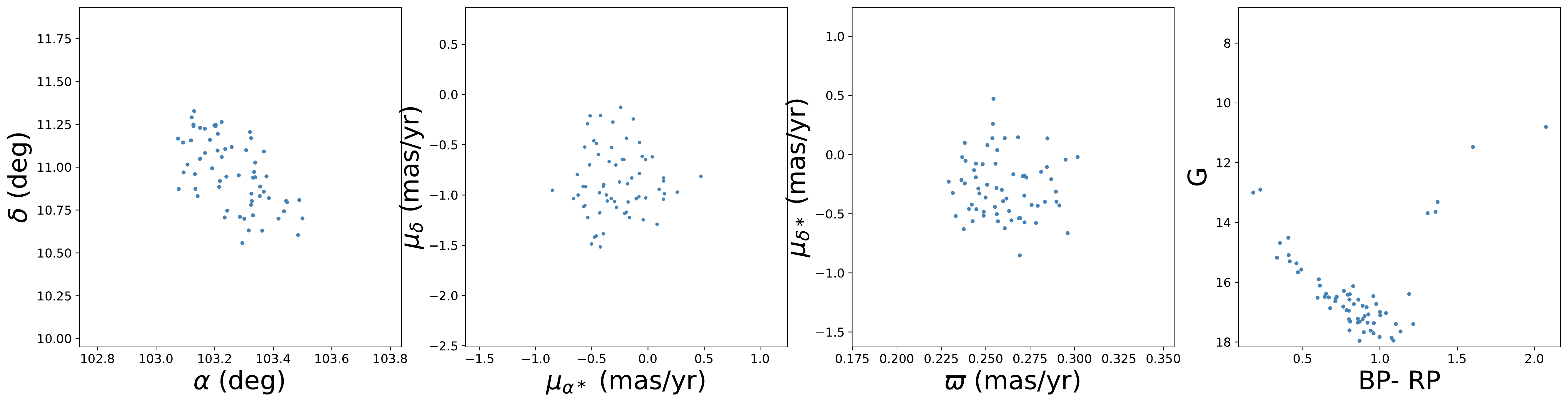}}
\caption{Visual Inspection of LISC III 1044, LISC III 2601, LISC III 1301 and LISC III 1762, respectively. Leftmost plots:
position of the OC in ($\alpha$, $\delta$). Inner left plots: ($\varpi$ , $\mu_{a^{*}}$ ) distribution, whilst inner right plots: ($\mu_{a^{*}}$, $\mu_{\delta}$ ) distribution. Rightmost plots: CMD of OC.}
\label{fig:1046}
\end{figure*}

A real OC should follow a certain pattern in an isochrone. Following the method of \cite{Liu&Pang2019, Hao2022}, we fit the CMDs using the PARSEC theoretical isochrone models \citep{Bressan2012}.
And then, we divided new candidates into three classes according to isochrone fitting results and selected which has good isochrone fitting and a relatively clean main sequence.
After this, we get 184 likely open clusters that need further confirmation.
Furthermore, we performed a visual inspection among those candidates in their positional distributions (PDs), vector point diagrams (VPDs), colour-magnitude diagrams (CMDs) and $\varpi$ versus $\sigma_{\mu_{a^{*}}}$ distributions. An example is shown in Figure~\ref{fig:1046}.
Finally,83 candidates were considered real OCs.
In Figure~\ref{fig:universe}, we present the sky distribution of OCs compared with  \citet{CG2020} (CG20).

\subsection{Determination of Stellar Population Properties by CMD-fitting}
% If an OC
% candidate can be considered as a potentially real OC , it could be shown to clearly pass its members, distance, age, or velocity
% relative to the field stars, as well as the density of the background
% stellar distribution, for instance.
Many previous works on determining the stellar populations of OCs in Gaia took the isochrone fitting technique.
To further reduce the uncertainty of estimation of the main parameters of OCs, it is necessary to consider that binary stars, rotating stars, and multiple populations exist in OCs.

\begin{figure*}
\begin{center}
\subfloat {
\includegraphics[width=1.8in,height=1.7in]{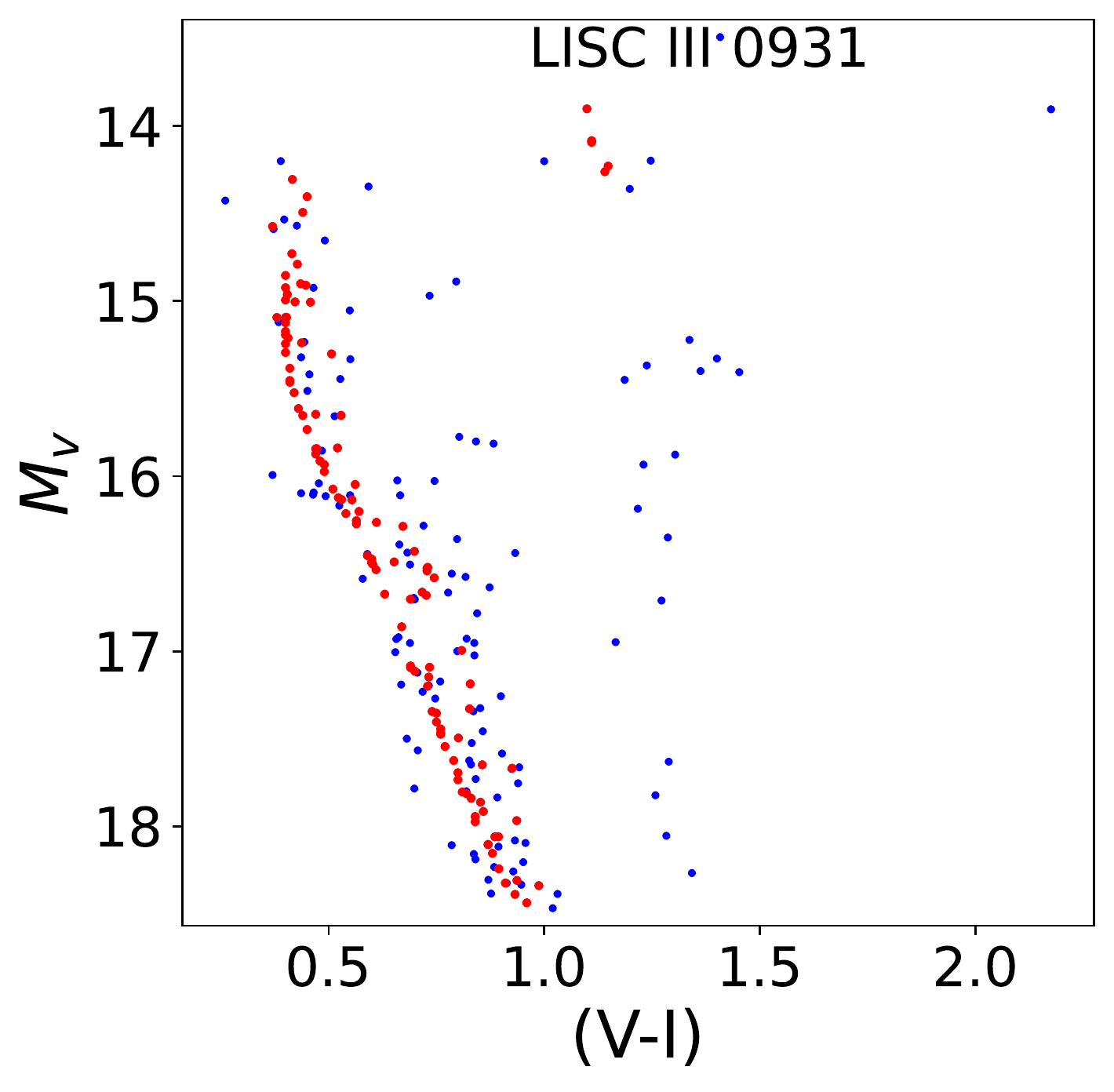}
}
\subfloat  {
\includegraphics[width=1.8in,height=1.7in]{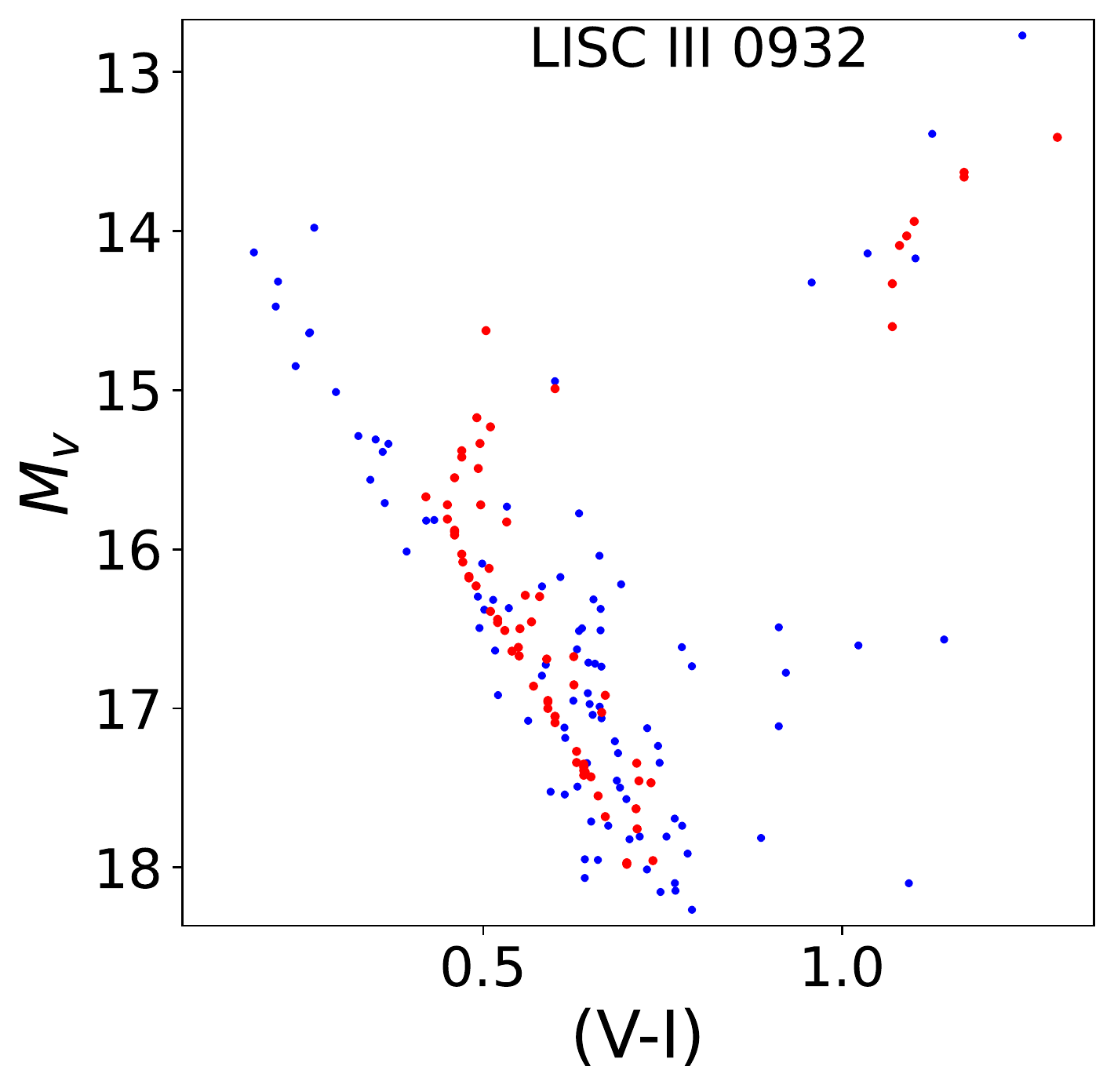}
}
\subfloat {
\includegraphics[width=1.8in,height=1.7in]{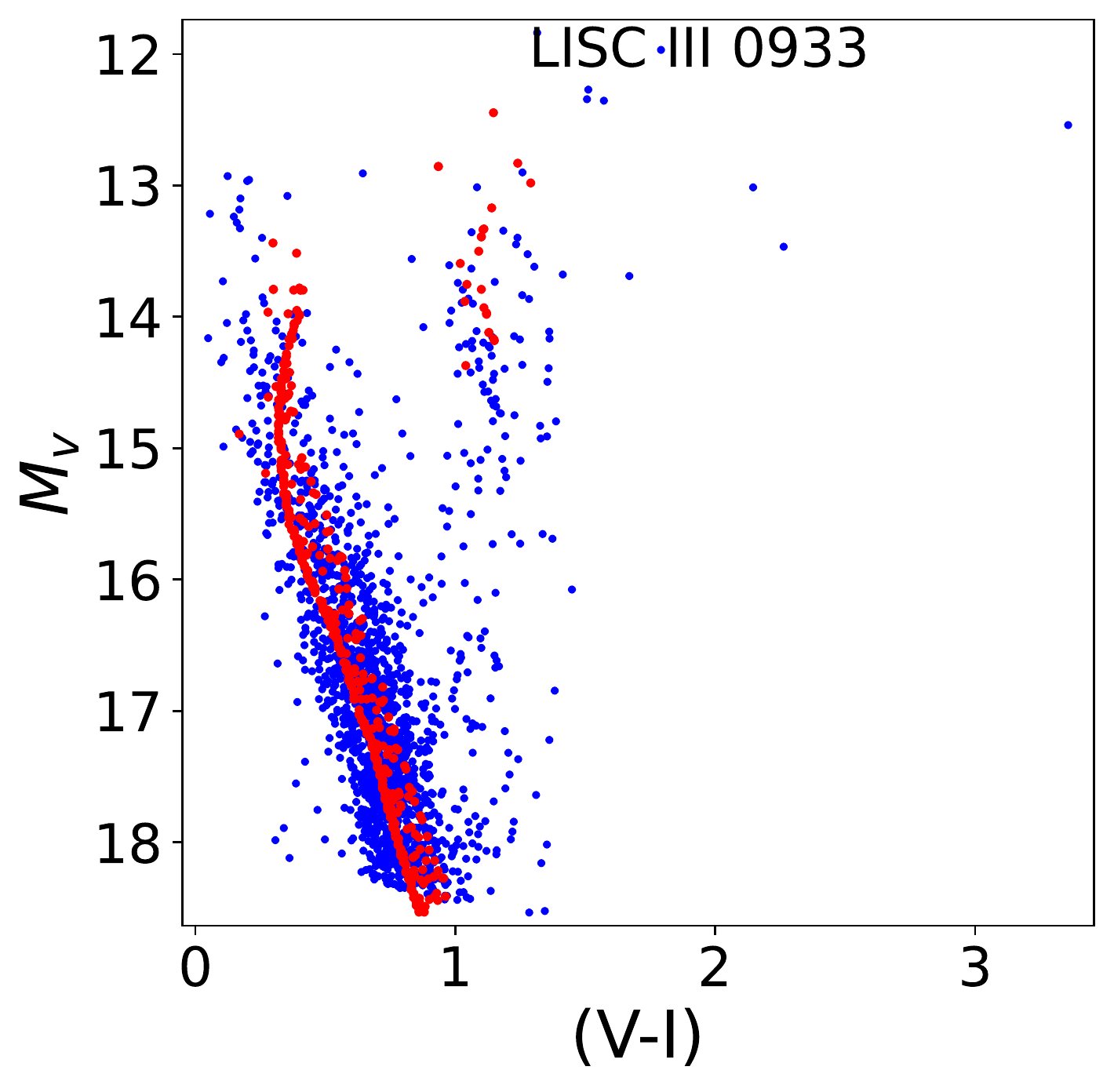}
}
\end{center}

\begin{center}
\subfloat {
\includegraphics[width=1.8in,height=1.7in]{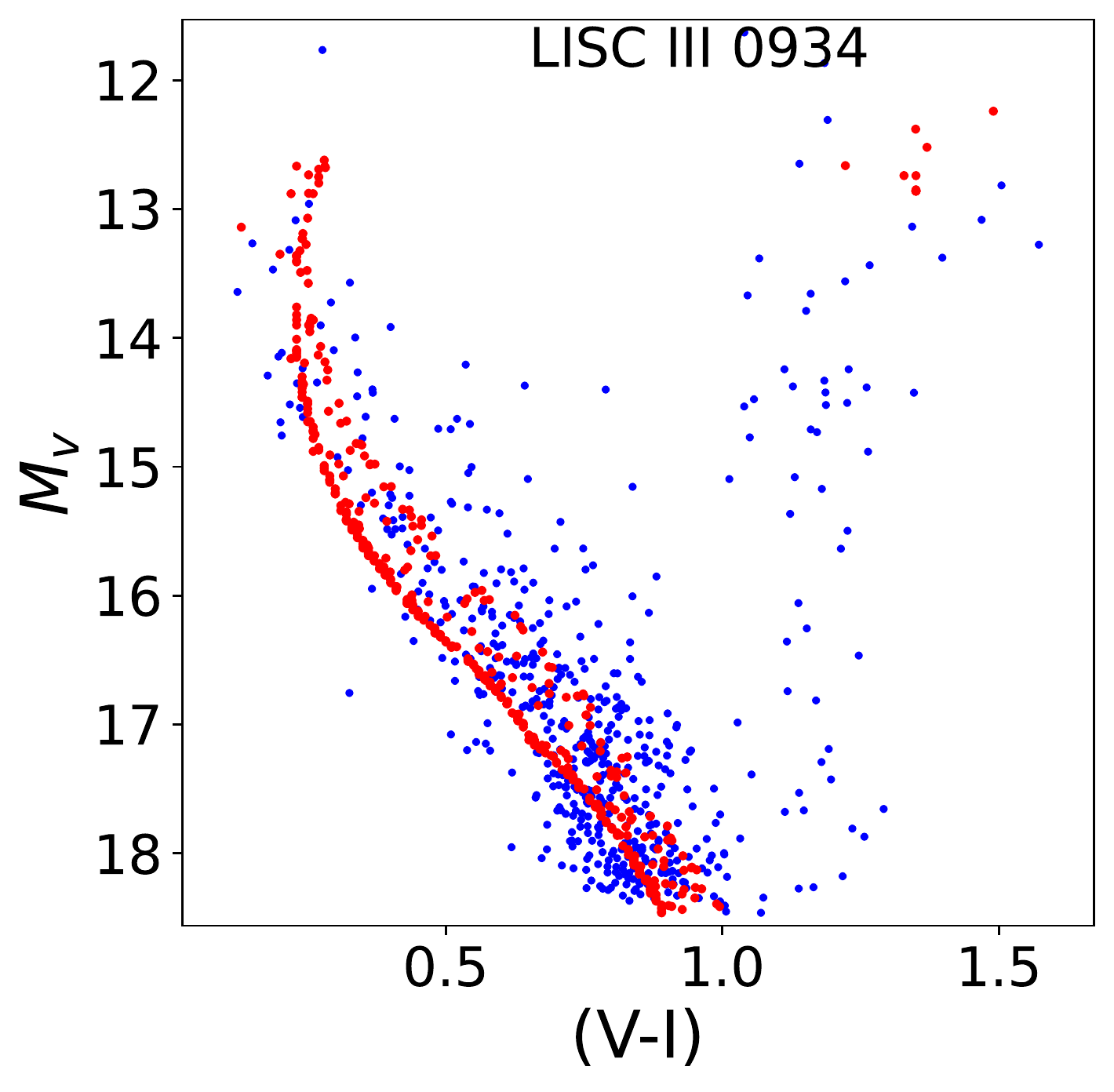}
}
\subfloat  {
\includegraphics[width=1.8in,height=1.7in]{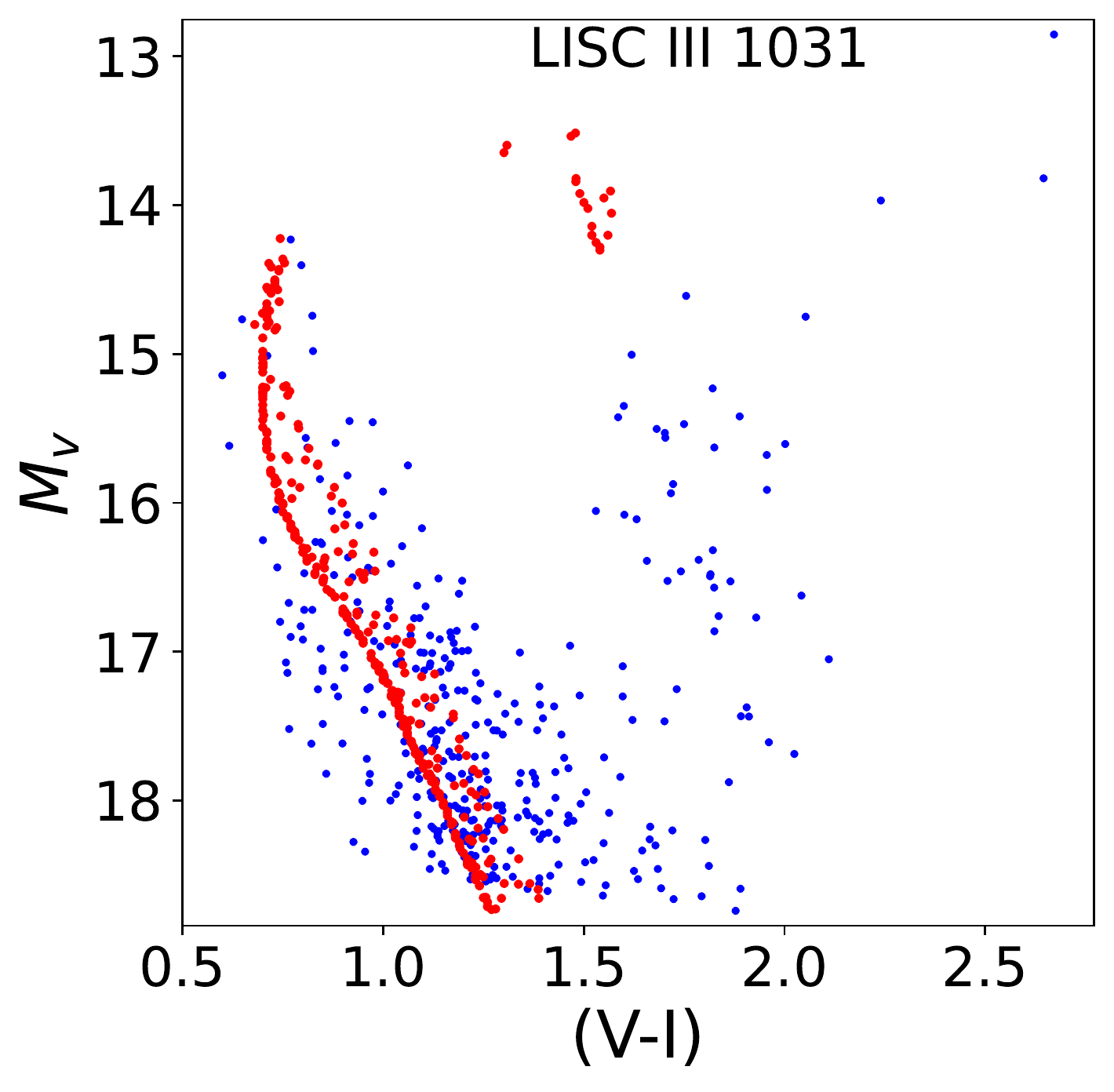}
}
\subfloat {
\includegraphics[width=1.8in,height=1.7in]{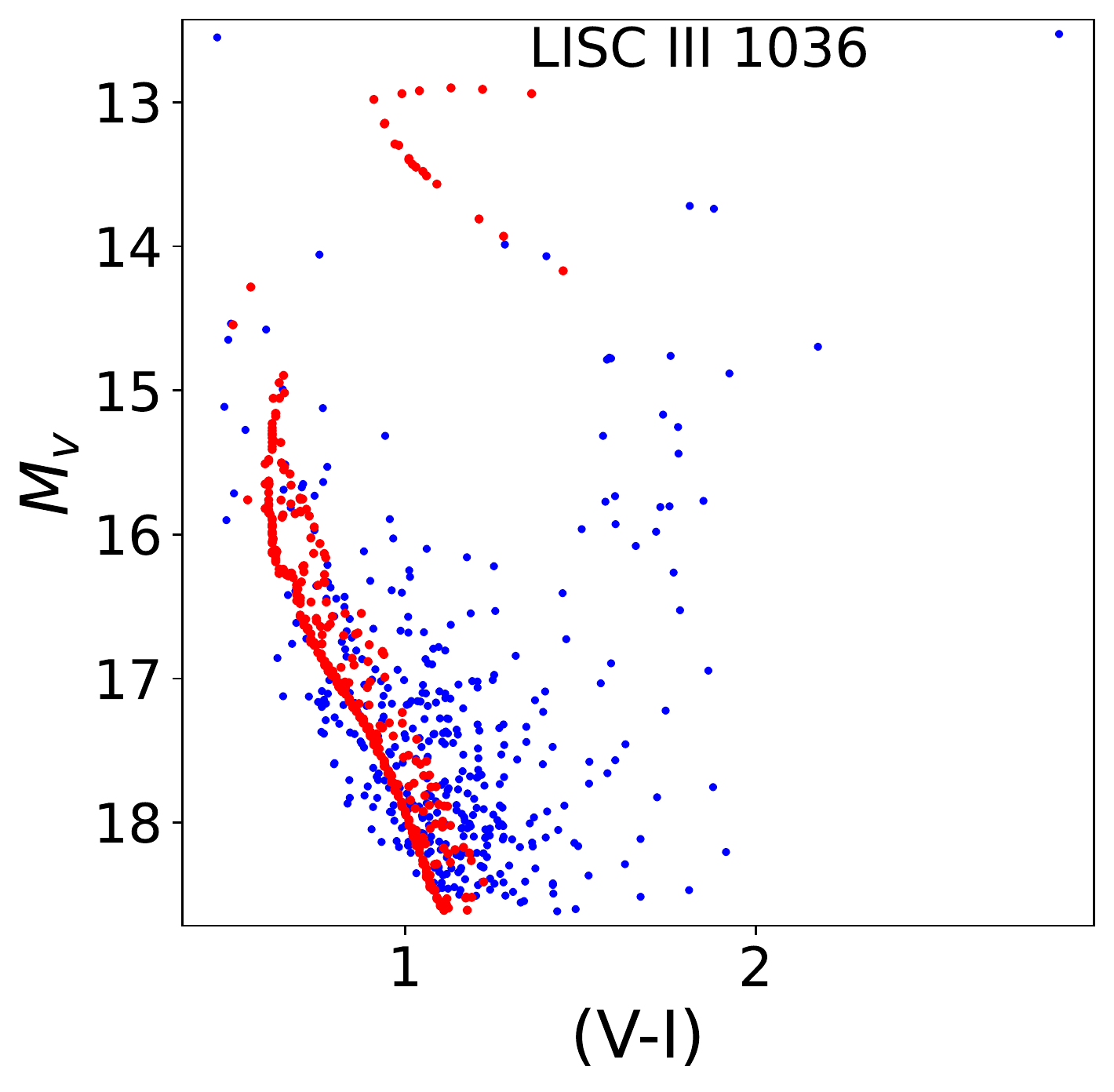}
}
\end{center}
\begin{center}
\subfloat {
\includegraphics[width=1.8in,height=1.7in]{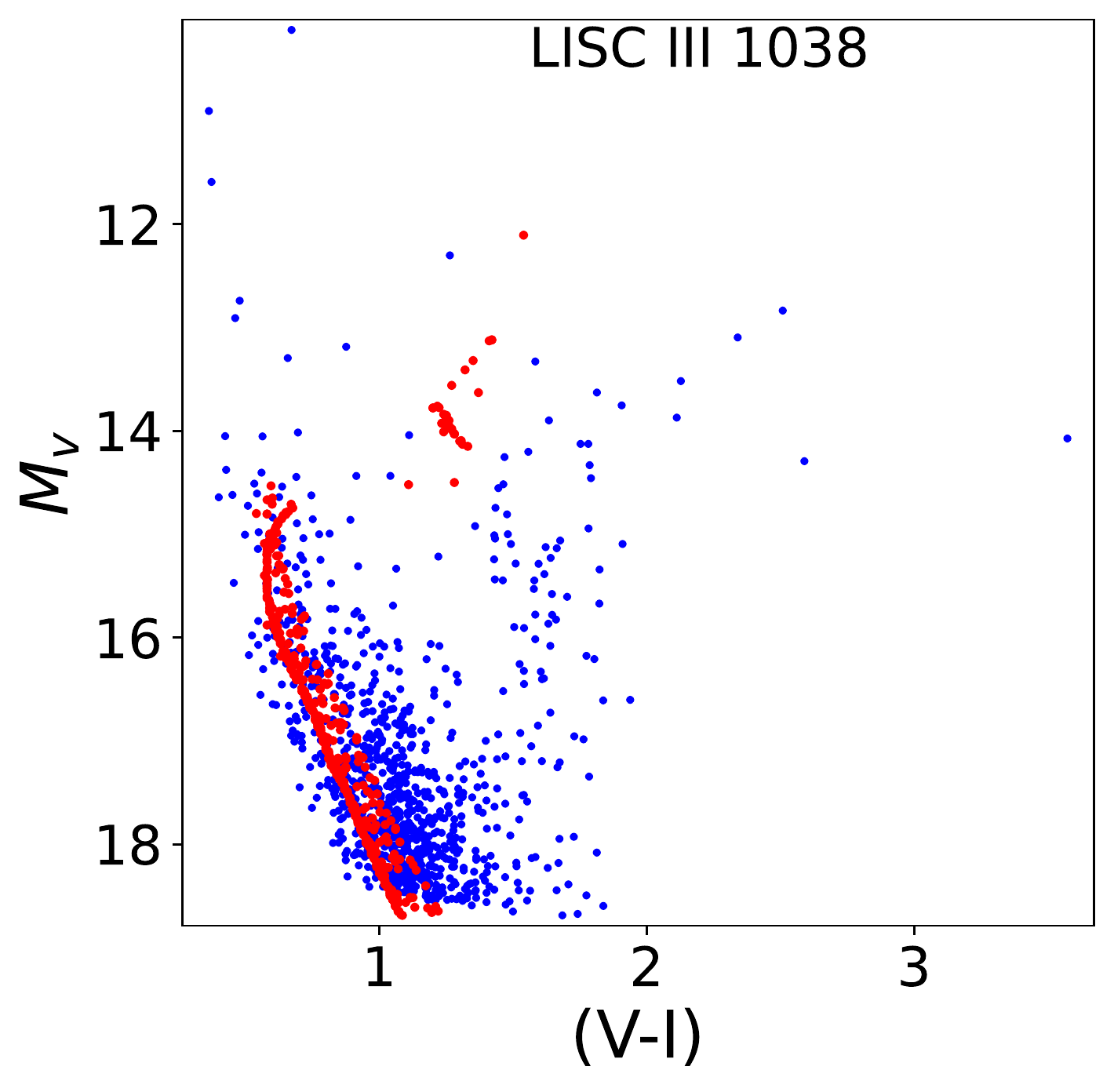}
}
\subfloat  {
\includegraphics[width=1.8in,height=1.7in]{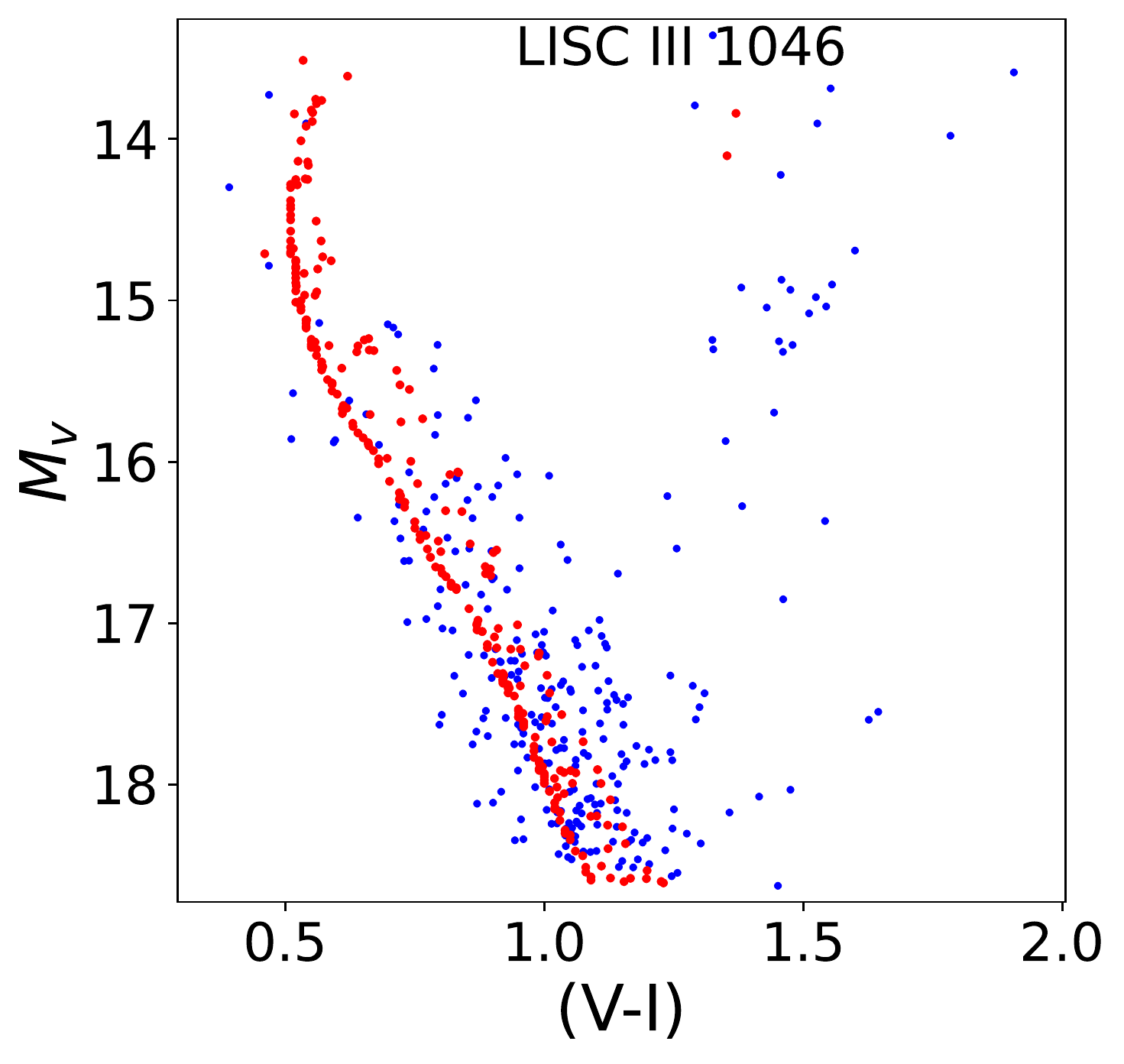}
}
\subfloat  {
\includegraphics[width=1.8in,height=1.7in]{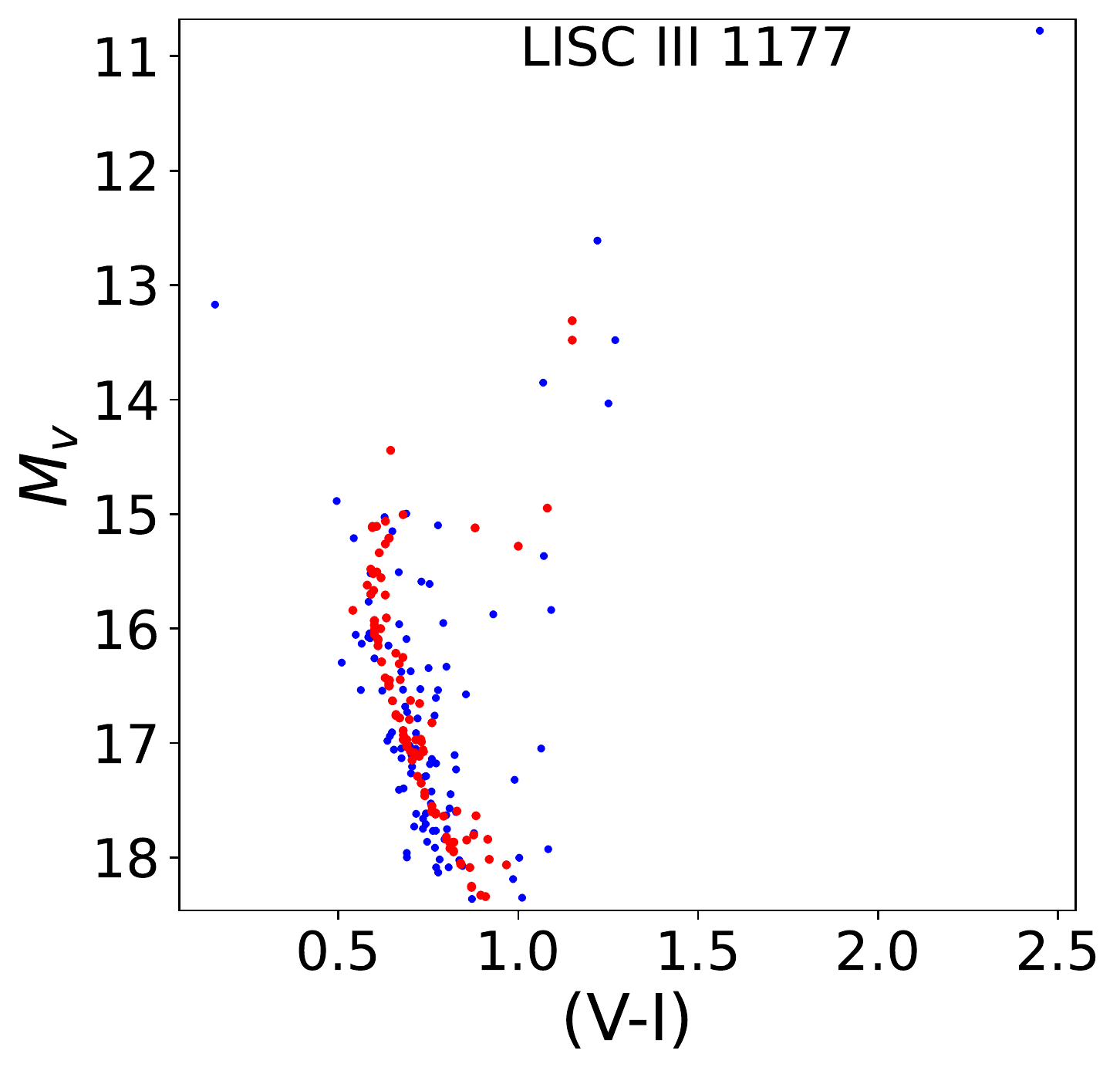}
}
\end{center}

\begin{center}
\subfloat {
\includegraphics[width=1.8in,height=1.7in]{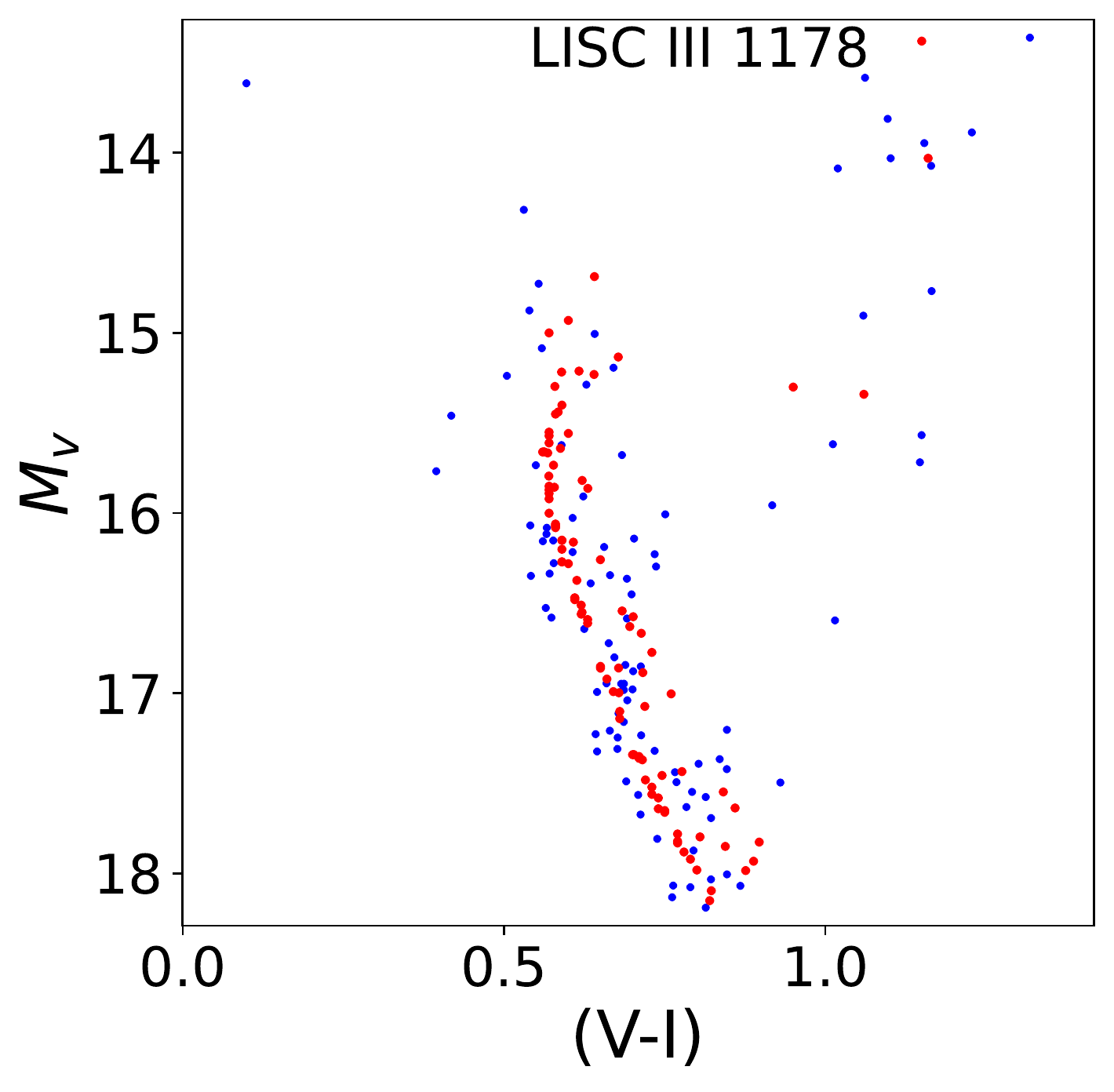}
}
\subfloat  {
\includegraphics[width=1.8in,height=1.7in]{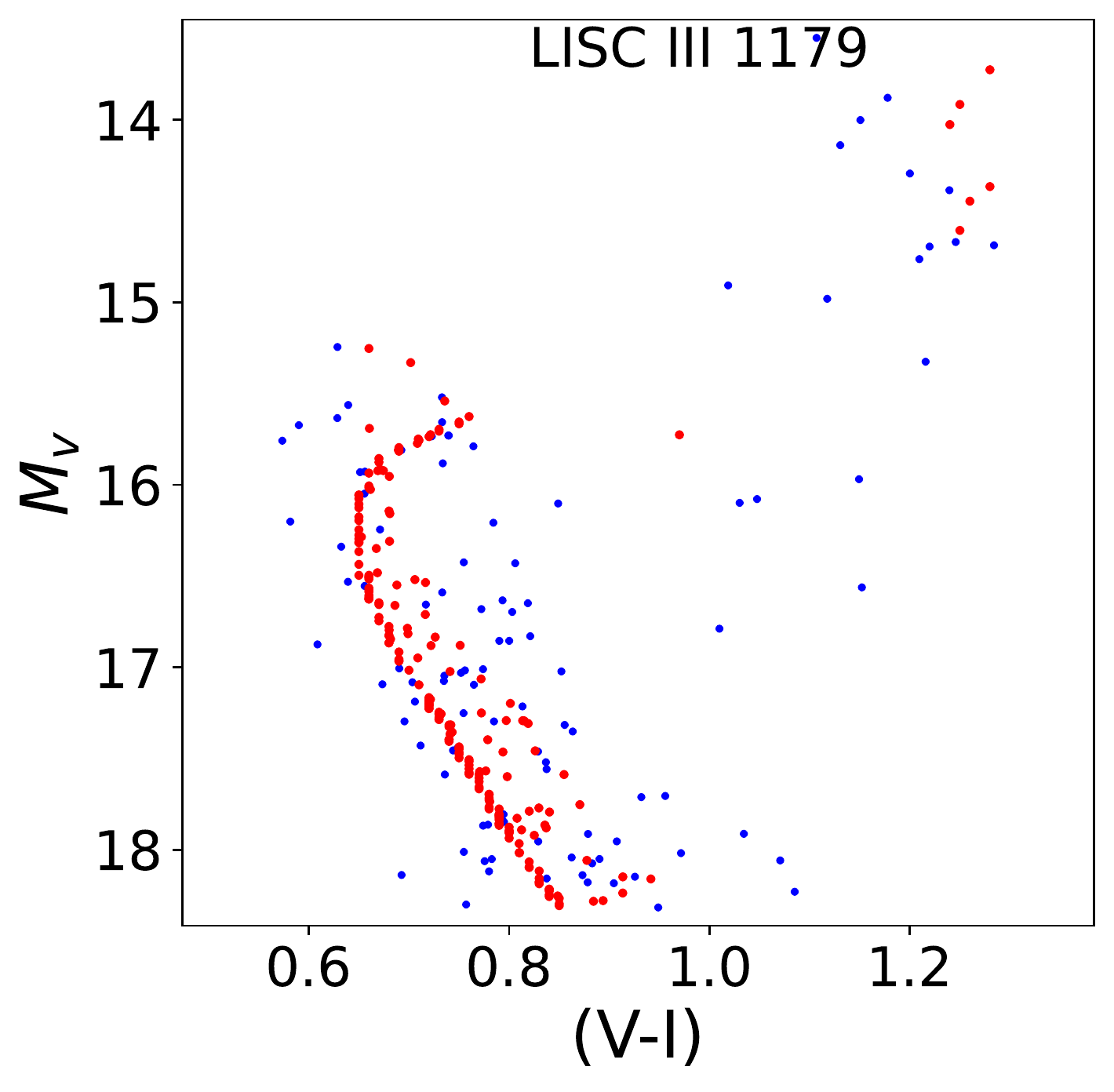}
}
\subfloat  {
\includegraphics[width=1.8in,height=1.7in]{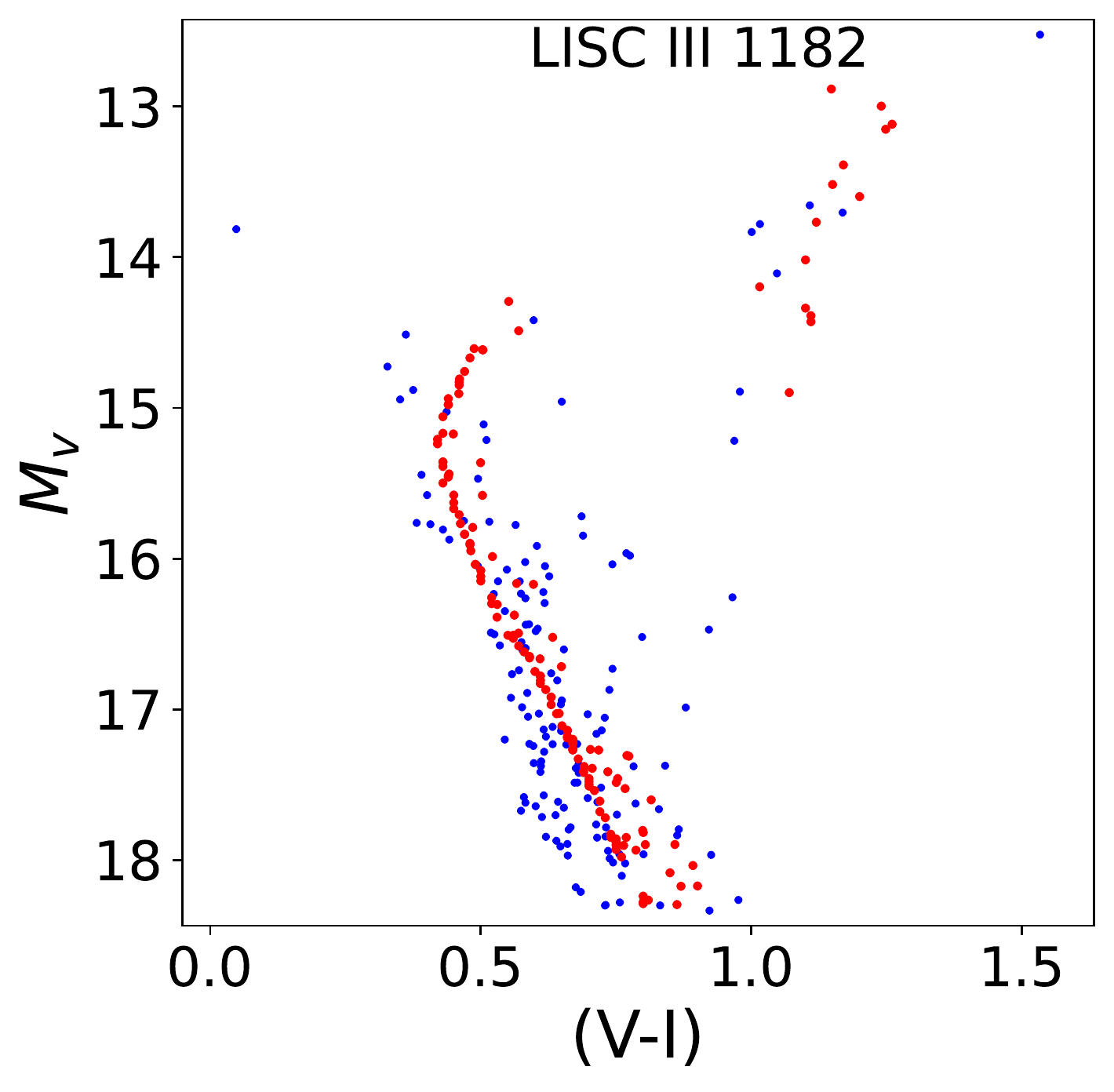}
}
\end{center}
\begin{center}
\subfloat {
\includegraphics[width=1.8in,height=1.7in]{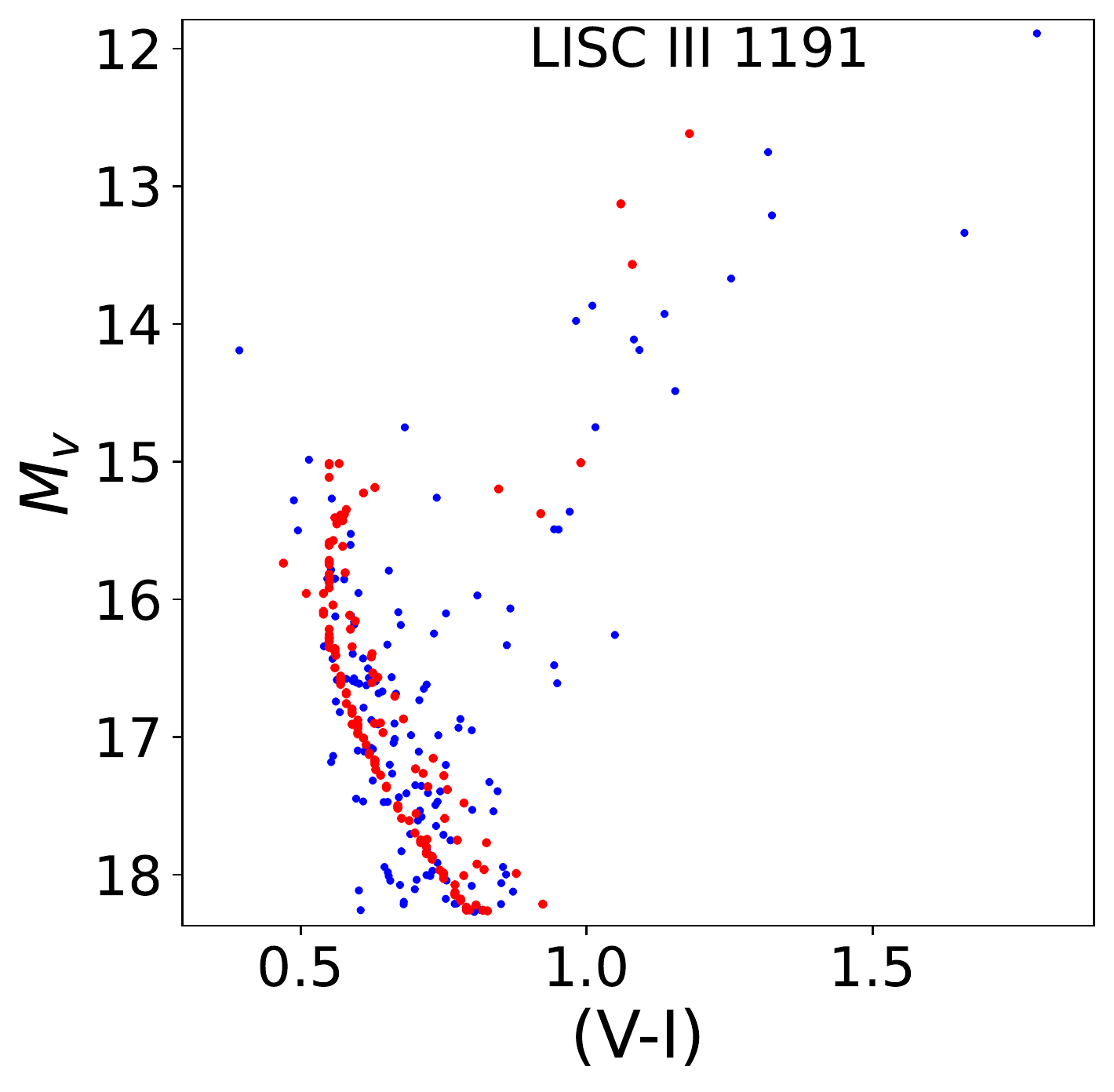}
}
\subfloat  {
\includegraphics[width=1.8in,height=1.7in]{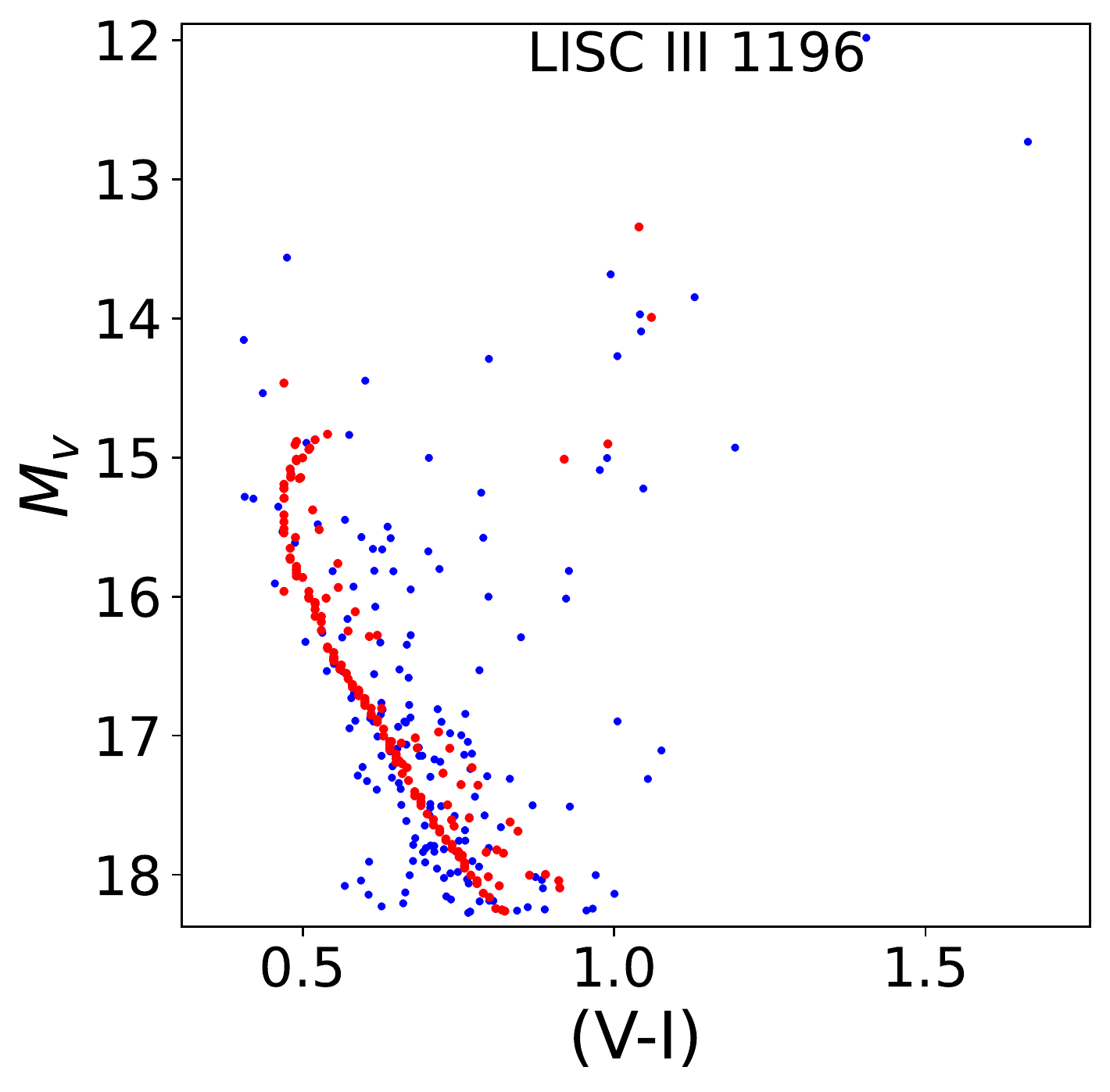}
}
\subfloat  {
\includegraphics[width=1.8in,height=1.7in]{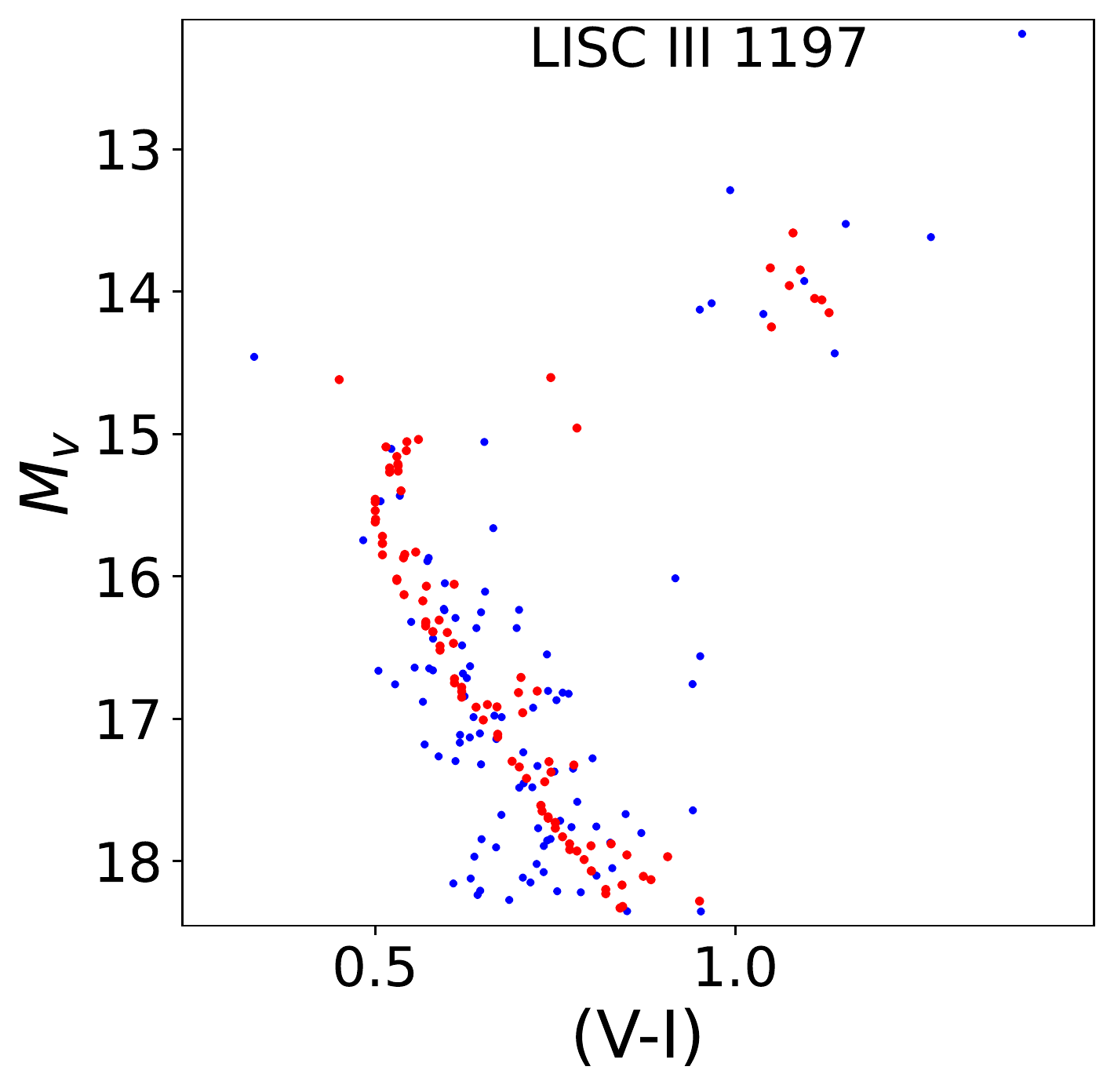}
}
\end{center}

\begin{center}

\caption{Comparison of best-fit and observed CMDs of 15 new identified open clusters with clear main sequence. Blue
points are for observed stars and red ones are for best-fit stellar populations. The complete figure set (83 images) is available on line. }
    \label{fig:CMD_1}
\end{center}
\end{figure*}

\begin{figure*}
\begin{center}
\subfloat {
\includegraphics[width=1.8in,height=1.7in]{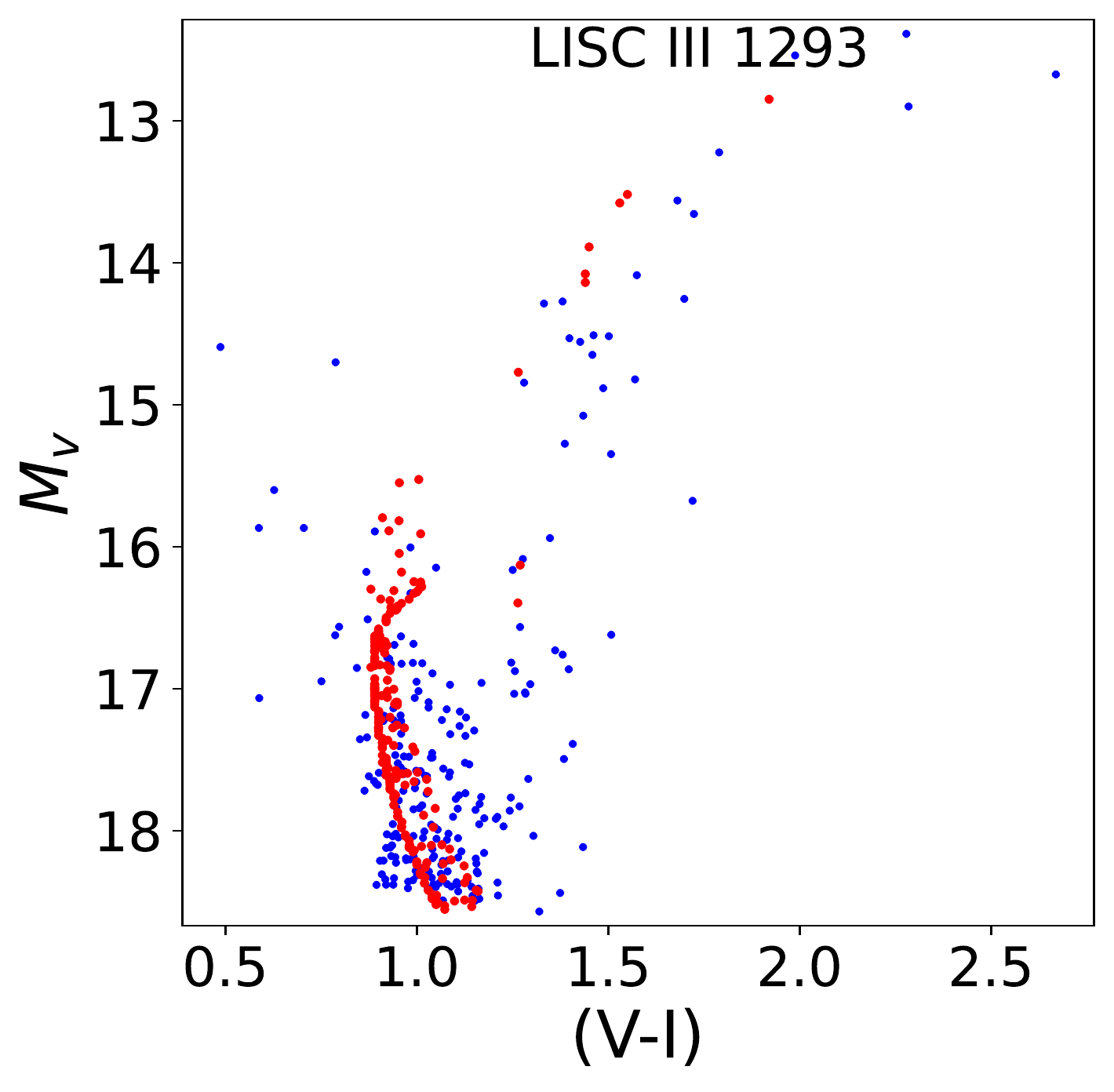}
}
\subfloat  {
\includegraphics[width=1.8in,height=1.7in]{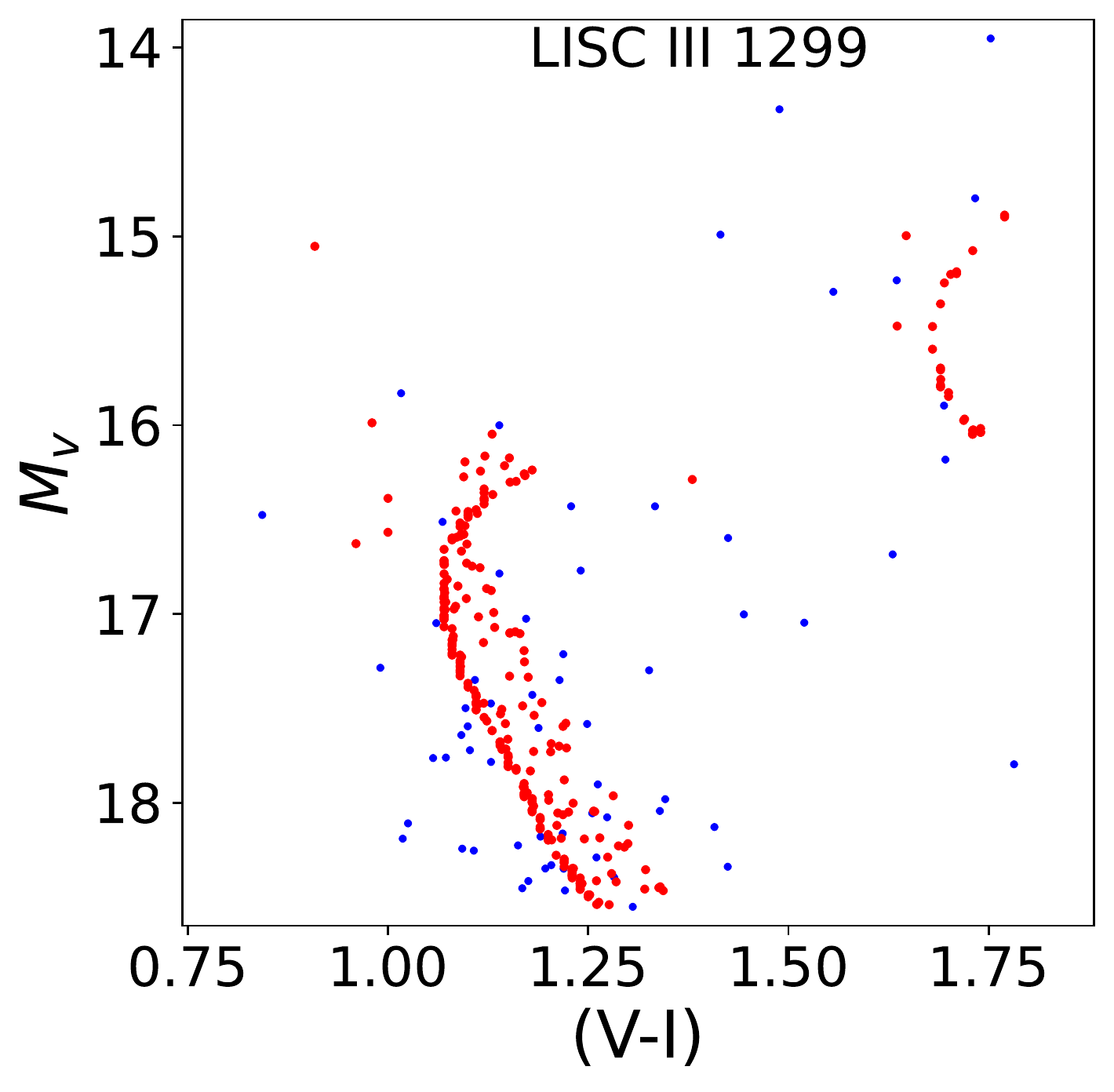}
}
\subfloat {
\includegraphics[width=1.8in,height=1.7in]{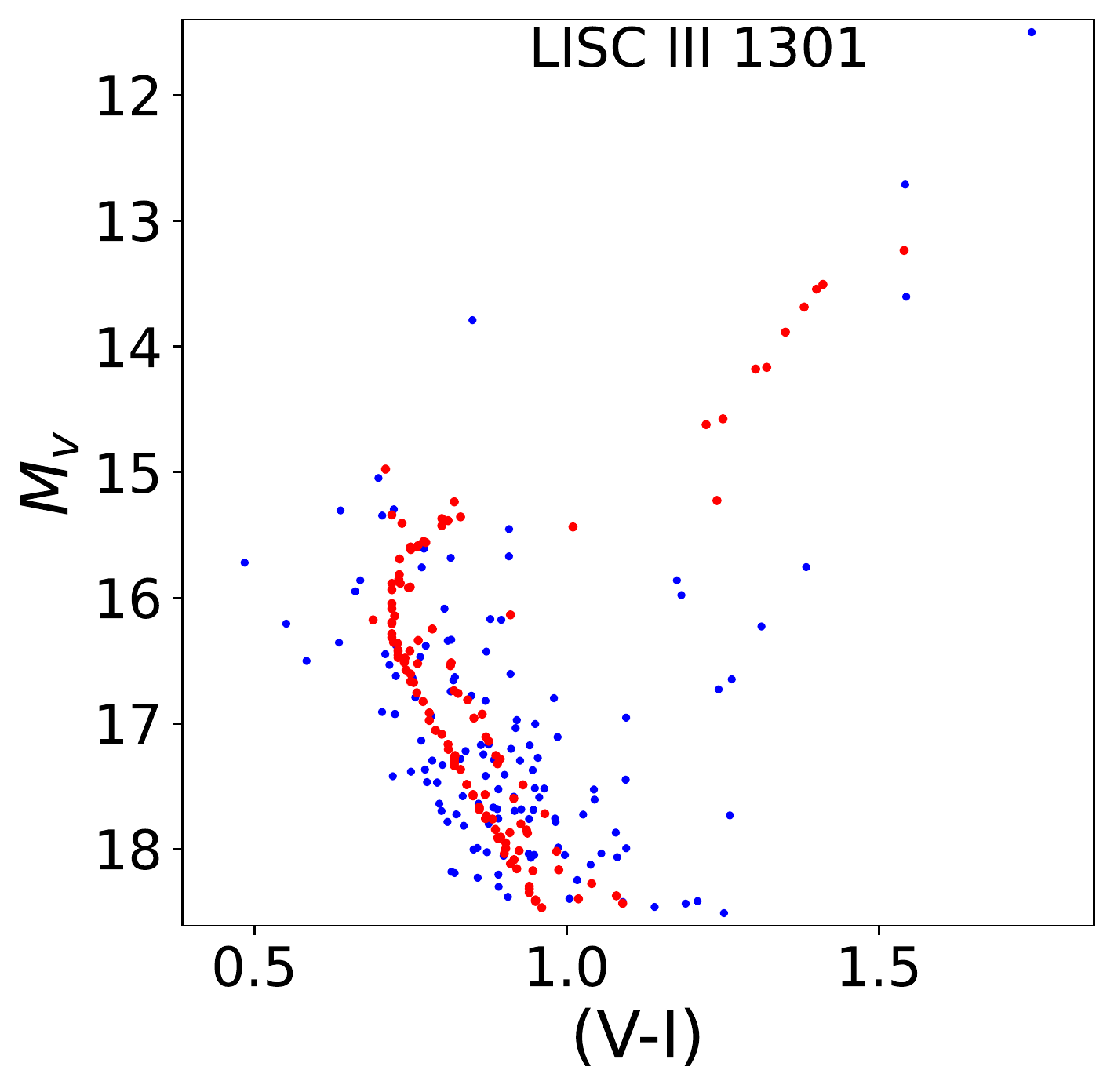}
}
\end{center}

\begin{center}
\subfloat {
\includegraphics[width=1.8in,height=1.7in]{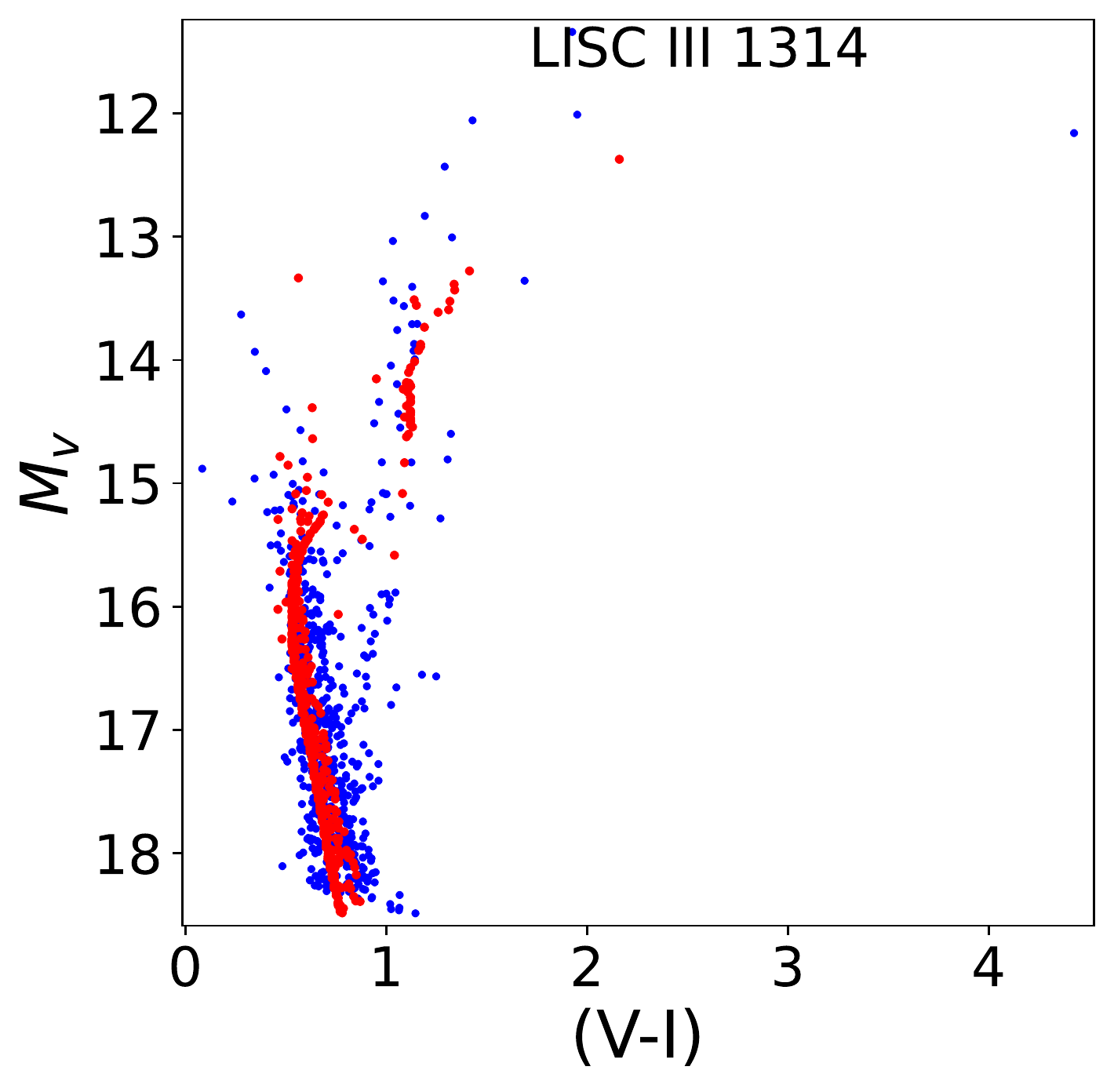}
}
\subfloat  {
\includegraphics[width=1.8in,height=1.7in]{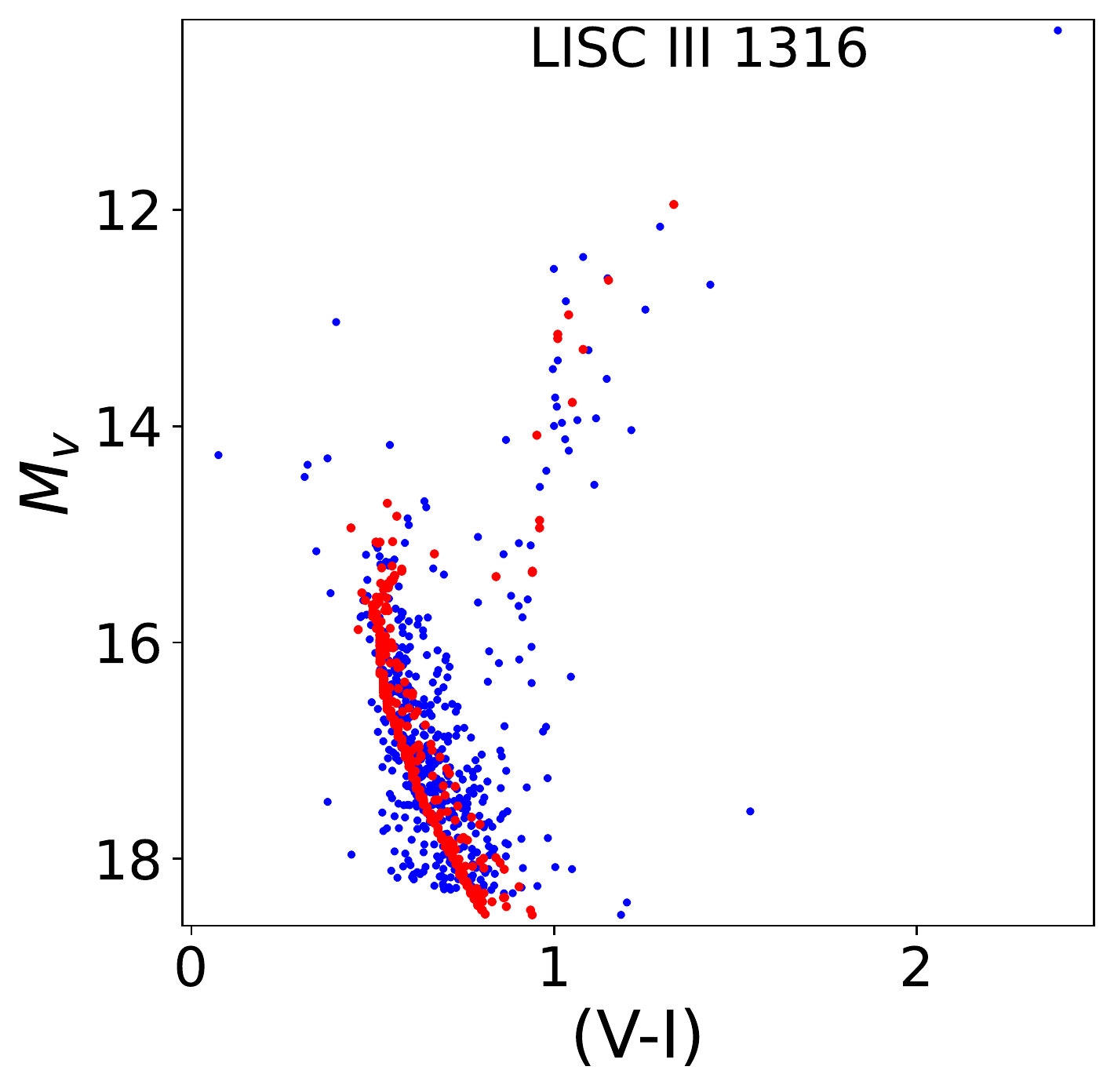}
}
\subfloat {
\includegraphics[width=1.8in,height=1.7in]{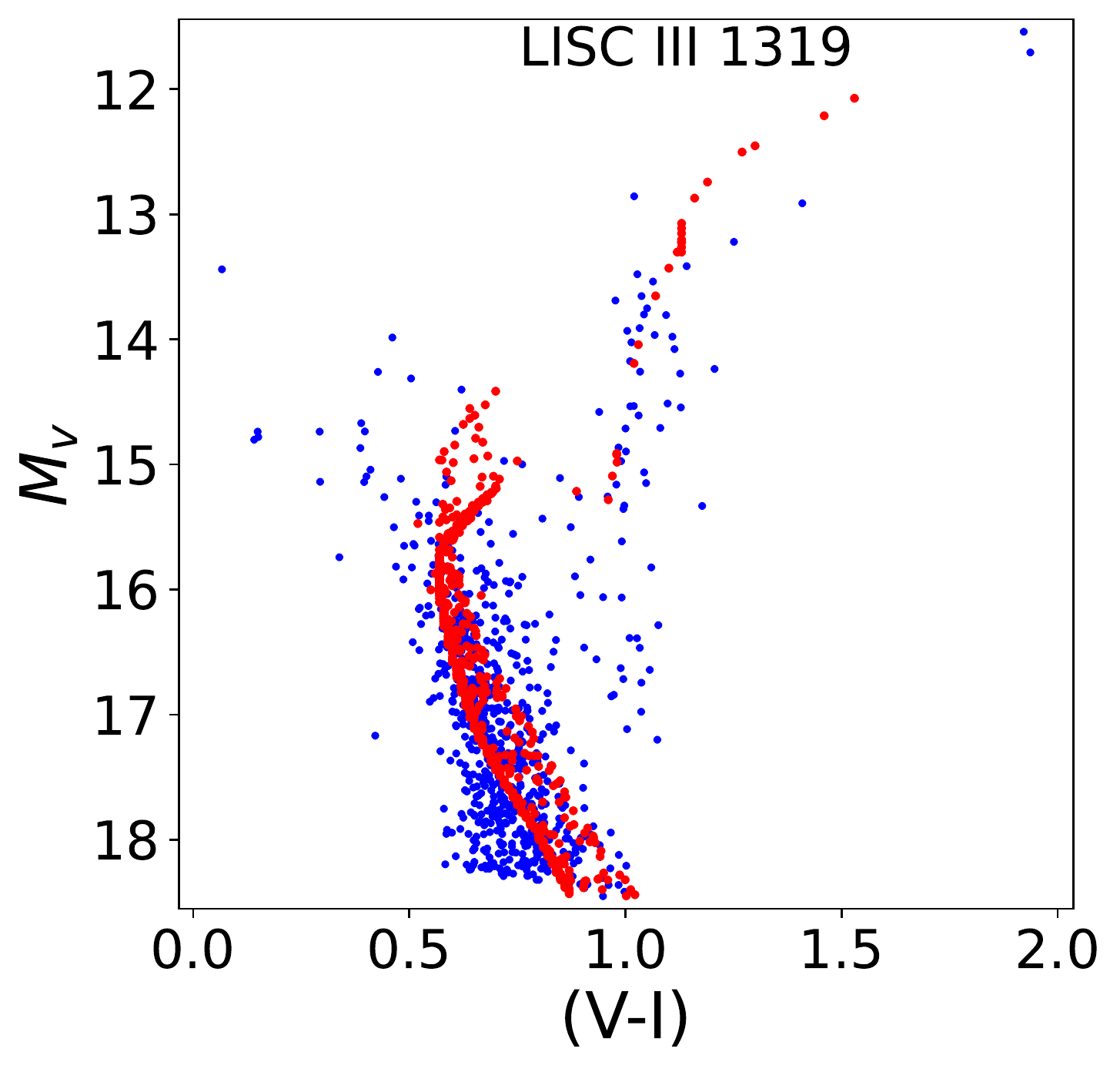}
}
\end{center}
\begin{center}
\subfloat {
\includegraphics[width=1.8in,height=1.7in]{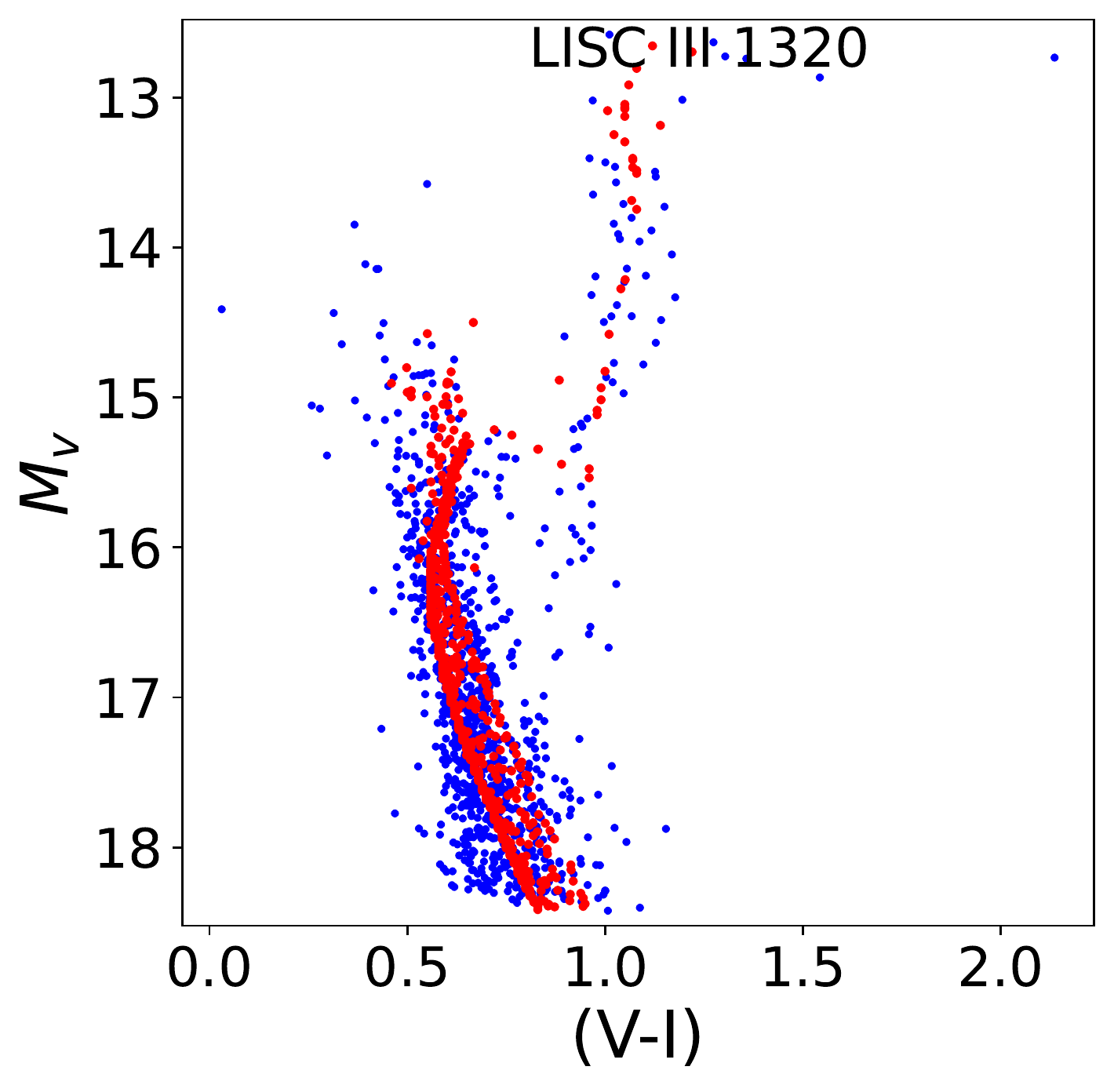}
}
\subfloat  {
\includegraphics[width=1.8in,height=1.7in]{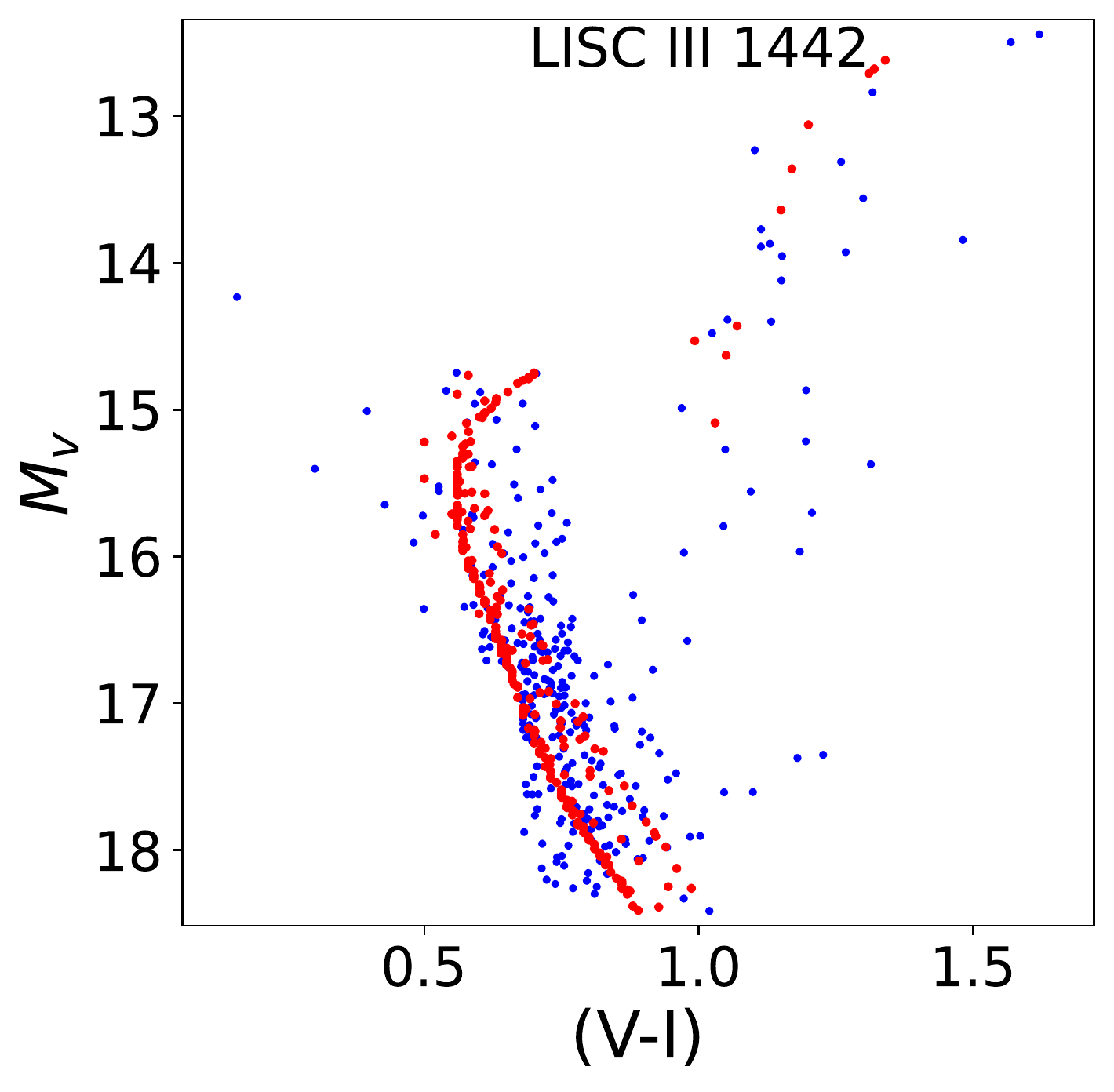}
}
\subfloat  {
\includegraphics[width=1.8in,height=1.7in]{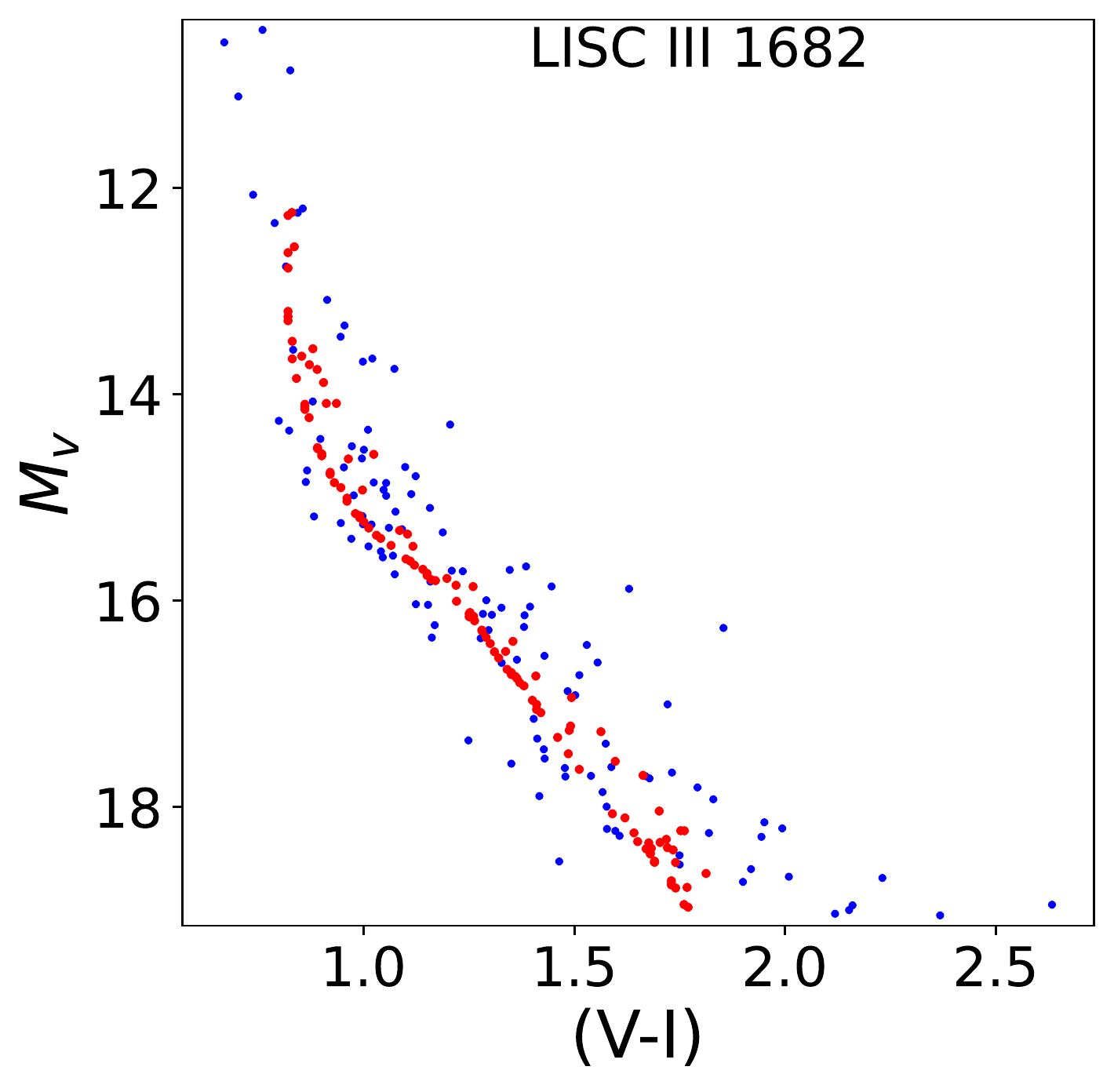}
}
\end{center}

\begin{center}
\subfloat {
\includegraphics[width=1.8in,height=1.7in]{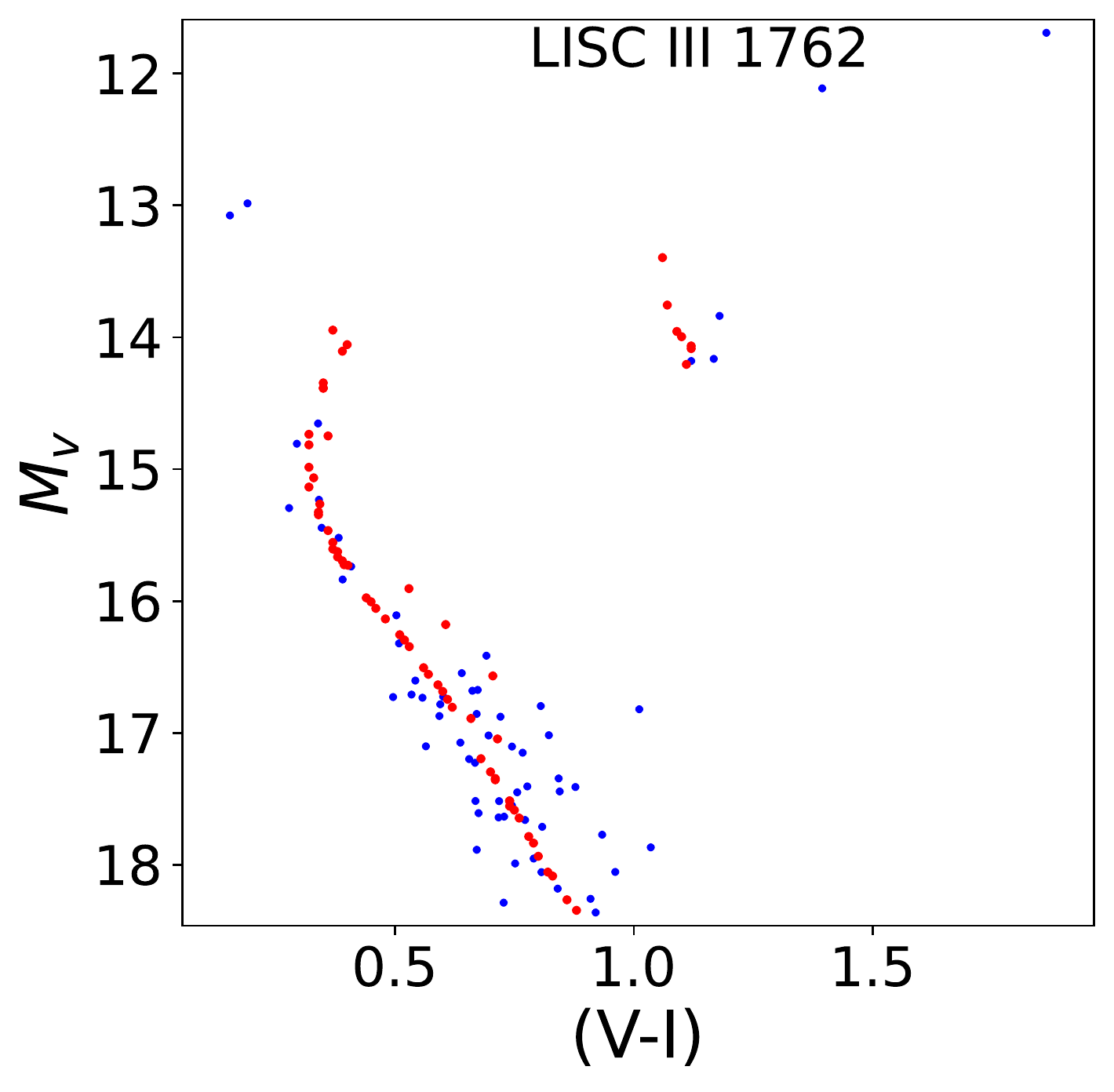}
}
\subfloat  {
\includegraphics[width=1.8in,height=1.7in]{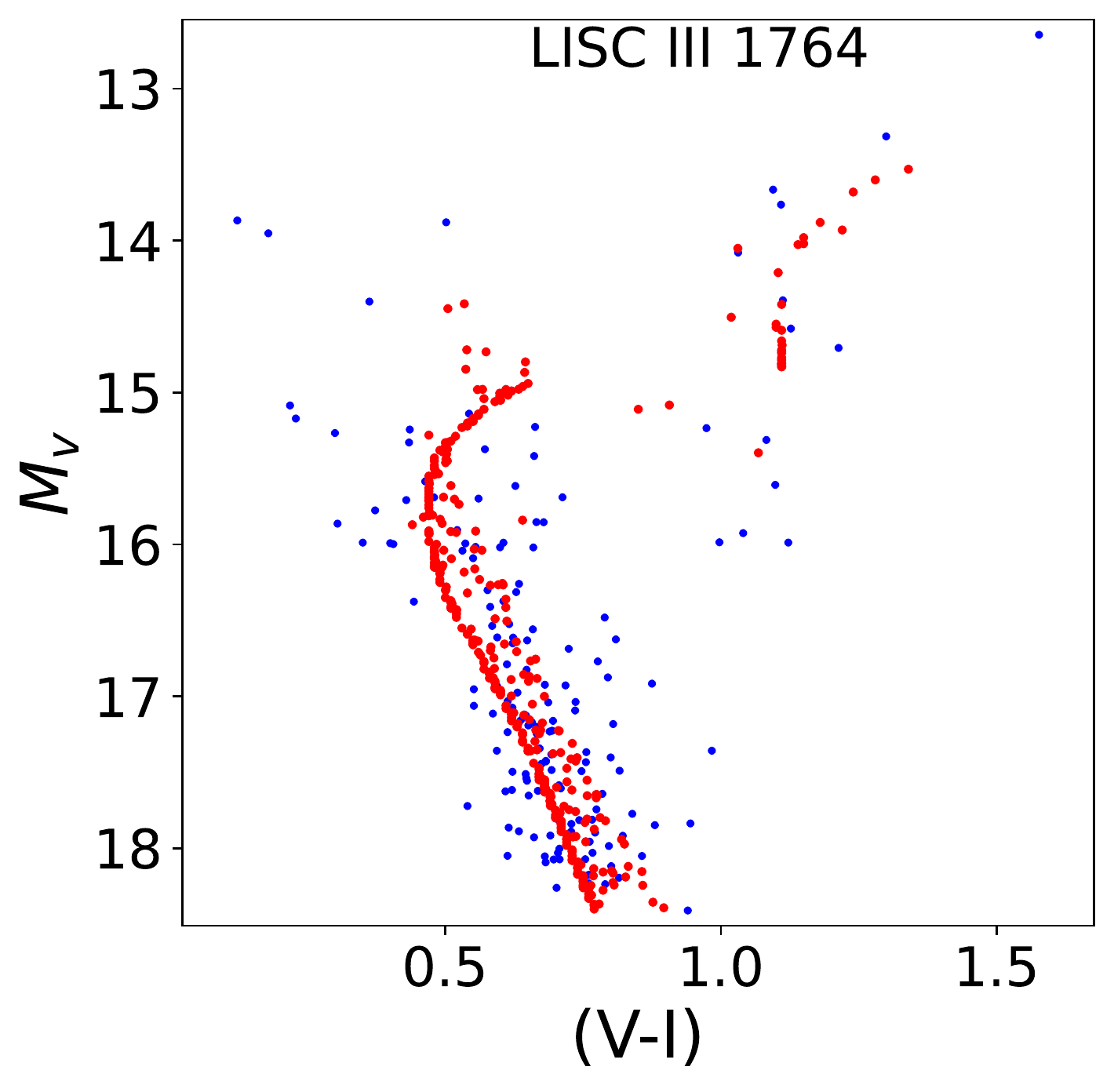}
}
\subfloat  {
\includegraphics[width=1.8in,height=1.7in]{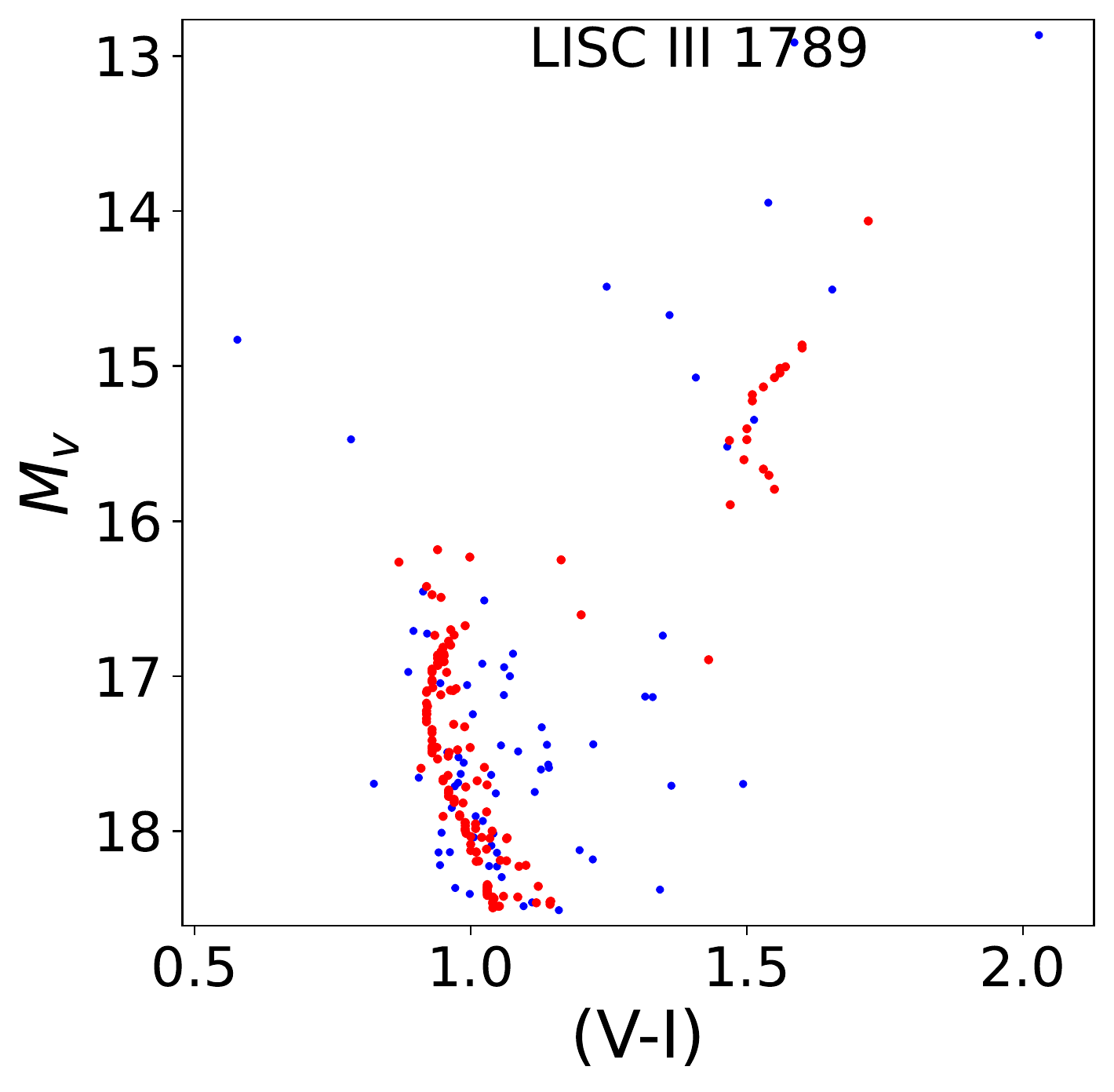}
}
\end{center}
\begin{center}
\subfloat {
\includegraphics[width=1.8in,height=1.7in]{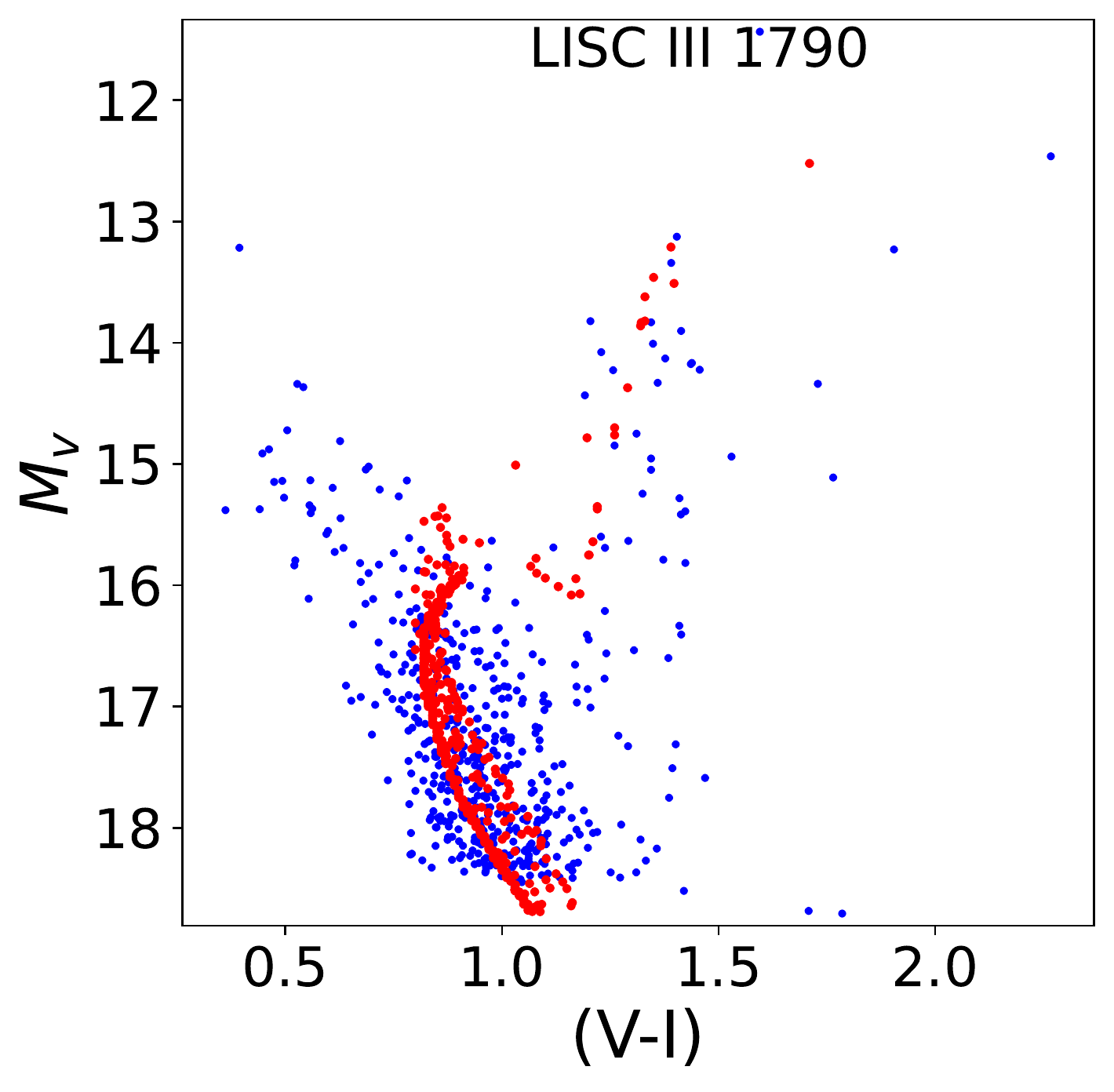}
}
\subfloat  {
\includegraphics[width=1.8in,height=1.7in]{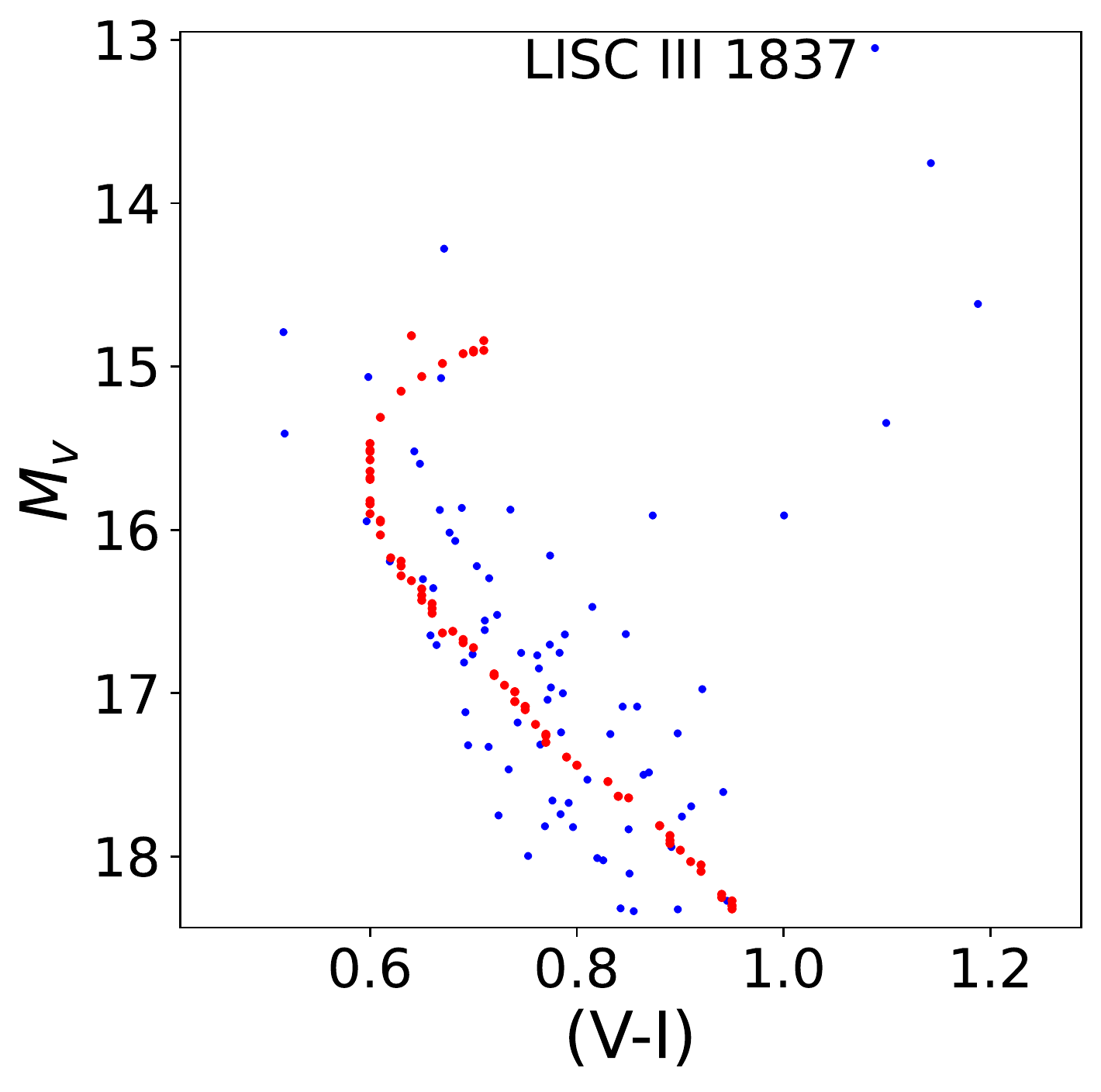}
}
\subfloat  {
\includegraphics[width=1.8in,height=1.7in]{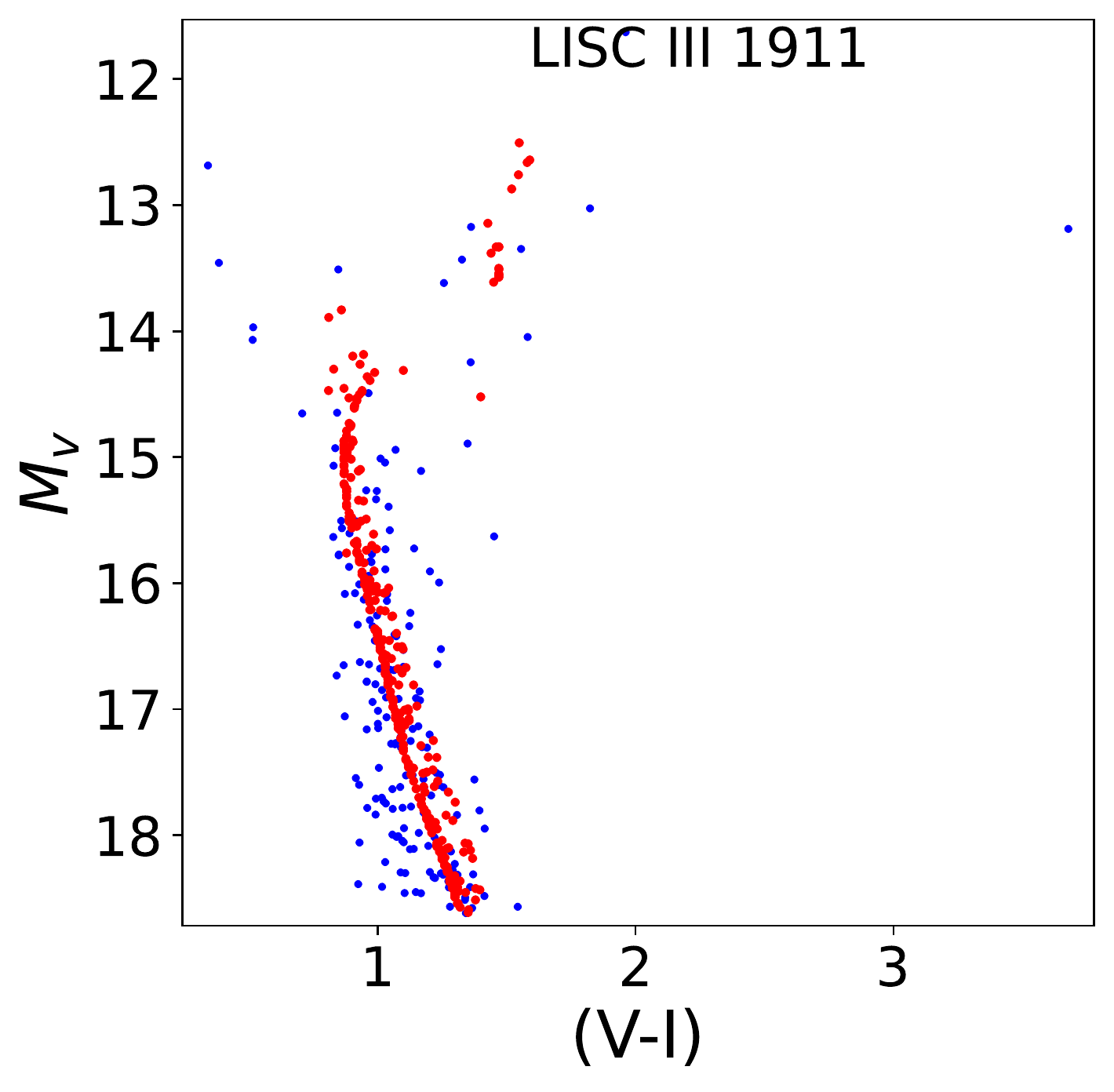}
}

\end{center}

\begin{center}

\caption{Same as Fig.~\ref{fig:CMD_1} but for other 15 OCs.}
\label{fig:CMD_2}
\end{center}
\end{figure*}

\begin{figure*}
\begin{center}
\subfloat {
\includegraphics[width=1.8in,height=1.7in]{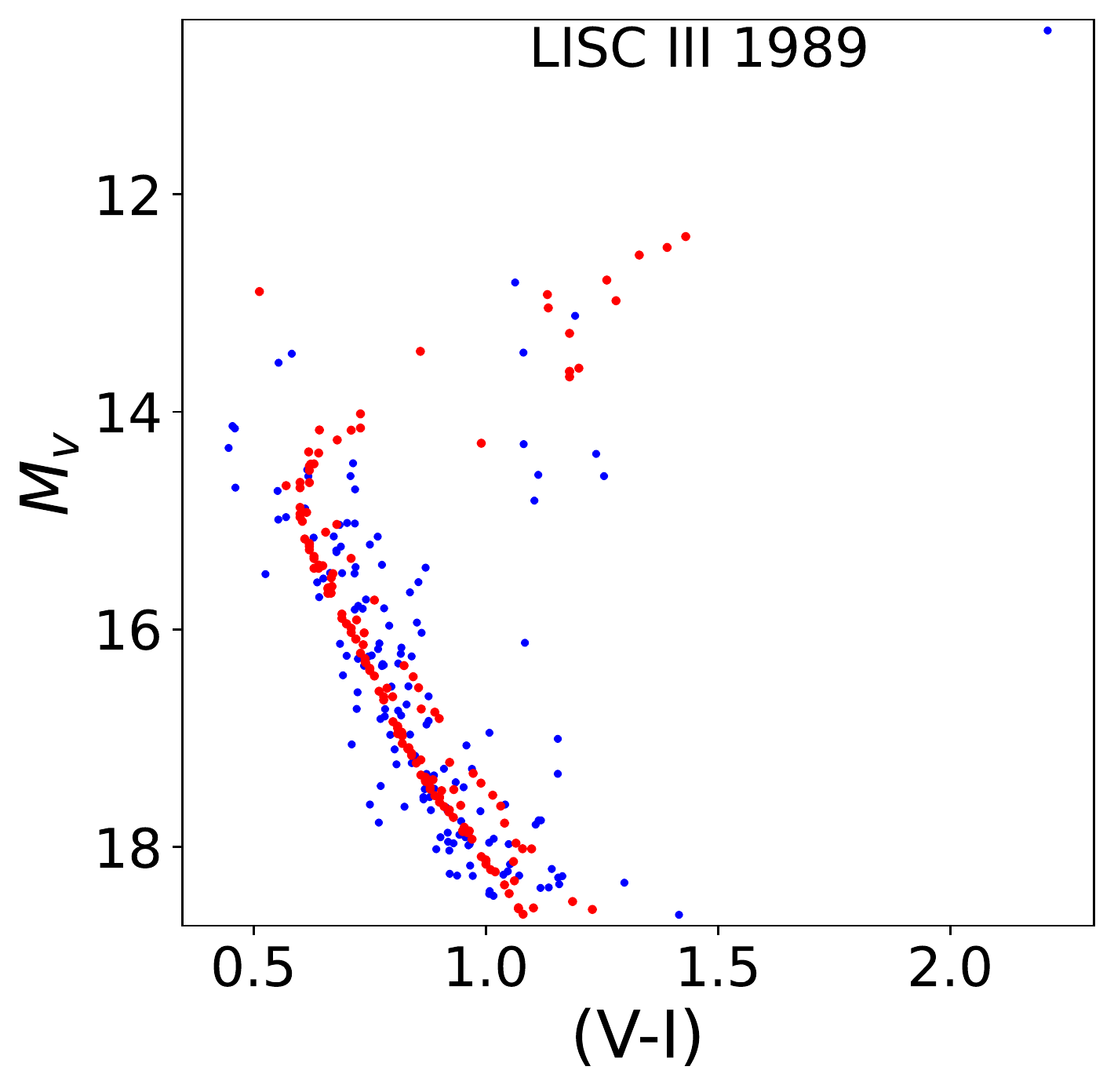}
}
\subfloat  {
\includegraphics[width=1.8in,height=1.7in]{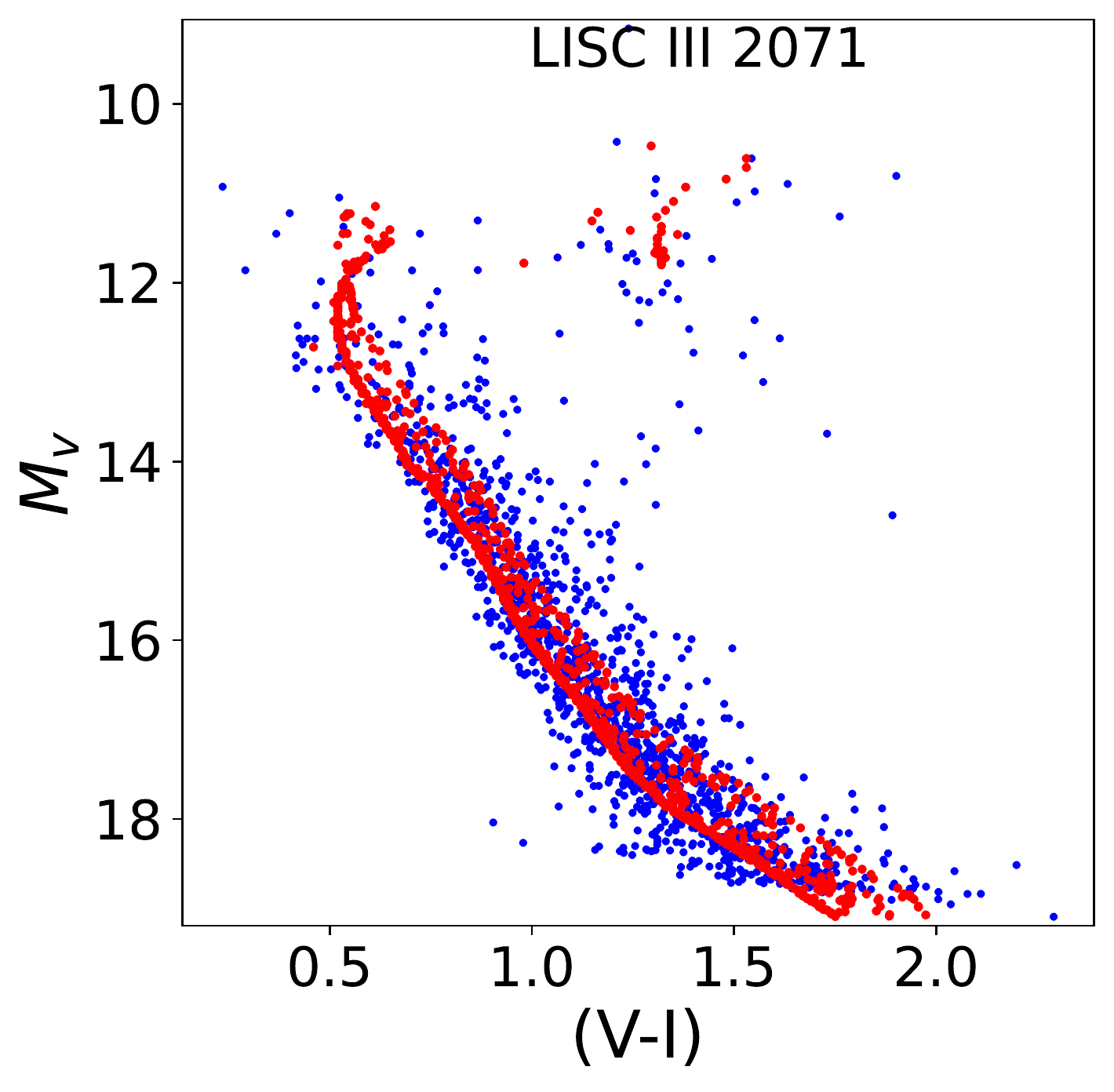}
}
\subfloat {
\includegraphics[width=1.8in,height=1.7in]{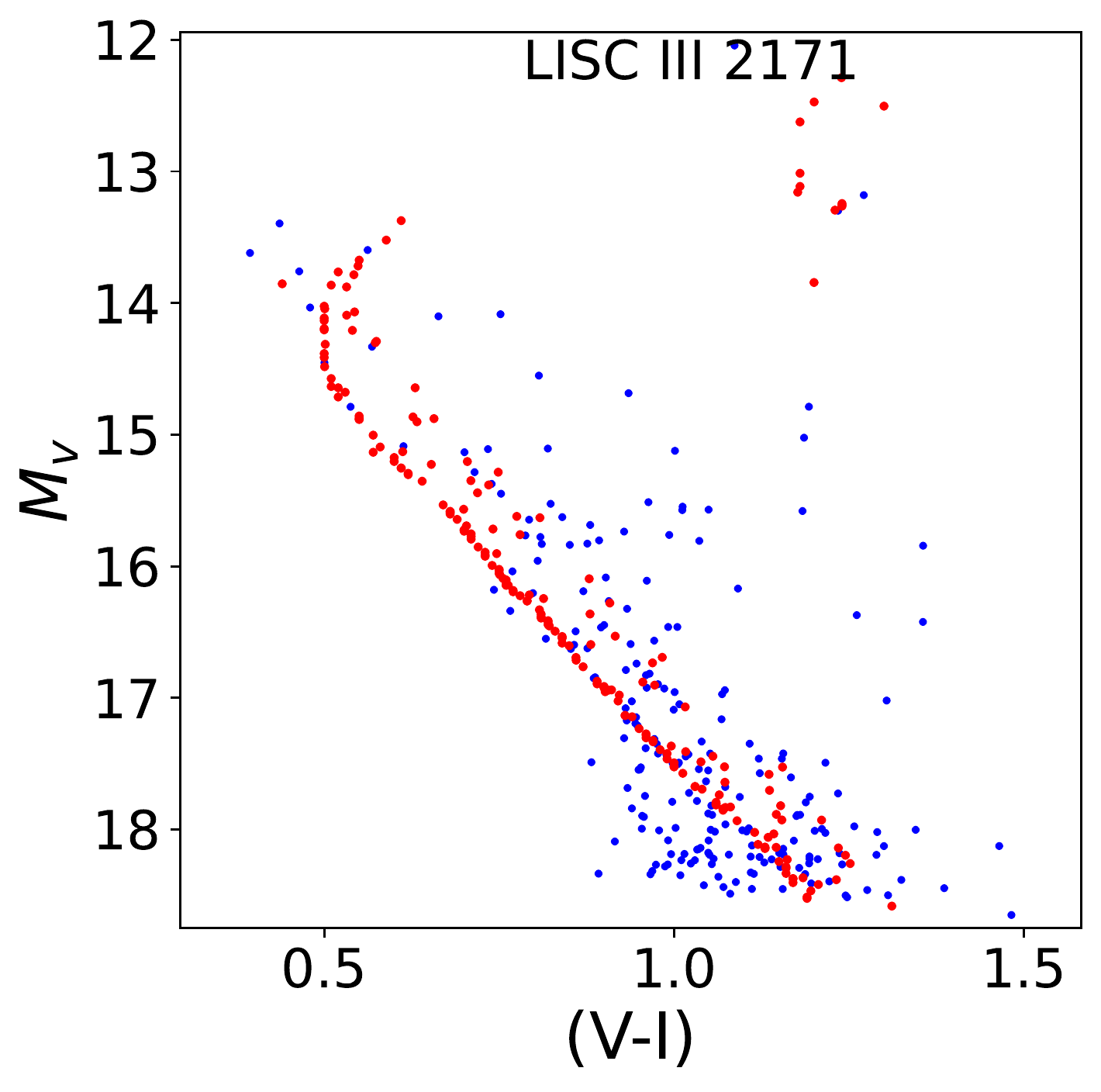}
}
\end{center}

\begin{center}
\subfloat {
\includegraphics[width=1.8in,height=1.7in]{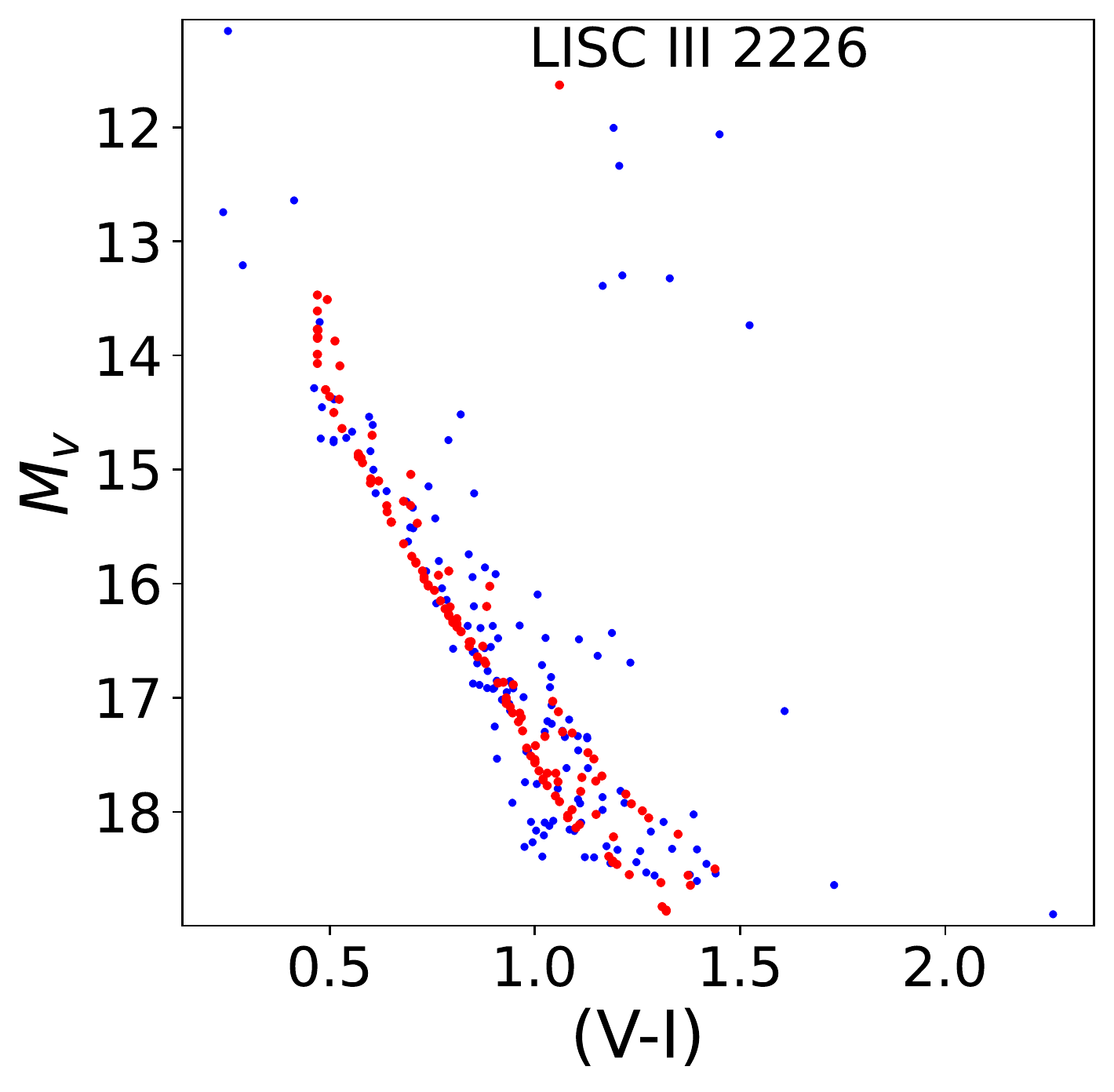}
}
\subfloat  {
\includegraphics[width=1.8in,height=1.7in]{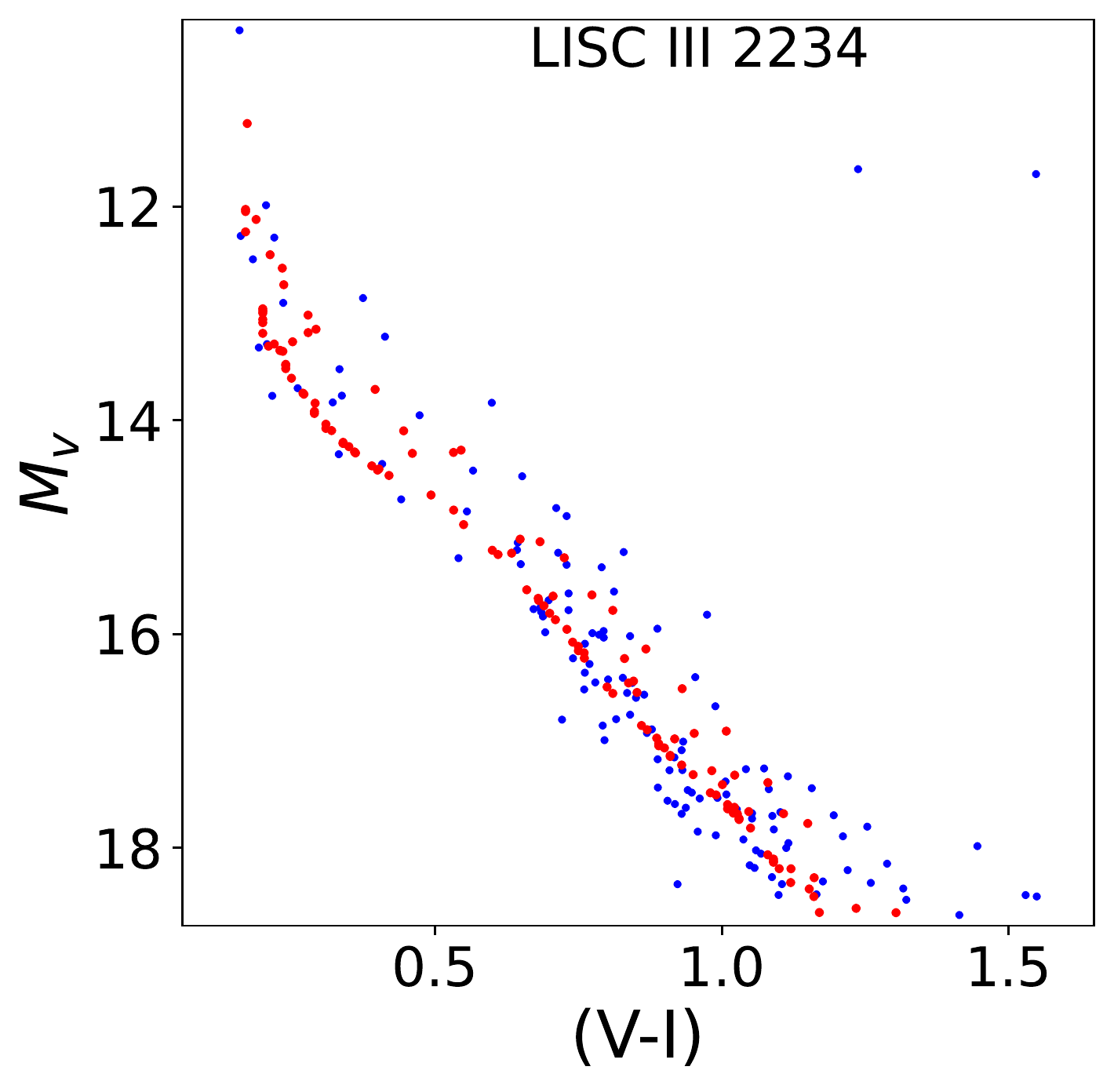}
}
\subfloat {
\includegraphics[width=1.8in,height=1.7in]{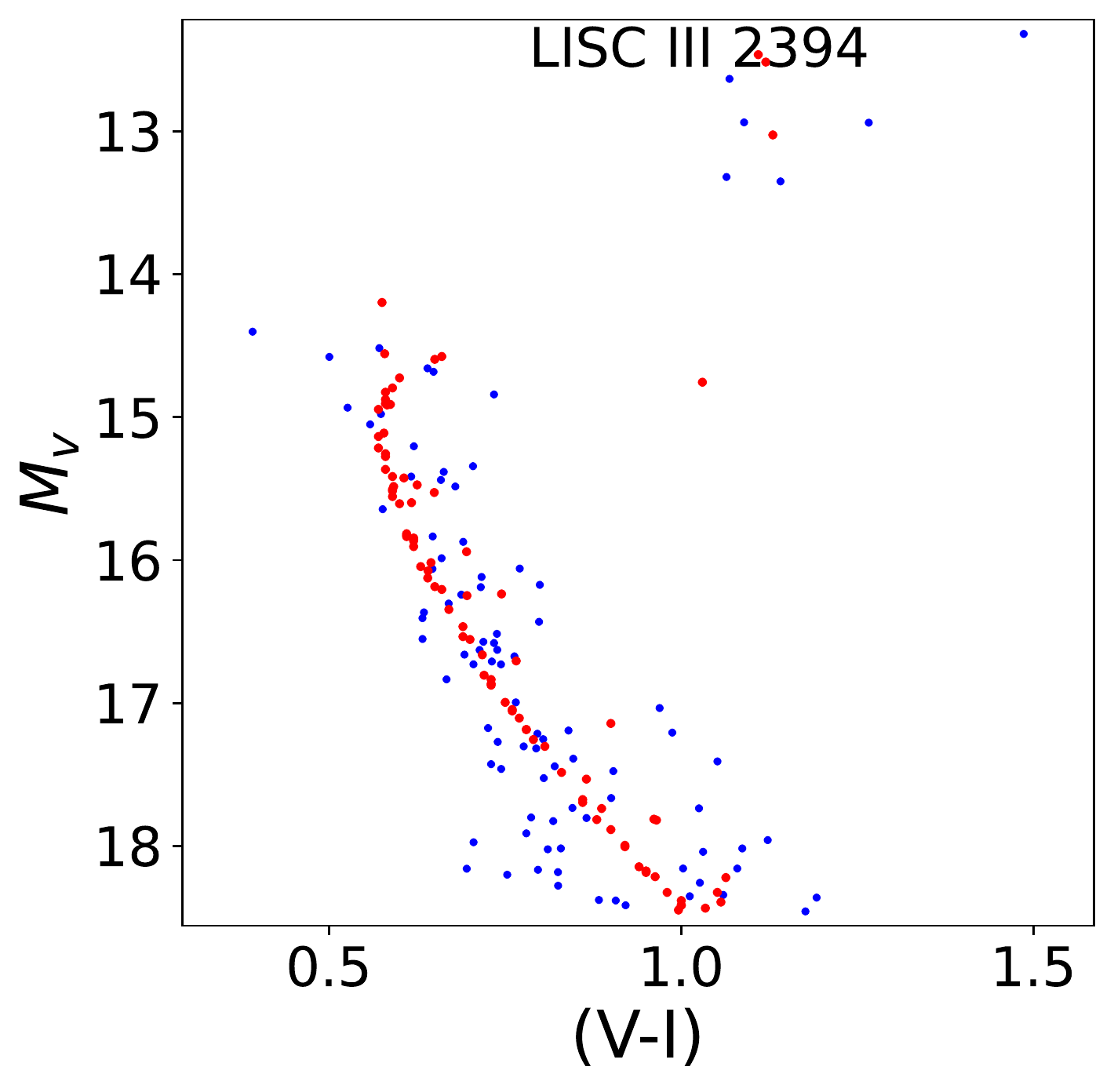}
}
\end{center}
\begin{center}
\subfloat {
\includegraphics[width=1.8in,height=1.7in]{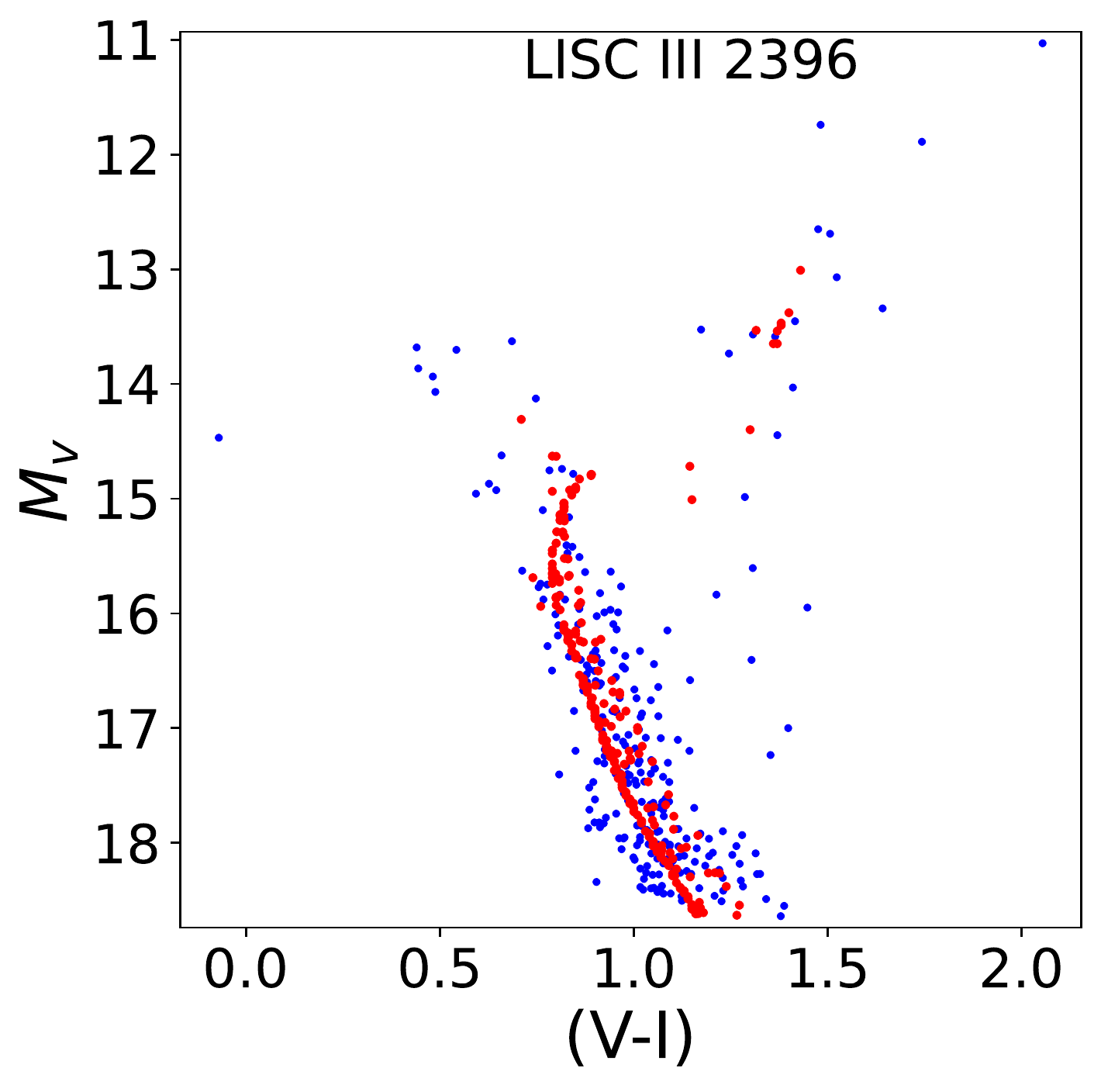}
}
\subfloat  {
\includegraphics[width=1.8in,height=1.7in]{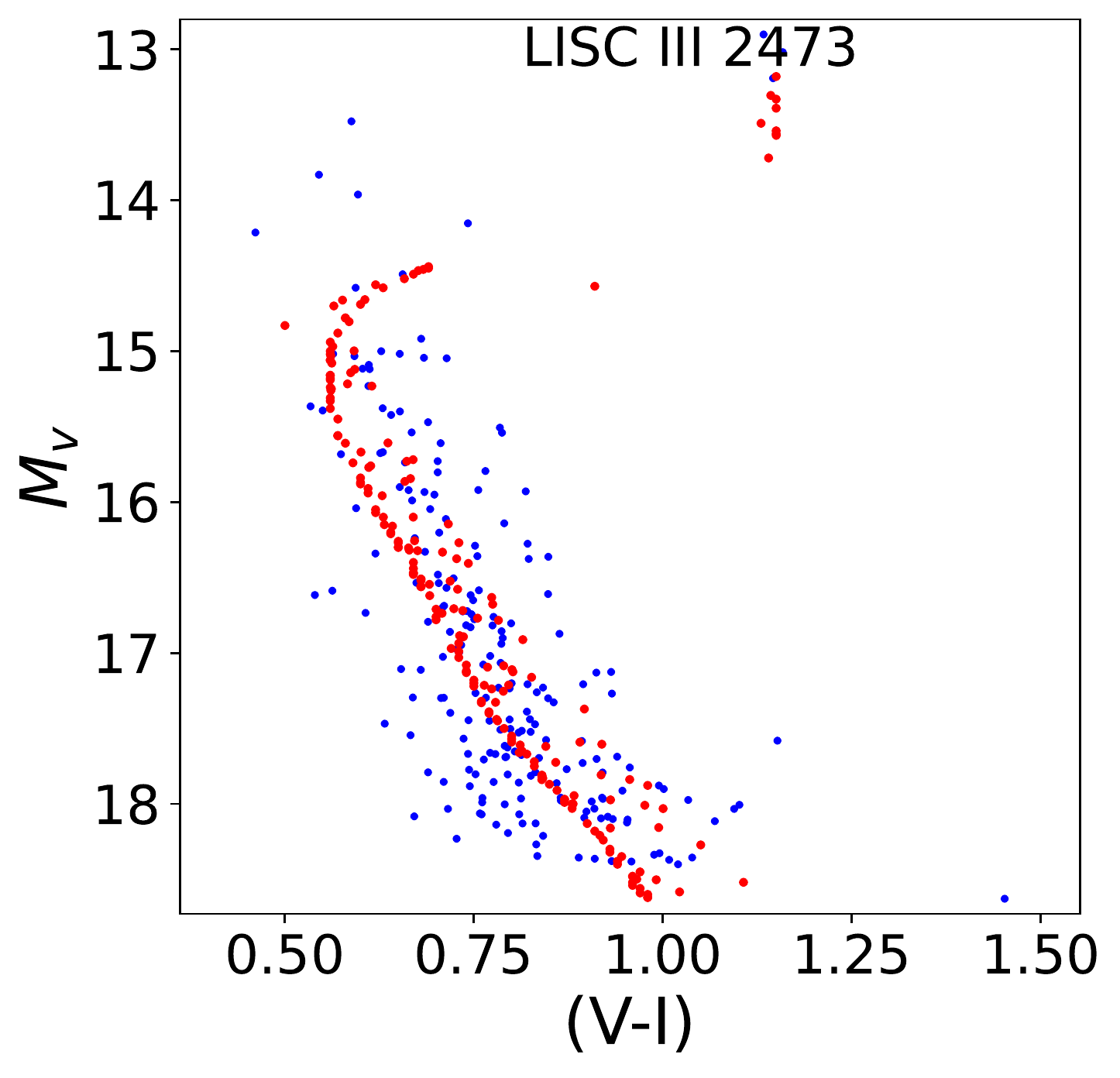}
}
\subfloat  {
\includegraphics[width=1.8in,height=1.7in]{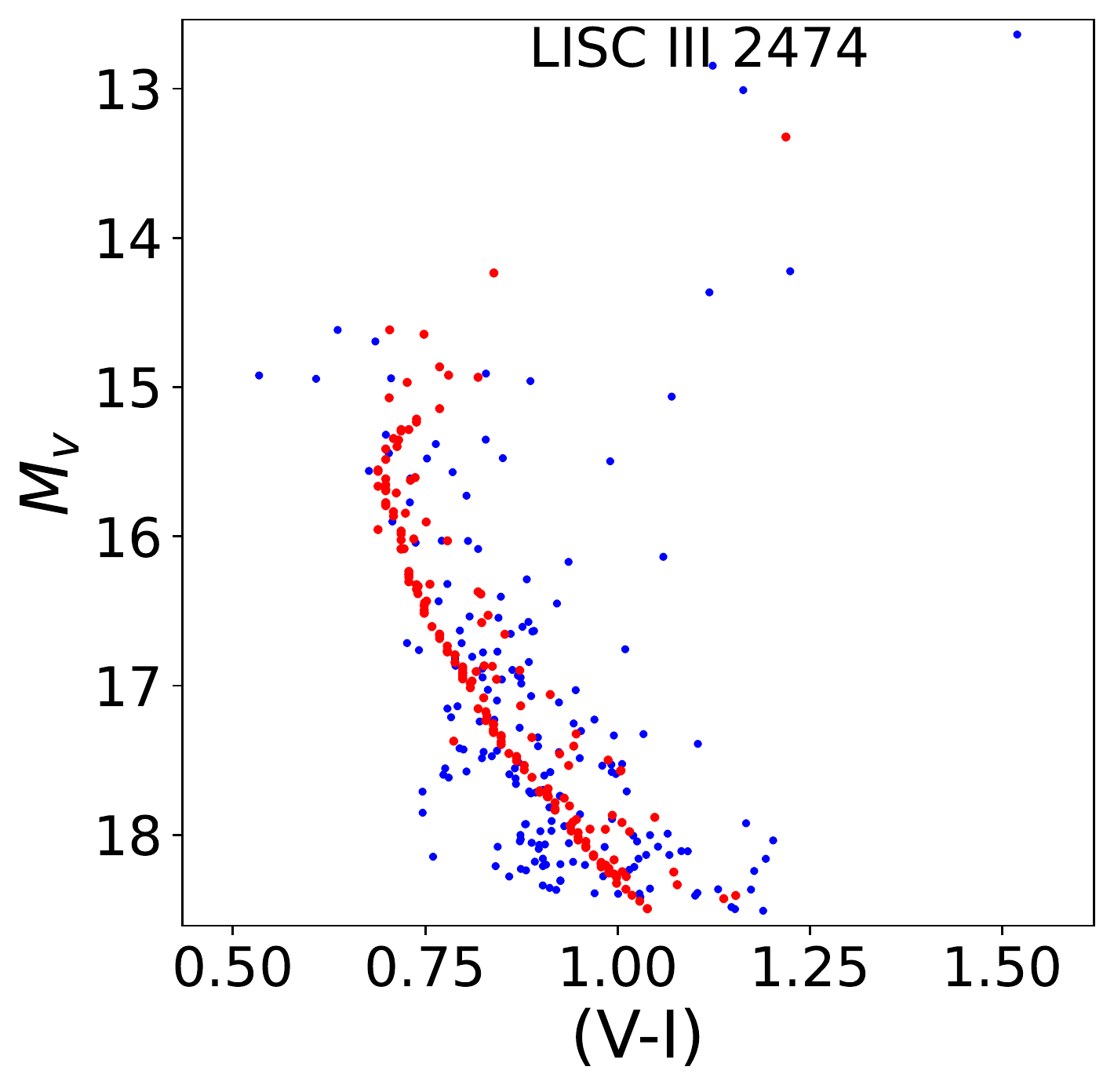}
}
\end{center}

\begin{center}
\subfloat {
\includegraphics[width=1.8in,height=1.7in]{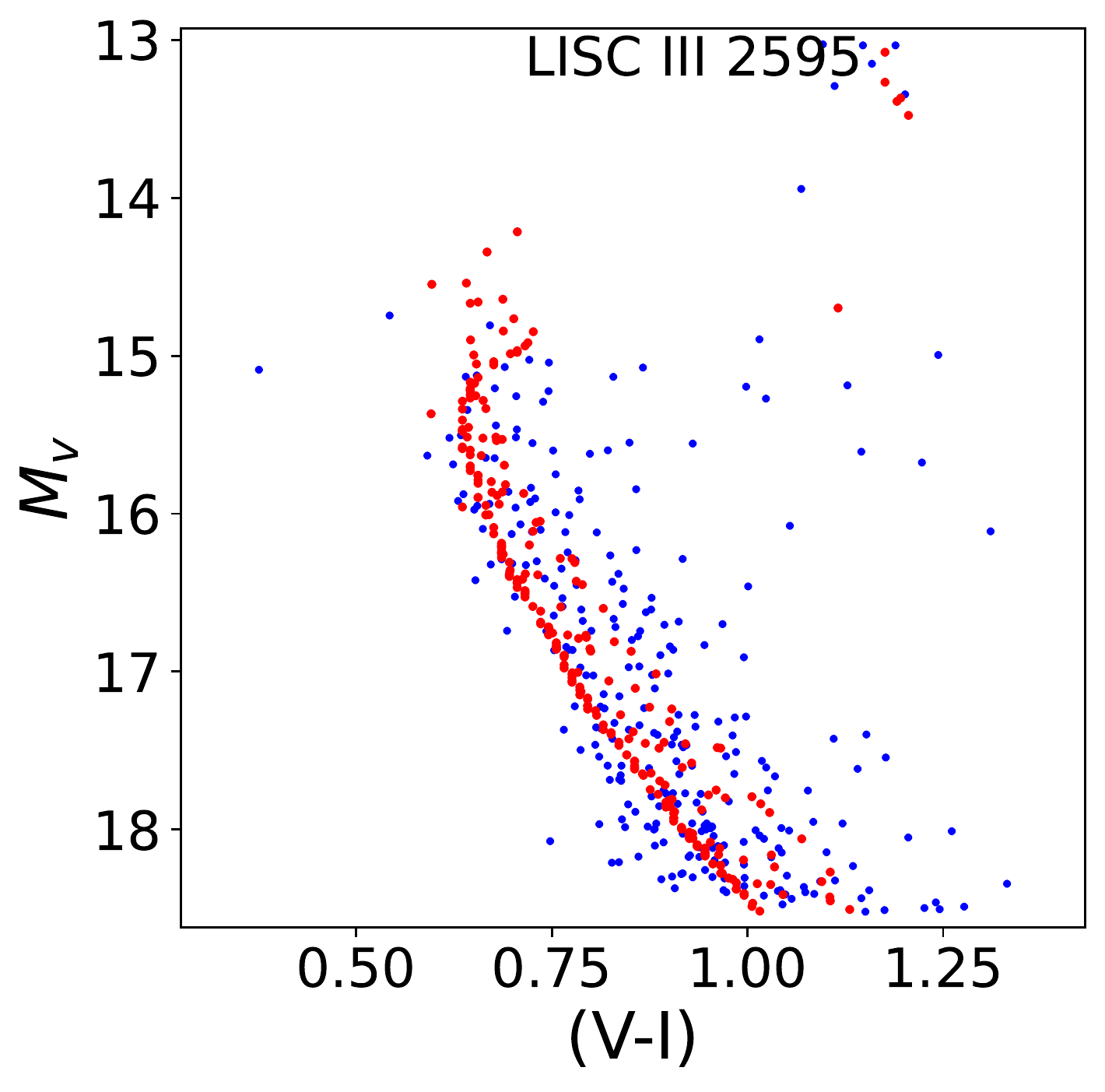}
}
\subfloat  {
\includegraphics[width=1.8in,height=1.7in]{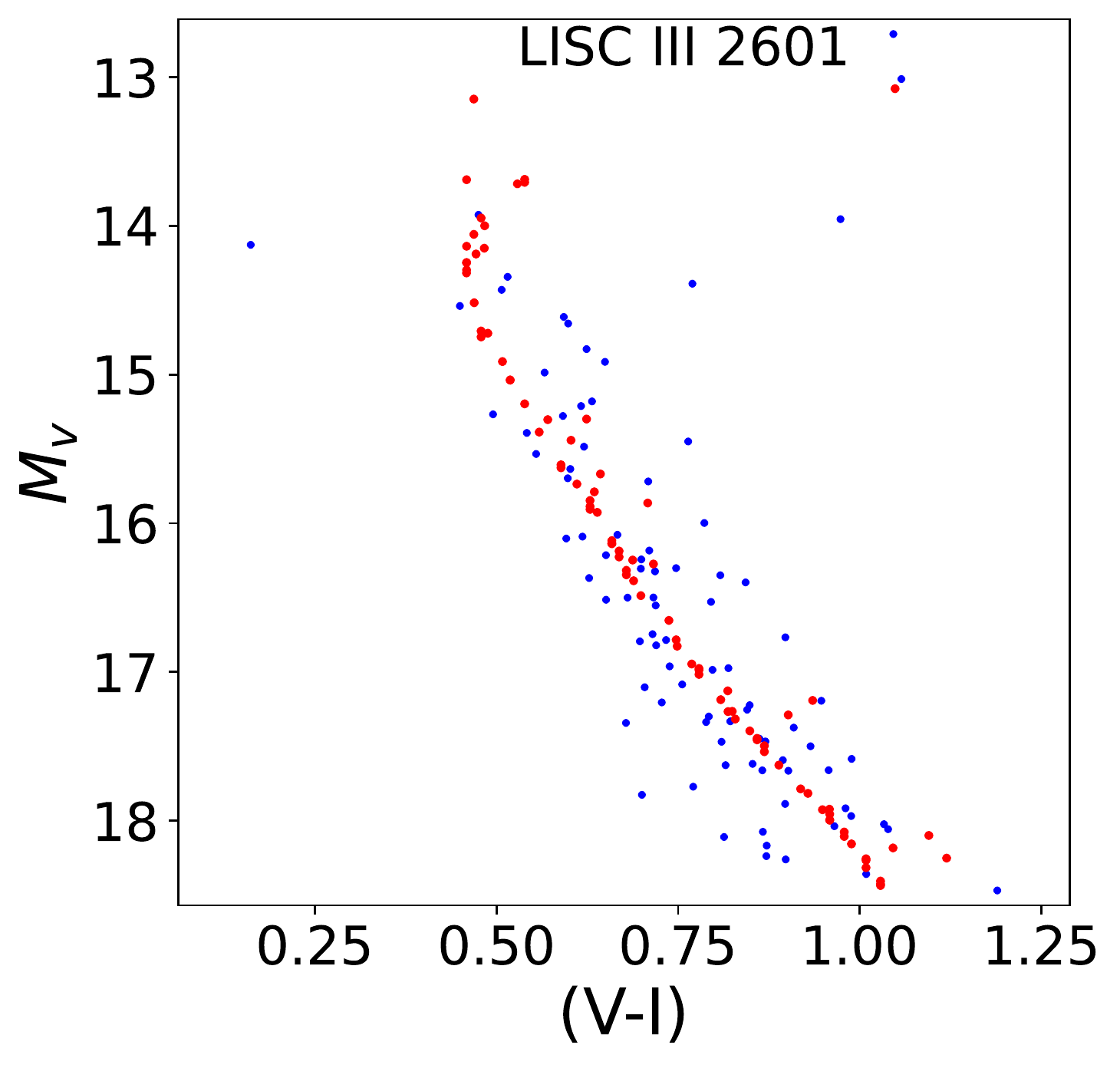}
}
\subfloat  {
\includegraphics[width=1.8in,height=1.7in]{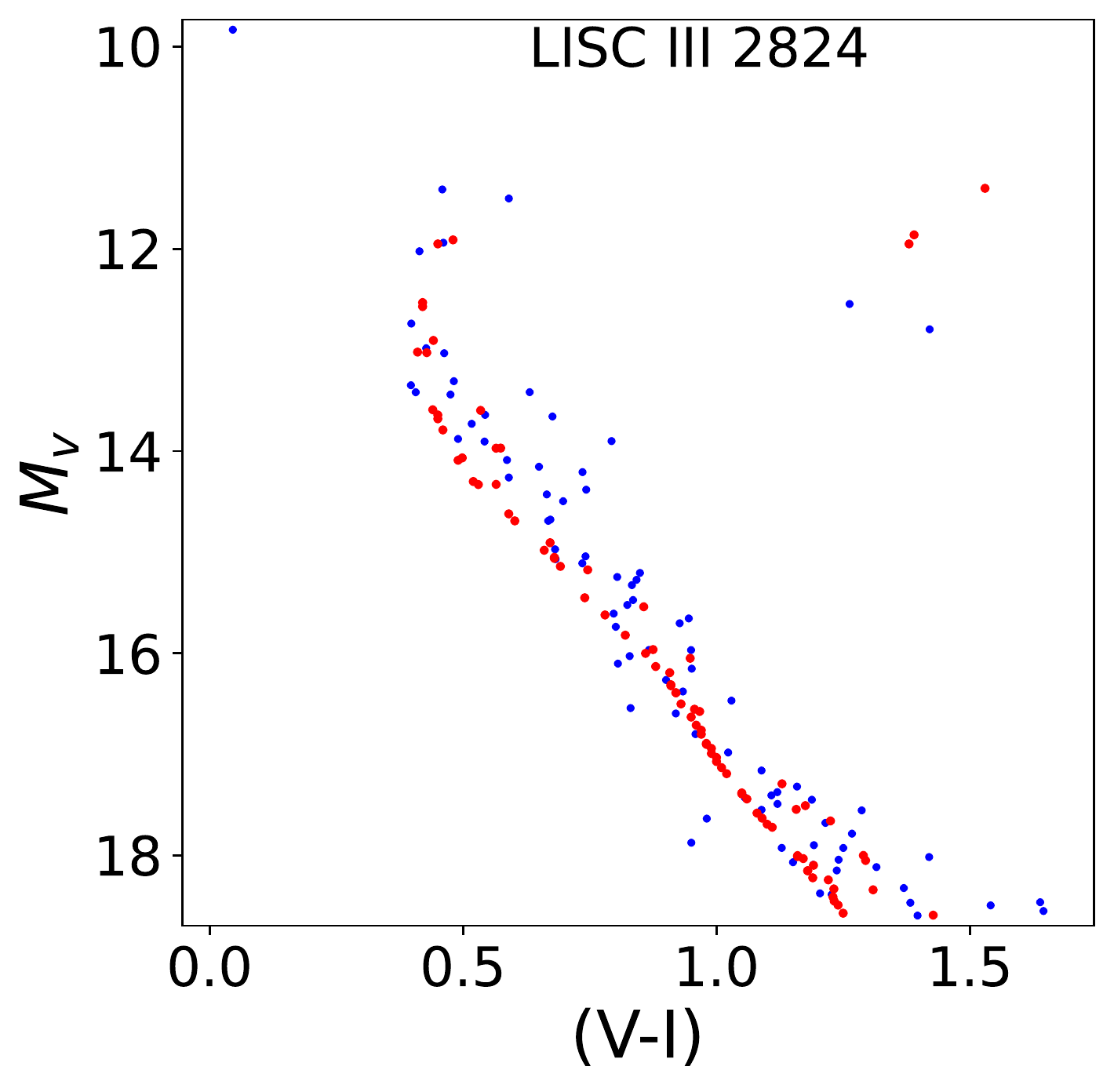}
}
\end{center}
\begin{center}
\subfloat {
\includegraphics[width=1.8in,height=1.7in]{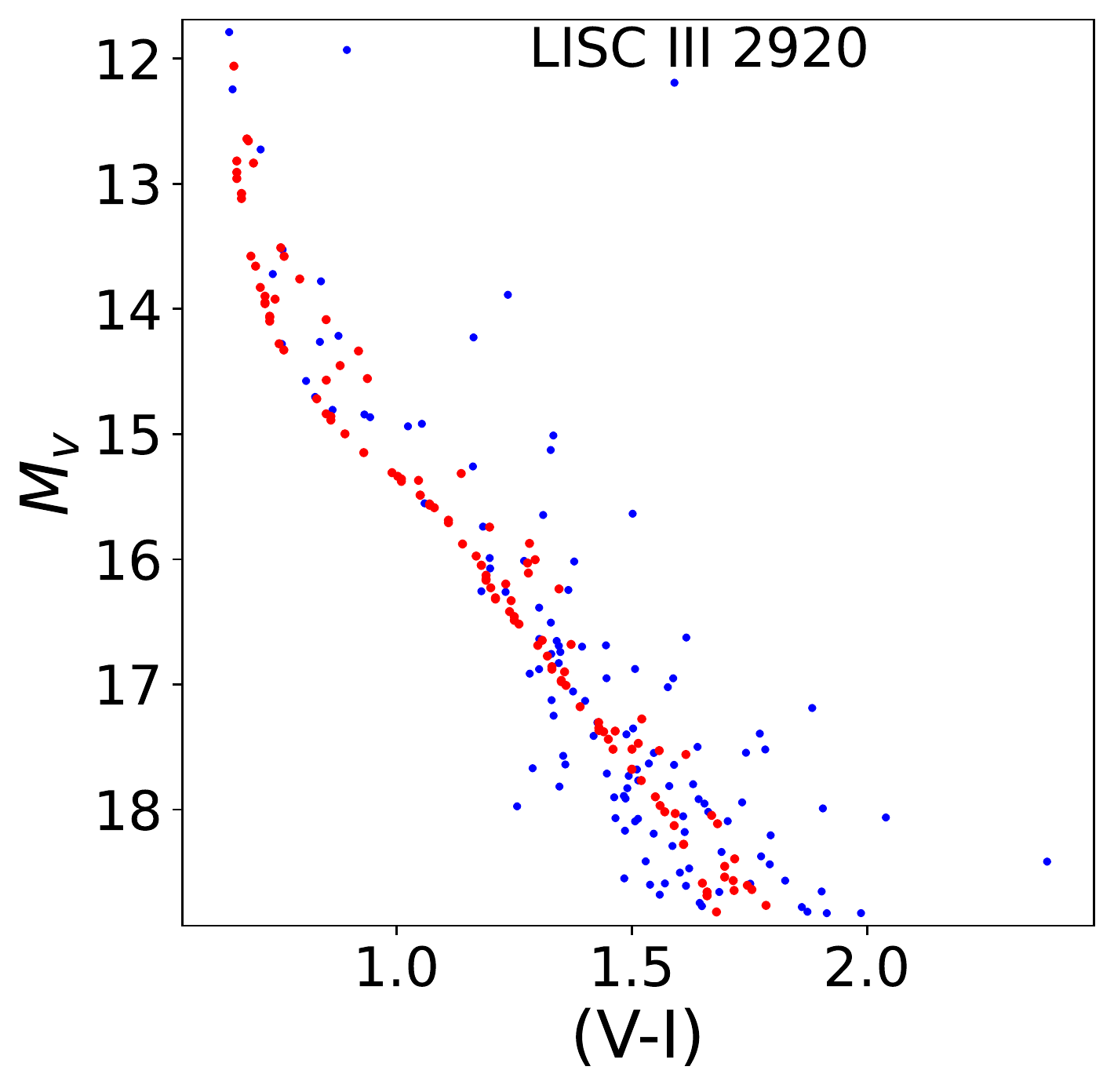}
}
\subfloat  {
\includegraphics[width=1.8in,height=1.7in]{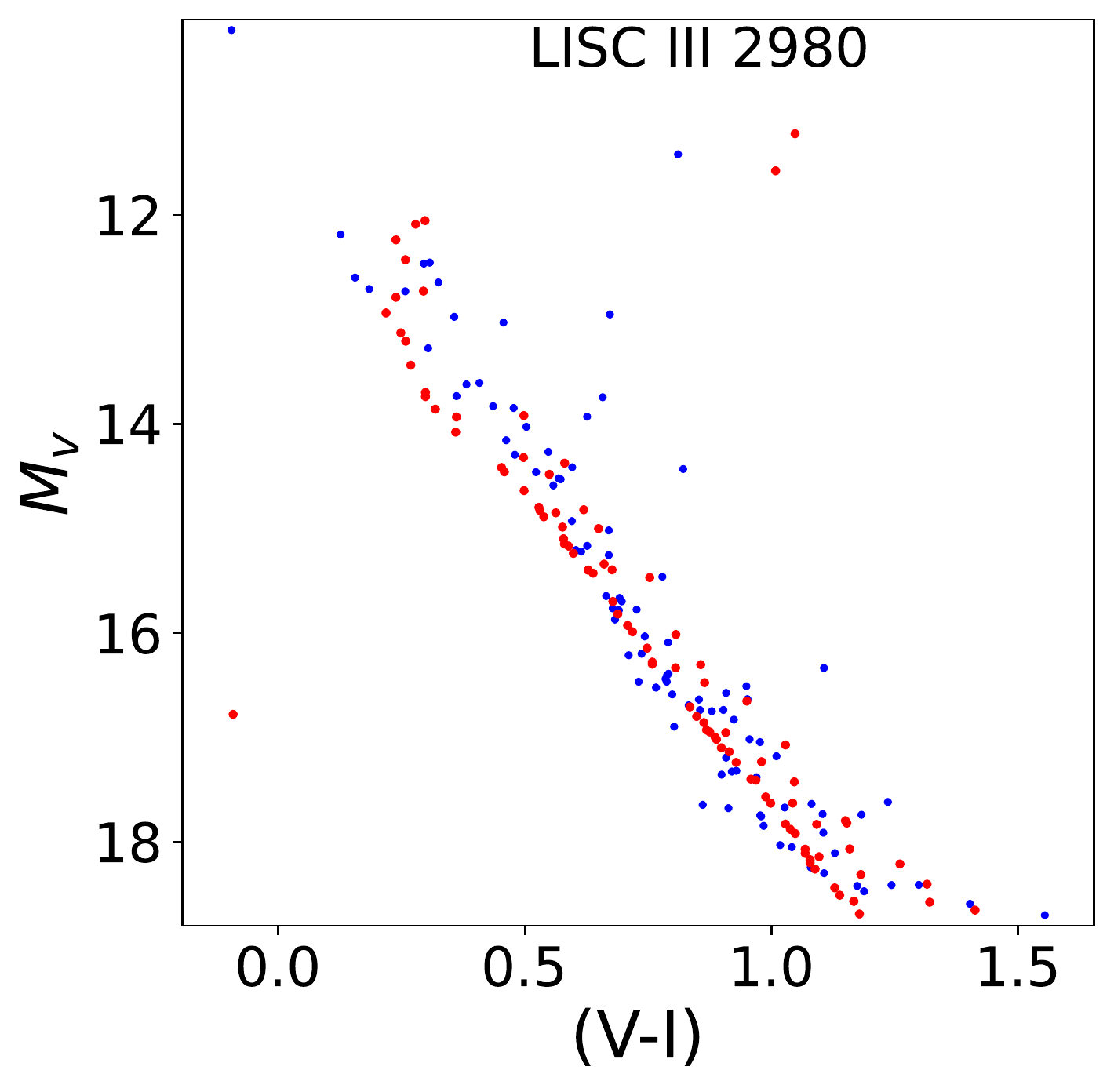}
}
\subfloat  {
\includegraphics[width=1.8in,height=1.7in]{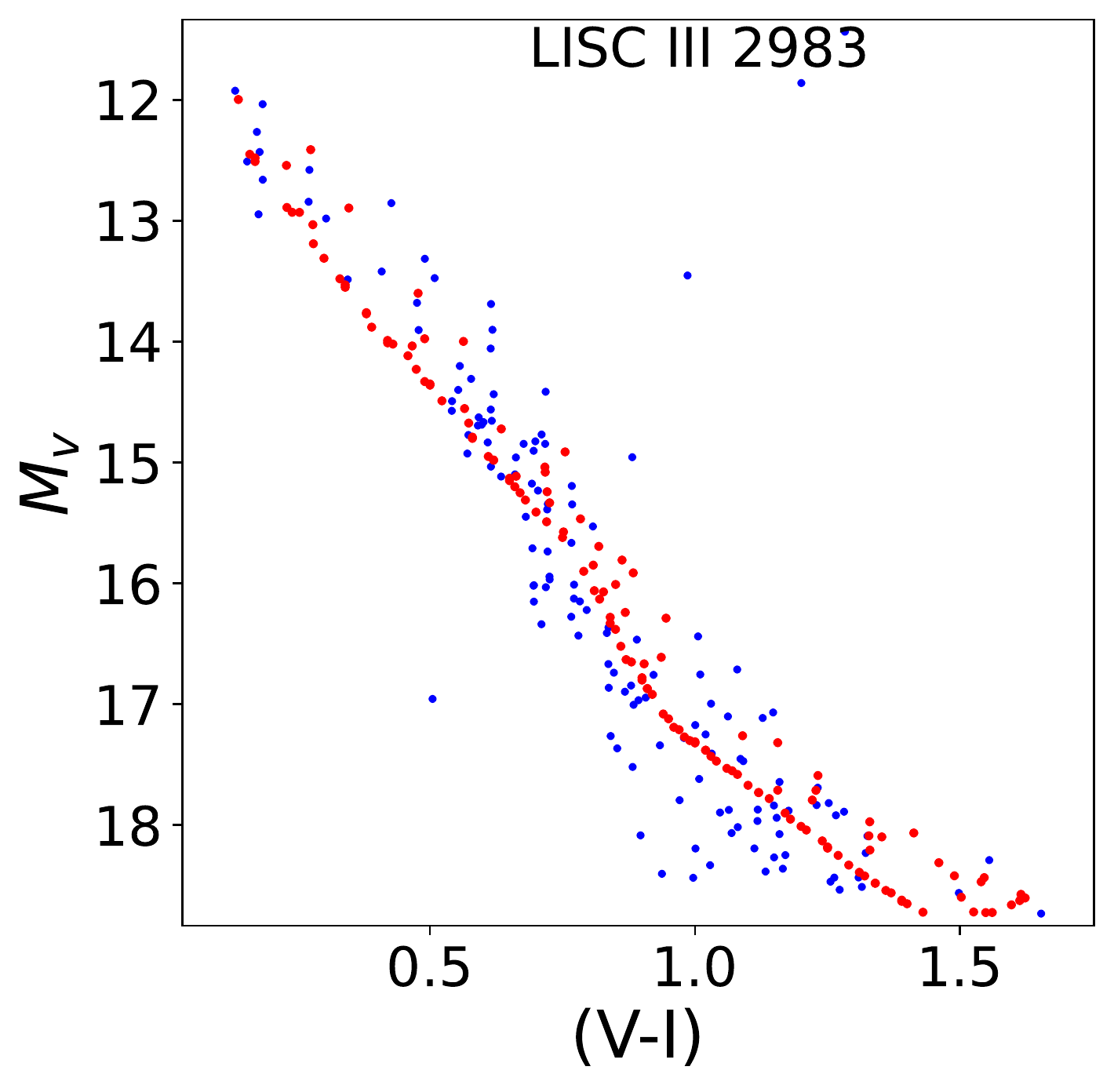}
}
\end{center}

\begin{center}

\caption{Same as Fig.~\ref{fig:CMD_1} but for other 15 OCs.}
\label{fig:CMD_3}
\end{center}
\end{figure*}

This work employs the advance stellar population synthesis model (ASPS) \citep{Li2012} for CMD fitting using  $Powerful\_CMD$  tool \citep{Li2015,Li2016A}. Since CMD fitting is not the focus of this paper,  one can read our recent paper, \citet{Li2022}, for more details.
Seven parameters, i.e., distance modulus $(m $--$ M )$,  color excess $E(V $--$ I)$, 
young stellar age ${t}$, age spread  $t_{sp}$,  binary fraction $f_{bin}$ and rotating star fraction $f_{rot}$ , are determined for each cluster.
Some best fitting CMDs are given in Figure~\ref{fig:CMD_1},~\ref{fig:CMD_2},~\ref{fig:CMD_3} and the best-fitting  parameters  in Table~\ref{tab:param_tab}. 
We show the colour-magniutde diagrams in V-I vs Mv plane
because the Powerful CMD code takes only B, V and I magnitudes, the photometry obtained from Gaia bandpasses are transformed to V, I photometry.
Refer to the~\cite{Riello2021}, $G_{BP}$ and $G_{RP}$ transform to $V-I$ and $M_{V}$ by equation~\ref{equ:6}.

\begin{equation}
\label{equ:6}
   \left\{\begin{array}{l}
M_{V}-G_{B P}-0.02696+0.1086 (V-I)-0.009148 (V-I)^2+0.004715 (V-I)^3=0 \\
G_{R P}-0.01612+1.274 (V-I)-0.08143 (V-I)^2-G_{B P}=0
\end{array}\right.
\end{equation}

The isochrones that were calculated by~\citet{Li2017}.
Using the rapid stellar evolution code of~\citet{Hurley2002} are used for CMD fitting. The best isochrone is chosen according to the statistical difference between the theoretical and observed CMDs. In detail, the CMD is divided into some grids and the difference is computed via 
\begin{equation}
    \mathrm{WAD}=\Sigma[|F_{\mathrm{ob}}-F_{\mathrm{th}}\|]
\end{equation}
where $F_{ob}$ and $F_{th}$ are star fractions of the observed stars and simulated stars in a grid. In the CMD fitting metallicity Z and stellar age are set to free parameter.

% 83 candidates CMDs

% 83 candidates CMDs group 3

% Parameters of the final 83 New OCs

\begin{table*}[htbp]
\center
\caption{Parameters of the final 83 New OCs. Some best-fit and observed CMDs are compared in Figure~\ref{fig:CMD_1}.}
\resizebox{\textwidth}{!}{
\begin{tabular}{lcccccccccccccccccc}
\toprule
  &    ID &    l &  l\_std &       b &  b\_std &    plx &  plx\_std &    pmra &  pmra\_std &   pmdec &  pmdec\_std &     N\_member &  $Z$ &  $m -M$ &  E $( V-I)$ &  t/$t_{sp}$ &  $f_{bin}$ &  $f_{rot}$ \\
 & & [deg] & [deg] & deg] & [deg]  &[mas]  & [mas] & [$mas~yr^{-1}$ ] & [$mas~yr^{-1}$] & [$mas ~yr^{-1}$] &[$mas~yr^{-1}$]& & & [mas] &  [mas]& [Gyr] & &  \\
\midrule
0  &   931 &  203.643 &  0.412 &   4.020 &  0.108 &  0.239 &    0.016 &  -0.308 &     0.266 &  -0.828 &      0.293 &   116 &  0.0040 &  12.7200 &  0.2900 &  1.1 &  0.53 &   0.0 \\
1  &   932 &  204.163 &  0.245 &   5.734 &  0.113 &  0.245 &    0.016 &  -0.385 &     0.269 &  -0.602 &      0.304 &    95 &  0.0300 &  13.4000 &  0.0800 &  1.2 &  0.34 &   0.0 \\
2  &   933 &  205.529 &  1.188 &   4.789 &  0.631 &  0.247 &    0.020 &  -0.374 &     0.289 &  -0.682 &      0.318 &  1633 &  0.0100 &  13.1300 &  0.2400 &  0.7 &  0.55 &   0.0 \\
3  &   934 &  206.504 &  0.728 &   3.699 &  0.332 &  0.250 &    0.019 &  -0.374 &     0.278 &  -0.597 &      0.311 &   542 &  0.0300 &  13.5000 &  0.2800 &  0.2 &  0.55 &   0.0 \\
4  &  1031 &  157.044 &  0.448 &   5.264 &  0.544 &  0.258 &    0.026 &   0.504 &     0.272 &  -0.907 &      0.372 &   361 &  0.0040 &  12.4000 &  0.6800 &  0.7 &  0.54 &   0.0 \\
5  &  1036 &  151.713 &  0.534 &   5.769 &  0.315 &  0.251 &    0.020 &   0.394 &     0.288 &  -0.522 &      0.349 &   401 &  0.0003 &  12.5000 &  0.6700 &  0.9 &  0.53 &   0.0 \\
6  &  1038 &  149.444 &  0.569 &   5.704 &  0.339 &  0.257 &    0.026 &   0.348 &     0.320 &  -0.427 &      0.377 &   925 &  0.0040 &  12.6000 &  0.4300 &  1.3 &  0.55 &   0.0 \\
7  &  1046 &  180.344 &  0.364 &   6.652 &  0.359 &  0.280 &    0.027 &   0.181 &     0.340 &  -1.671 &      0.371 &   268 &  0.0200 &  12.7000 &  0.4000 &  0.6 &  0.60 &   0.0 \\
8  &  1177 &  180.903 &  0.411 &  11.987 &  0.136 &  0.277 &    0.025 &   0.220 &     0.331 &  -1.718 &      0.290 &   106 &  0.0100 &  12.7200 &  0.2000 &  1.8 &  0.51 &   0.0 \\
9  &  1178 &  184.312 &  0.554 &  11.758 &  0.123 &  0.283 &    0.018 &   0.170 &     0.343 &  -1.681 &      0.315 &    98 &  0.0100 &  12.8900 &  0.2000 &  1.7 &  0.55 &   0.0 \\
10 &  1179 &  180.577 &  0.490 &  11.363 &  0.142 &  0.262 &    0.018 &   0.267 &     0.347 &  -1.774 &      0.307 &   107 &  0.0100 &  13.2700 &  0.3000 &  1.6 &  0.38 &   0.0 \\
11 &  1182 &  195.679 &  0.480 &  11.966 &  0.159 &  0.294 &    0.026 &  -0.225 &     0.388 &  -1.434 &      0.401 &   149 &  0.0200 &  13.2000 &  0.1500 &  1.0 &  0.46 &   0.0 \\
12 &  1191 &  224.412 &  0.428 &  11.083 &  0.203 &  0.269 &    0.021 &  -1.561 &     0.398 &   0.404 &      0.436 &   146 &  0.0100 &  12.8600 &  0.1100 &  2.0 &  0.50 &   0.0 \\
13 &  1196 &  229.813 &  0.560 &  11.790 &  0.207 &  0.279 &    0.029 &  -1.981 &     0.430 &   0.877 &      0.492 &   171 &  0.0080 &  12.8600 &  0.1700 &  1.5 &  0.46 &   0.0 \\
14 &  1197 &  226.638 &  0.369 &  11.571 &  0.281 &  0.277 &    0.029 &  -1.704 &     0.432 &   0.701 &      0.441 &   100 &  0.0080 &  13.0000 &  0.2300 &  1.4 &  0.48 &   0.0 \\
15 &  1293 &  115.502 &  0.313 &  12.028 &  0.283 &  0.263 &    0.021 &  -2.031 &     0.588 &  -0.467 &      0.468 &   231 &  0.0300 &  12.7000 &  0.3100 &  2.8 &  0.55 &   0.0 \\
16 &  1299 &  149.949 &  0.309 &  12.246 &  0.197 &  0.271 &    0.024 &   0.662 &     0.290 &  -1.145 &      0.388 &    62 &  0.0100 &  13.2000 &  0.8200 &  1.2 &  0.55 &   0.0 \\
17 &  1301 &  159.115 &  0.316 &  11.887 &  0.334 &  0.267 &    0.027 &   0.745 &     0.346 &  -1.394 &      0.368 &   132 &  0.0200 &  13.0000 &  0.3000 &  1.5 &  0.53 &   0.0 \\
18 &  1314 &  218.772 &  0.407 &  13.309 &  0.713 &  0.281 &    0.031 &  -1.360 &     0.433 &  -0.115 &      0.462 &   567 &  0.0300 &  13.4500 &  0.0900 &  1.4 &  0.52 &   0.0 \\
19 &  1316 &  224.037 &  0.547 &  12.856 &  0.358 &  0.285 &    0.031 &  -1.492 &     0.461 &   0.252 &      0.534 &   477 &  0.0080 &  13.0000 &  0.1200 &  2.0 &  0.46 &   0.0 \\
20 &  1319 &  243.166 &  1.007 &  12.843 &  0.342 &  0.279 &    0.029 &  -2.763 &     0.552 &   1.887 &      0.633 &   737 &  0.0300 &  12.5500 &  0.0100 &  2.4 &  0.51 &   0.0 \\
21 &  1320 &  238.885 &  1.349 &  12.733 &  0.291 &  0.285 &    0.030 &  -2.589 &     0.553 &   1.648 &      0.681 &   931 &  0.0100 &  12.8000 &  0.1000 &  2.3 &  0.50 &   0.0 \\
22 &  1442 &  173.901 &  0.462 &  12.892 &  0.399 &  0.278 &    0.029 &   0.442 &     0.360 &  -1.839 &      0.303 &   287 &  0.0300 &  12.7000 &  0.0800 &  1.6 &  0.53 &   0.0 \\
23 &  1682 &  208.639 &  0.157 &  -3.006 &  0.123 &  0.632 &    0.031 &  -2.079 &     0.472 &  -0.410 &      0.418 &   120 &  0.0040 &  10.7500 &  0.9300 &  0.2 &  0.55 &   0.0 \\
24 &  1762 &  203.323 &  0.213 &   5.328 &  0.085 &  0.260 &    0.018 &  -0.275 &     0.256 &  -0.871 &      0.329 &    64 &  0.0040 &  12.8000 &  0.2600 &  0.9 &  0.19 &   0.0 \\
25 &  1764 &  244.071 &  0.333 &   5.684 &  0.260 &  0.255 &    0.023 &  -2.466 &     0.369 &   2.337 &      0.334 &   154 &  0.0300 &  13.5000 &  0.1200 &  1.1 &  0.55 &   0.0 \\
26 &  1789 &  163.055 &  0.196 &   8.818 &  0.235 &  0.276 &    0.024 &   0.676 &     0.320 &  -1.433 &      0.379 &    70 &  0.0080 &  13.7000 &  0.6500 &  1.4 &  0.54 &   0.0 \\
27 &  1790 &  180.195 &  0.094 &   8.216 &  0.784 &  0.308 &    0.035 &   0.300 &     0.647 &  -1.780 &      0.566 &   500 &  0.0200 &  12.6000 &  0.2800 &  2.7 &  0.50 &   0.0 \\
28 &  1837 &  154.865 &  0.293 &  17.445 &  0.556 &  0.356 &    0.028 &   1.095 &     0.389 &  -1.980 &      0.485 &    77 &  0.0300 &  13.6000 &  0.4000 &  0.7 &  1.00 &   0.0 \\
29 &  1911 &  181.198 &  1.060 & -12.631 &  0.221 &  0.472 &    0.019 &   0.882 &     0.517 &  -1.827 &      0.468 &   182 &  0.0200 &  11.7000 &  0.4800 &  1.4 &  0.52 &   0.0 \\
30 &  1989 &  128.696 &  0.596 & -12.842 &  0.476 &  0.477 &    0.022 &  -1.147 &     0.763 &  -1.575 &      0.608 &   153 &  0.0300 &  12.1500 &  0.1900 &  1.3 &  0.51 &   0.0 \\
31 &  2071 &  193.767 &  4.119 &  -9.211 &  0.861 &  0.935 &    0.039 &   1.017 &     0.787 &  -3.846 &      0.911 &  1294 &  0.0300 &  10.5000 &  0.3200 &  0.7 &  0.53 &   0.0 \\
32 &  2171 &  304.203 &  0.360 &  -5.247 &  0.266 &  0.462 &    0.014 &  -7.058 &     0.790 &  -1.233 &      0.512 &   213 &  0.0040 &  11.6500 &  0.3700 &  1.2 &  0.60 &   0.0 \\
33 &  2226 &  232.338 &  0.281 &  -1.304 &  0.322 &  0.467 &    0.015 &  -1.400 &     0.463 &   1.154 &      0.666 &   141 &  0.0040 &  10.8000 &  0.1200 &  2.7 &  0.51 &   0.0 \\
34 &  2234 &  243.248 &  0.274 &  -1.754 &  0.136 &  0.460 &    0.014 &  -2.160 &     0.497 &   2.294 &      0.573 &   128 &  0.0080 &  11.4500 &  0.2600 &  0.2 &  0.51 &   0.0 \\
35 &  2394 &  240.714 &  1.059 &   8.231 &  0.170 &  0.467 &    0.015 &  -2.932 &     0.586 &   1.612 &      0.646 &    92 &  0.0080 &  12.2200 &  0.2200 &  1.8 &  0.49 &   0.0 \\
36 &  2396 &  256.331 &  1.286 &   6.358 &  0.336 &  0.465 &    0.016 &  -3.915 &     0.640 &   2.713 &      0.643 &   259 &  0.0200 &  12.2000 &  0.3500 &  1.6 &  0.49 &   0.0 \\
37 &  2473 &  185.104 &  1.048 &  12.008 &  0.363 &  0.463 &    0.014 &  -0.044 &     0.670 &  -2.012 &      0.534 &   206 &  0.0300 &  12.5000 &  0.1200 &  1.4 &  0.52 &   0.0 \\
38 &  2474 &  181.315 &  0.901 &  10.702 &  0.251 &  0.461 &    0.016 &   0.217 &     0.654 &  -2.256 &      0.593 &   181 &  0.0080 &  12.4138 &  0.3286 &  1.9 &  0.50 &   0.0 \\
39 &  2595 &  139.851 &  0.821 &  17.152 &  0.933 &  0.463 &    0.015 &   0.591 &     0.632 &  -0.428 &      0.756 &   281 &  0.0080 &  12.4138 &  0.2857 &  1.8 &  0.55 &   0.0 \\
40 &  2601 &  194.069 &  0.491 &  19.241 &  0.606 &  0.467 &    0.016 &  -0.612 &     0.614 &  -2.152 &      0.650 &    88 &  0.0100 &  11.9138 &  0.1786 &  1.3 &  0.41 &   0.0 \\
41 &  2824 &  272.186 &  0.520 &  -6.925 &  0.201 &  0.695 &    0.018 &  -4.134 &     0.747 &   6.307 &      0.599 &    86 &  0.0300 &  11.6900 &  0.3600 &  0.4 &  0.58 &   0.0 \\
42 &  2920 &  133.222 &  0.758 &  -0.719 &  0.191 &  0.679 &    0.023 &  -0.758 &     0.708 &  -0.547 &      0.679 &   120 &  0.0010 &  10.3200 &  0.8200 &  0.1 &  0.45 &   0.0 \\
43 &  2980 &  208.073 &  1.021 &   8.282 &  0.325 &  0.684 &    0.017 &  -0.693 &     0.632 &  -2.592 &      0.537 &    96 &  0.0080 &  11.3276 &  0.2086 &  0.6 &  0.53 &   0.0 \\
44 &  2983 &  229.383 &  1.271 &   8.249 &  0.438 &  0.691 &    0.022 &  -2.330 &     0.601 &   0.295 &      0.605 &   141 &  0.0010 &   9.8000 &  0.0000 &  2.3 &  0.44 &   0.0 \\
45 &  3249 &  330.295 &  2.754 & -14.026 &  1.046 &  1.995 &    0.119 &  -0.713 &     1.059 &  -8.323 &      1.208 &   412 &  0.0200 &   9.3000 &  0.2000 &  0.0 &  0.45 &   0.0 \\
46 &  3263 &   34.712 &  0.264 & -11.549 &  0.238 &  1.672 &    0.072 &  -0.253 &     0.623 &  -4.617 &      0.963 &   144 &  0.0003 &   8.0000 &  0.4500 &  1.9 &  0.37 &   0.0 \\
47 &  3286 &  180.626 &  1.076 &  -8.961 &  0.574 &  0.926 &    0.032 &   1.463 &     0.682 &  -4.744 &      0.814 &   167 &  0.0100 &  11.2000 &  0.7000 &  0.9 &  0.42 &   0.0 \\
48 &  3312 &  107.309 &  2.886 &  -9.992 &  0.353 &  1.828 &    0.083 &   2.677 &     0.961 &  -2.107 &      0.968 &   295 &  0.0080 &   8.7800 &  0.2400 &  0.0 &  0.54 &   0.0 \\
49 &  3323 &  260.589 &  0.867 &  -7.483 &  0.147 &  1.677 &    0.070 &  -6.240 &     0.578 &   7.108 &      0.444 &    57 &  0.0001 &   8.1100 &  0.1000 &  0.0 &  0.55 &   0.0 \\
50 &  3344 &  217.277 &  0.672 &  -4.718 &  0.254 &  0.915 &    0.041 &  -0.418 &     0.730 &  -1.022 &      0.837 &   253 &  0.0040 &  10.0500 &  0.2300 &  0.4 &  0.52 &   0.0 \\
51 &  3345 &  229.477 &  0.468 &  -3.382 &  0.261 &  0.933 &    0.030 &  -2.024 &     0.990 &   1.085 &      0.877 &    80 &  0.0100 &  10.0900 &  0.2200 &  0.2 &  0.49 &   0.0 \\
52 &  3372 &  121.932 &  1.318 &  -5.190 &  0.390 &  0.947 &    0.036 &   0.324 &     1.054 &  -1.897 &      0.647 &   116 &  0.0300 &  12.0000 &  0.7500 &  0.1 &  0.51 &   0.0 \\
53 &  3373 &  114.189 &  1.387 &  -4.546 &  0.377 &  0.955 &    0.040 &  -0.500 &     1.081 &  -1.575 &      0.688 &   294 &  0.0200 &  10.1900 &  0.2800 &  0.2 &  0.54 &   0.0 \\
54 &  3411 &  318.974 &  0.460 &   1.480 &  0.294 &  0.940 &    0.065 &  -3.754 &     0.892 &  -3.030 &      0.689 &   200 &  0.0300 &  11.5000 &  0.8000 &  0.0 &  0.54 &   0.0 \\
55 &  3431 &  158.929 &  1.345 &   1.741 &  0.369 &  0.940 &    0.032 &   0.762 &     0.844 &  -3.722 &      1.043 &   251 &  0.0200 &  11.0200 &  0.6000 &  0.0 &  0.54 &   0.0 \\
56 &  3434 &  182.507 &  0.918 &   6.993 &  0.365 &  0.924 &    0.031 &   0.853 &     0.766 &  -4.916 &      0.809 &    74 &  0.0080 &  12.3900 &  1.0000 &  0.4 &  0.51 &   0.0 \\
57 &  3449 &   37.349 &  2.276 &   5.380 &  0.626 &  2.034 &    0.139 &   0.835 &     1.106 &  -4.849 &      1.083 &   998 &  0.0080 &   8.2000 &  0.1800 &  1.2 &  0.49 &   0.7 \\
58 &  3492 &  239.518 &  1.990 &  12.466 &  0.468 &  0.954 &    0.038 &  -4.599 &     0.987 &   1.181 &      0.902 &   160 &  0.0010 &  11.0000 &  0.5800 &  0.5 &  0.46 &   0.7 \\
59 &  3503 &    2.765 &  1.414 &  12.353 &  0.321 &  0.949 &    0.035 &  -0.982 &     0.797 &  -3.278 &      0.880 &    98 &  0.0080 &  10.5000 &  0.4500 &  1.3 &  0.54 &   0.0 \\
60 &  3517 &  154.397 &  1.886 &  12.945 &  0.799 &  2.008 &    0.127 &  -2.632 &     1.173 &  -3.336 &      1.255 &   439 &  0.0200 &   9.0000 &  0.5500 &  0.0 &  0.55 &   0.0 \\
61 &  3611 &  191.806 &  0.794 & -18.879 &  0.680 &  2.540 &    0.117 &   0.891 &     0.593 &  -4.546 &      0.631 &    90 &  0.0080 &   9.0000 &  0.6000 &  0.0 &  0.55 &   0.0 \\
62 &  3613 &  203.583 &  0.435 & -24.004 &  0.556 &  2.636 &    0.107 &   1.126 &     0.426 &  -0.804 &      0.337 &    91 &  0.0080 &   6.3500 &  0.2000 &  0.0 &  0.54 &   0.0 \\

63 &  3629 &  333.667 &  2.491 & -18.349 &  0.869 &  1.895 &    0.086 &  -0.223 &     1.365 &  -8.290 &      2.045 &   136 &  0.0010 &   8.3500 &  0.2000 &  0.6 &  0.53 &   0.0 \\
64 &  3630 &  322.244 &  2.296 & -17.662 &  0.611 &  1.881 &    0.072 &  -2.512 &     1.335 &  -8.968 &      1.664 &    70 &  0.0080 &   9.5000 &  0.3800 &  0.9 &  0.54 &   0.0 \\
65 &  3645 &  186.027 &  1.205 & -11.107 &  0.257 &  2.696 &    0.120 &   0.039 &     0.686 &  -5.714 &      0.709 &    73 &  0.0200 &   8.3000 &  0.3200 &  0.3 &  0.44 &   0.0 \\
66 &  3646 &  204.026 &  1.236 & -12.201 &  0.269 &  2.442 &    0.105 &  -1.294 &     0.935 &  -1.902 &      1.008 &   107 &  0.0040 &   7.0900 &  0.2000 &  0.0 &  0.52 &   0.0 \\
67 &  3649 &  200.728 &  1.324 & -12.402 &  1.181 &  1.877 &    0.079 &   0.896 &     1.125 &  -4.970 &      1.179 &    84 &  0.0200 &   9.0500 &  0.2600 &  1.1 &  0.44 &   0.0 \\
68 &  3661 &  132.378 &  2.325 & -13.541 &  1.389 &  1.865 &    0.073 &   0.919 &     2.218 &  -4.755 &      1.305 &   386 &  0.0100 &   8.5200 &  0.0600 &  2.9 &  0.47 &   0.0 \\
69 &  3664 &  180.910 &  0.634 & -12.323 &  1.115 &  2.002 &    0.116 &   1.348 &     1.420 &  -6.050 &      1.522 &    67 &  0.0040 &   8.2000 &  0.5000 &  0.2 &  0.52 &   0.0 \\

70 &  3668 &  260.326 &  1.890 &  -7.975 &  0.612 &  1.794 &    0.059 &  -6.223 &     0.690 &   7.063 &      0.954 &   217 &  0.0200 &   8.9000 &  0.0100 &  0.2 &  0.36 &   0.0 \\

71 &  3673 &  322.474 &  4.682 &  -8.374 &  0.914 &  1.876 &    0.106 &  -3.538 &     1.844 &  -6.227 &      1.681 &   849 &  0.0300 &   8.6000 &  0.0100 &  0.3 &  0.51 &   0.0 \\
72 &  3682 &  234.725 &  5.033 &  -6.576 &  1.207 &  1.826 &    0.070 &  -4.547 &     1.398 &   2.989 &      1.894 &   290 &  0.0100 &   9.1000 &  0.2400 &  0.6 &  0.45 &   0.0 \\
73 &  3715 &  325.816 &  1.271 &   2.812 &  0.966 &  1.863 &    0.054 &  -5.655 &     1.297 &  -6.435 &      1.392 &   142 &  0.0300 &   8.9000 &  0.0500 &  0.4 &  0.47 &   0.0 \\

74 &  3723 &   83.173 &  2.052 &   6.428 &  0.711 &  2.903 &    0.137 &   3.468 &     0.836 &   1.766 &      0.717 &   503 &  0.0080 &   7.3900 &  0.1050 &  0.0 &  0.48 &   0.0 \\
75 &  3728 &  329.039 &  5.760 &   9.203 &  1.238 &  1.872 &    0.084 &  -6.424 &     2.298 &  -5.233 &      1.565 &  1260 &  0.0100 &   8.7600 &  0.1100 &  0.5 &  0.53 &   0.0 \\
76 &  3732 &   63.094 &  1.590 &   4.935 &  0.499 &  1.886 &    0.061 &   0.792 &     1.087 &  -2.900 &      1.550 &    71 &  0.0300 &  10.7931 &  0.7000 &  0.0 &  0.31 &   0.0 \\
77 &  3737 &  306.166 &  7.076 &  12.106 &  0.479 &  1.875 &    0.060 & -12.042 &     1.777 &  -2.801 &      2.040 &   104 &  0.0001 &   9.5000 &  0.8000 &  0.0 &  0.48 &   0.0 \\

78 &  3749 &  310.208 &  4.242 &  11.992 &  0.762 &  1.887 &    0.061 & -12.021 &     1.830 &  -3.862 &      1.386 &   104 &  0.0300 &  10.0000 &  0.4500 &  0.3 &  0.49 &   0.0 \\
79 &  3755 &   54.403 &  0.836 &  17.142 &  0.575 &  2.810 &    0.145 &  -0.147 &     1.163 &  -5.460 &      0.898 &   200 &  0.0200 &   8.0000 &  0.2000 &  0.0 &  0.47 &   0.0 \\
80 &  3759 &  138.326 &  1.395 &  11.615 &  0.777 &  1.890 &    0.089 &  -1.238 &     1.425 &  -1.246 &      1.663 &    96 &  0.0080 &   8.5500 &  0.4000 &  0.6 &  0.42 &   0.0 \\

81 &  3783 &  256.616 &  2.702 &  12.670 &  0.557 &  3.848 &    0.126 & -20.561 &     1.316 &  15.861 &      1.607 &    59 &  0.0040 &   8.0000 &  0.5200 &  0.2 &  0.23 &   0.0 \\
82 &  3785 &   48.633 &  3.613 &  20.382 &  0.577 &  3.553 &    0.128 &  -1.400 &     1.090 &  -7.091 &      1.731 &    98 &  0.0040 &   7.1000 &  0.3000 &  0.0 &  0.47 &   0.0 \\
\bottomrule
\end{tabular}
}
\begin{tablenotes}
	\tiny	
	\item
	\begin{flushleft}
Note: For each cluster, seven parameters, i.e., distance modulus $(m - M)$,
color excess $E(V - I)$,
young stellar age $t$, age spread
$t_{sp}$
, binary fraction $f_{bin}$ and rotating star fraction $f_{rot}$, are determined.  The $n_{member}$ means the total number of member star in each cluster.
	\end{flushleft}
	\end{tablenotes}
\label{tab:param_tab}
\end{table*}

\section{Conclusions}
\label{sec:conclusion}

In this work, we presented an improved method-based HDBSCAN algorithm, e-HDBSCAN, for open cluster hunting.
This method uses only the astrometric measurements
from Gaia eDR3 to blind search for clusters. This technique consists of three steps described in Section ~\ref{sec:method}, which agrees with most other works.
When we apply this method to the data of Gaia eDR3, 3787 potential OCs are found, of which 83  are newly detected.
It suggests that e-HDBSCAN is more effective than DBSCAN in searching for Galactic OCs.
In addition, the fundamental parameters of newly detected clusters are determined by fitting CMDs with the ASPS model.
Besides distance modulus, colour excess, age , metallicity and the fractions of binary stars and rotating stars are obtained in this work.
Although e-HDBSCAN found some new OCs, further confirmation is needed.
For example, these OC candidates can be confirmed with radial velocities if their member stars are available in Gaia DR3.

The results are summarised as follows:
\begin{enumerate}
    \item  In the work, our method achieves promising results  and suggests that the proposed three-step clustering method (e-HDBSCAN)  is efficient in star cluster searches.
% is metal-poor ?
    \item More than half of the newly discovered OCs are young clusters (age $<$ 0.5Gyr), and most of them are metal-poor ( metallicity  Z $<$ 0.03).

    \item Although more than 3,000 OCs have been reported so far, the search for Gaia disk clusters needs to continue, especially with the release of updated Gaia data.

    \item It is of great interest to try different methods to search for OCs since different methods have their own merits and shortcomings.
\end{enumerate}

\begin{acknowledgements}
This work is supported by the National SKA Program of China No 2020SKA0110300, Joint Research Fund in Astronomy (U1831204) under cooperative agreement between the National Natural Science Foundation of China (NSFC) and the Chinese Academy of Sciences (CAS),  the  National Key Research and Development Program of China (2018YFA0404603), the National Natural Science Foundation of China (Nos. 11863002) the National Natural Science Foundation of China (Nos. 11961141001).  The authors acknowledge that this work has been supported by the Yunnan Academician Workstation of Wang Jingxiu (No. 202005AF150025), China Manned Space Project with NO.CMS-CSST-2021-A08 and Sino-German Cooperation Project (No. GZ 1284).

This work has made use of data from the European Space Agency (ESA) mission Gaia
\href{https://www.cosmos.esa.int/gaia}{https://www.cosmos.esa.int/gaia}, processed by the Gaia Data Processing and Analysis Consortium (DPAC,
 \href{https://www.cosmos.esa.int/web/gaia/dpac/consortium}{https://www.cosmos.esa.int/web/gaia/dpac/consortium}). Funding for the DPAC has been provided by national
institutions, in particular the institutions participating in the Gaia Multilateral Agreement.
\end{acknowledgements}

\label{lastpage}
\bibliographystyle{raa}
\bibliography{context}
\end{CJK*}
\end{document}